\documentclass[12 pt]{article}
\usepackage{amssymb}
\usepackage{latexsym}
\usepackage{makeidx}  
\usepackage{amsfonts}
\usepackage{amsmath}
\usepackage{graphicx}
\usepackage{subfigure}

\begin{document}

\markboth{Yang} {Prospect Agents and Volatility Smile}

\title{Prospect Agents and the Feedback Effect on Price Fluctuations}

\author{Yipeng Yang\thanks{Department of Mathematics, University of
Missouri-Columbia, USA,  yangyip@missouri.edu} \and Allanus Tsoi
\thanks{Department of Mathematics, University of Missouri-Columbia,
USA,  tsoia@missouri.edu}}

\maketitle

\begin{abstract}
A microeconomic approach is proposed to derive the fluctuations of
risky asset price, where the market participants are modeled as
prospect trading agents. As asset price is generated by the
temporary equilibrium between demand and supply, the agents' trading
behaviors can affect the price process in turn, which is called the
feedback effect. The prospect agents make actions based on their
reactions to gains and losses, and as a consequence of the feedback
effect, a relationship between the agents' trading behavior and the
price fluctuations is constructed, which explains the implied
volatility skew and smile observed in actual market.
\end{abstract}

\section{Introduction}
In financial mathematics, the widely used model of the price
fluctuation for the underlying risky asset is the Geometric Brownian
motion, also referred to as the Black-Scholes-Merton model,
\begin{equation}\label{gbm} dP_t=\mu P_tdt+\sigma P_tdW_t, \end{equation} where $P_t$ is
the asset price, $\mu$ and $\sigma$ are drift and volatility
respectively, and $W_t$ is a standard Brownian Motion. Given a set
of historical data, the parameters of this diffusion process are
estimated through calibration, where statistical analysis is used to
test the goodness of fit. This model builds the fundamental of an
important branch of financial study, and facilitates the research of
many financial problems, like option pricing, portfolio
optimization, investment and hedging.

It is a broadly observed phenomenon that the fixed volatility in
(\ref{gbm}) alone can not explain the implied volatility smile or
skew observed from real market data. The volatility of actual market
return often shows a stochastic property with other observations,
for example, it is mean reverting but persisting, and it is usually
correlated with asset price shocks. These phenomena boost the study
of stochastic volatility, like the $GARCH(p,q)$ model in discrete
time \cite{Engle82} and stochastic volatility models in continuous
time \cite{Heston93}\cite{Taylor94} \cite{Fouque00}. Another major
direction to study the volatility smile/skew phenomenon is to  use
the regime switching models, see, e.g., \cite{Bollen98} \cite{Yao06}
\cite{Papanicolaou13}.

Mathematical models make the analytical solutions of some financial
problems possible, although they lack the explanation of why the
asset prices fluctuate in that way. For example, if the price
process is modeled as regime switching process due to the switching
of market or economic states, then it is reasonable to construct the
same model for price process before the year 1987 since market
states kept changing all the time. However, the volatility
smile/skew phenomenon is not seen from the market data before 1987.
Recently researchers have turned to a microeconomic approach to
model the asset price fluctuations by noticing the fact that prices
are after all generated by demand and supply of market participants,
and the fluctuations are due to the temporary imbalance between
demand and supply. Because volatility smile/skew is only seen from
the market data after 1987, it is reasonable to argue that the crash
in 1987 fundamentally changed the behaviors of investors. By
modeling different types of participants, different models for the
 price process can be derived. In \cite{Follmer93} and also in
\cite{Horst05}, the authors presented an agent based model. The
agent's (denoted $a$) excess demand $z_t^a(P)$ of a certain asset at
time $t$ is obtained by comparing the proposed price $P$ with some
individual reference level $P^a_t$ which the agent adopts for that
period, and it takes the form $z_t^a(P)=\log P_t^a-\log P$. The
market clearing condition states that the total excess demand equals
zero, i.e., $\sum_a z_t^a(P)=0$. In their work, the agents trading
behavior consists of a fundamental component and a trend chasing
component. The reference of a fundamentalist is given by
\begin{displaymath}
\log P_t^a=\log P_{t-1}+c_F(\log F-\log P_{t-1}),\ \ c_F>0,
\end{displaymath}
and the idea is that the agent believes that the asset price will
finally go to its fundamental value $F$.  While by the trend chasing
component, the agent believes that the asset price forms a trend and
hence the reference is given by
\begin{displaymath}
\log P_t^a=\log P_{t-1}+c_C(\log P_t-\log P_{t-1}),\ \ c_C>0.
\end{displaymath}
The final trading decision is a random combination of these two
components, and the price is thus determined by the market clearing
condition. A limiting diffusion model is derived, which is a random
regime switching dynamic process. An ergodicity theory of the price
process about this model is given in \cite{Follmer05}. However, the
properness of assuming these components still needs further
investigation.

The interaction between traders' behavior and the asset price is
usually called the \emph{feedback effect}, see, e.g.,
\cite{Subrahmanyam01} for a case where stock price levels affect the
firm's cash flow, \cite{Hirshleifer06} where traders make profit by
feedback effect, and \cite{Khanna04} and \cite{Goldstein08} for
cases where price plays a manipulation role. It is an interesting
research topic to explain the stochastic volatility of asset price
through the microeconomic approach, i.e., through the behavior of
the market participants. Frey and Stremme \cite{Frey97} modeled the
market participants as reference traders and program traders, where
the former make decisions based on their utility function, and
investigated the effect of dynamic hedging on volatility. They
showed that the heterogeneity of the distribution of hedged
contracts is one of the key determinants for the transformation of
volatility. Heemeijer {\it et al} \cite{Heemeijer09} showed that
when the economic agents have positive expectations about the market
development, due to the feedback effect, large fluctuations in
realized prices and persistent deviations from the fundamental are
likely; when they have negative expectations, prices converge
quickly to their equilibrium values. The price to volatility
feedback rate due to the traders' actions is discussed in
\cite{Barucci03}, and agent-based limiting models analyzed through
queueing theories can be found in \cite{Bayraktar07}. Ozdenoren and
Yuan \cite{Ozdenoren08} modeled the market participants by risk
neutral informed traders and risk averse uninformed traders, and
showed that feedback effects are a significant source of excess
volatility. According to the work of Danilova \cite{Danilova05}, the
main explanation of the volatility of asset returns is the volume of
trade. In order to explain the volatility smile and skewness effect,
Platen and Schweizer \cite{Platen98} assumed that the market
participants consist of arbitrage-based agents or speculators, and
hedgers or technical agents, and the volatility smile effect is
obtained through the Black-Scholes hedging of European call options
from the technical component. To be more specific, they assumed that
the cumulative demand of the asset up to time $t$ is given by
\begin{displaymath}
D(t,\log P_t,U_t)=U_t+\gamma(\log P_t-L_0)+\xi(t,\log P_t),\ \
\gamma\neq 0,
\end{displaymath} where $U_t=vW_t+mt$ is a Brownian motion with
drift $m$ and volatility $v$. The part $\gamma(\log P_t-L_0)$
corresponds to the demand of speculators where $L_0$ is a constant,
and $\xi(t,\log P_t)$ is the cumulative demand of hedgers. The
dynamics of $\log P_t$ was then obtained by differentiating, using
Ito's formula, the equation $D(t,\log P_t,U_t)=const.$, which is due
to the market clearing condition.

All the aforementioned work provide evidence that feedback effects
from the traders' behavior play an important role in the study of
price fluctuations, and the research through microeconomic approach
and feedback effects is very promising, although there is no
agreement yet what kind of market participants and what kind of
behavior there should be. It should be noticed that, besides the
noisy demand in the market, there are usually more than one types of
agents (traders or components) considered in the market. Observing
the phenomenon that, market participants often have expectations on
their investment, either arbitrage or hedging, and they behave
differently when they feel a gain vs a loss of their investments, we
present in this paper the \emph{prospect agents}. Through this model
we are able to derive a stochastic price process that explains the
volatility smile or skewness phenomena observed from  real market
data.

The rest of this paper is organized as follows: in Section
\ref{prosp} we introduce a kind of reference traders called
\emph{prospect traders} and model their trading behavior, then by
the feedback effect via market clearing condition, an ARCH model for
the yield process is derived in Section \ref{diff}. In Section
\ref{locreg} a local polynomial regression is performed using actual
S\&P 500 daily data, and the result is compared with our ARCH model.
In Section \ref{smile}, the volatility smile/skew
 effect is discussed with a numerical study
verifying the effectiveness of our model. This model is then applied
to foreign currency market in Section \ref{fcm} which could well
explain some phenomena observed from real market data, and Section
\ref{ending} summarizes our conclusions. In the Appendix we discuss
a numerical issue in computing the implied volatility through Monte
Carlo simulations.

\section{Price Process with Prospect Market Participants}
\subsection{Prospect Traders}\label{prosp} Asset price is determined by the temporary balance between
 demand and supply. More demand would push the price high and more
supply would push it down, with the market clearing condition being
held.  This effect is determined by the traders' actions. In this
approach, to understand the behavior of the interacting traders is
the key to model the price process.  In this section we shall derive
the asset price dynamics through feedback effect via market clearing
condition. Our method is different from \cite{Horst05} and
\cite{Platen98} in that we model the behavior of traders through a
different point of view.

\begin{figure}[h!]
\centerline{\includegraphics[height=2.5in]{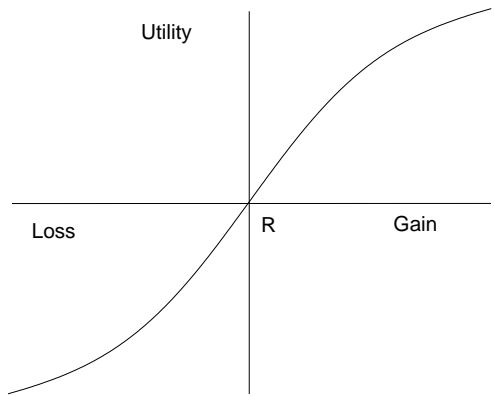}}
\vspace*{8pt} \caption{Prospect theory} \label{fu}
\end{figure}

Since Kahneman and Tversky \cite{Kahneman79} introduced the prospect
theory to economics, it has been playing an important role in
explaining many phenomena in economics and finance. The key idea of
this theory is that, in a risky situation, when people feel gains,
they show risk averse behavior and when people feel losses, they
show risk seeking behavior, see Figure \ref{fu}. And this theory can
be readily applied to investment.

It is a common phenomenon that the decisions of the investors - to
buy or to sell the asset -  depend on the gains or losses of this
asset. By means of gain and loss, we mean the difference between the
return $Y_t=\log P_t-\log P_{t-1}$ of the asset and a reference rate
of return, for example, a rate of return close to zero or comparable
to daily interest rate. That is, if $Y_t$ is bigger than a
threshold, investors feel gains of the investment, and if $Y_t$ is
less than this threshold, investors feel losses of their wealth.
These different feelings result in different trading behaviors. If
$Y_t$ is above the reference rate, the excessive demand of this
asset will increase. However, the investors who long the asset feel
a gain and some of them try to realize the profit by selling some
shares. In other words, they show a risk averse behavior. For
investors who short the asset, some of them feel that they have
already missed the chance to make profit, and so some of them do not
want to buy the asset. As a result, the excessive demand of this
asset does not increase linearly with $Y_t$ but shows a concave
pattern.

When $Y_t$ is lower than the reference return, the excessive
 demand of this asset will decrease. However, the investors will
show different behaviors. For holders, since they feel a loss, some
of them tend to hold rather than bail out. This can be explained by
their unwillingness to realize a loss, and so they hold in the hope
that they will have a gain later. That is, they show a risk seeking
behavior. For those who short the asset, some of them tend to buy
more of the asset due to the reduced price while they hold the hope
that the asset will gain soon. Consequently, the excessive demand
does not decrease linearly with $Y_t$ but shows a convex pattern.
Similar phenomenon has been observed in many articles, see, e.g.,
\cite{Shefrin85}. This phenomenon can be perfectly explained by the
prospect theory \cite{Kahneman79}. To be short, when the investors
feel a gain, they show risk averse behavior and when they feel a
loss, they show risk seeking behavior.

\begin{figure}[h!] \centering
{\includegraphics[width=2.5in, height=2in]{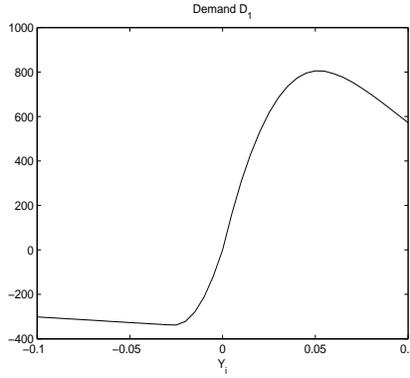}}
\caption{Excessive demand as a function of yield}\label{dem-p}
\end{figure}

We  shall extend the prospect theory to the extreme cases to model
the behavior of these traders. That is, when $Y_t$ is much bigger
(or much lower) than the threshold. If $Y_t$ is much bigger than the
threshold, for example, the stock price jumps by $8\%$ in one day,
then the investors show extreme risk averse behavior: more holders
tend to sell the shares to realize the profit, and less buyers are
willing to buy the shares since they feel having missed the best
chance. As a result, the net demand of this asset decreases. A
similar argument can be given to the case when the yield is way
below the threshold. Figure \ref{dem-p} shows a typical graph of the
excessive demand as a function of return. We used a piecewise
polynomial to represent this excessive demand function, $D_1(y)$, as
follows:
\begin{displaymath}
D_1(y)=\left\{\begin{aligned} &-352.1-504y, &\  y<-0.0286 \\
&2.63\times 10^4y+5.12\times 10^5y^2, &\ -0.0286\leqslant y < 0\\
&3.5\times 10^4y-4.5\times 10^5y^2+1.44\times 10^6y^3, & \ 0\leqslant y < 0.0648\\
&435.46+1.85\times 10^4y-2.66\times 10^5y^2+9.38\times 10^5y^3, & \
0.0648 \leqslant y
\end{aligned}\right.
\end{displaymath} Their cumulative demand is thus given by $\int_0^t D_1(Y_s)ds$.

To construct a discrete model, we use index $i$ to denote the sample
at time $t_i$. Then we let $D_1(Y_i)$ be the corresponding excessive
demand function at time $t_i$ of this type of traders.

\begin{figure}[h!] \centering
\subfigure[$D_2$]{\includegraphics[width=2.5in,height=2in]{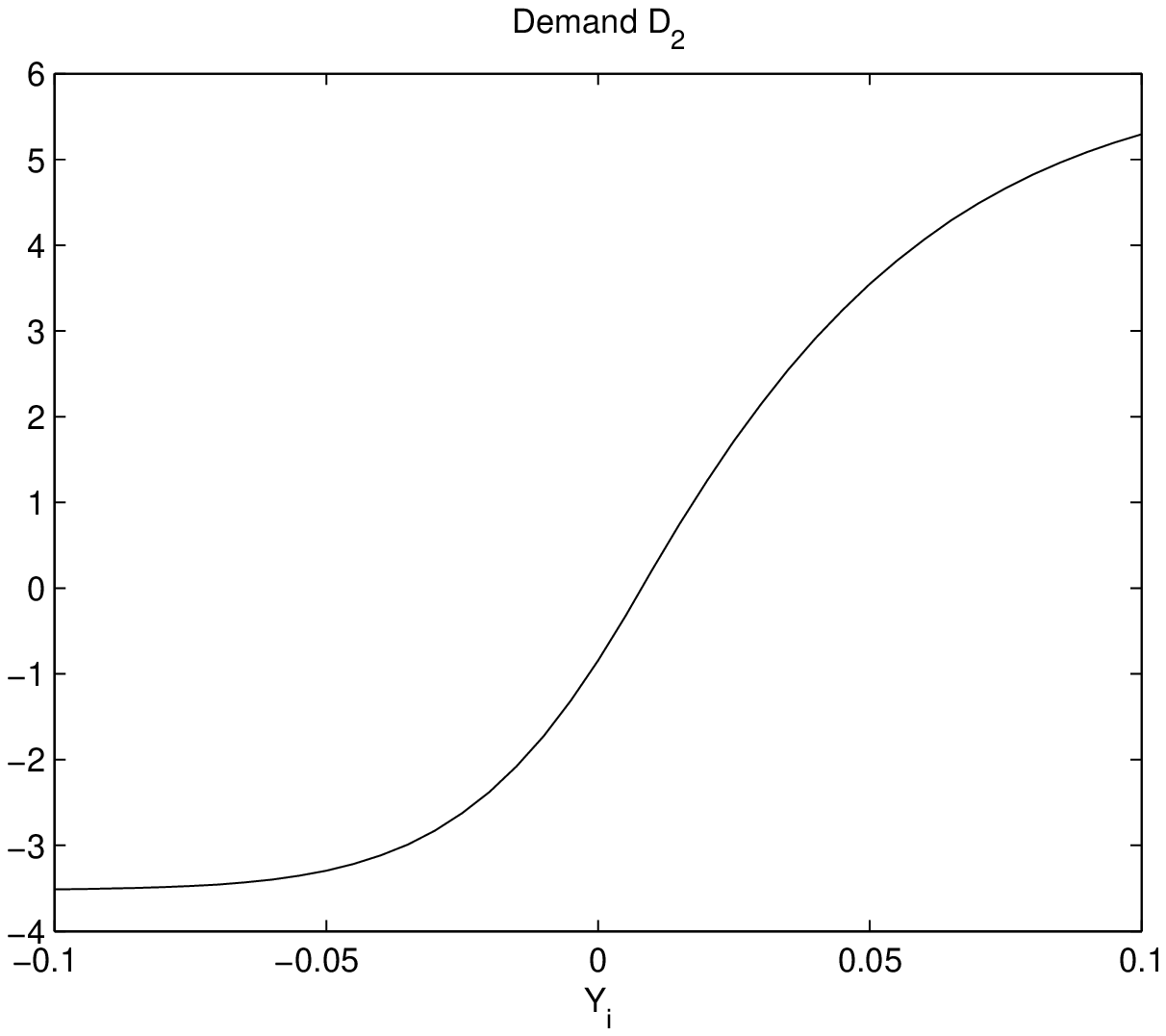}}
\subfigure[$D_2'$]{\includegraphics[width=2.5in,height=2in]{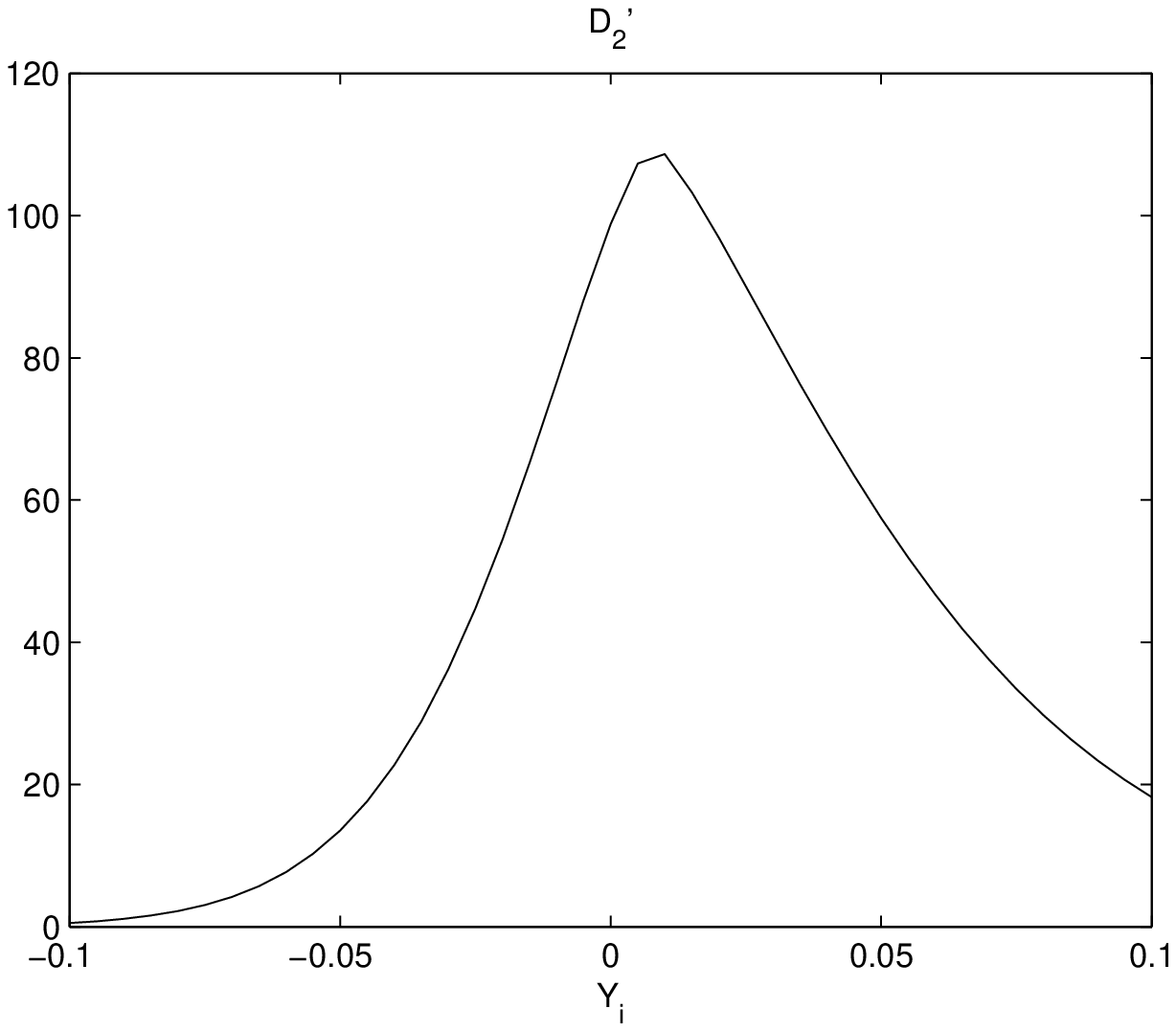}}
\caption{Cumulative demand as a function of yield}\label{D2}
\end{figure}

We assume that there is a second type of traders which are closely
related to the prospect theory. The cumulative demand of this type
of traders, $D_2$, is assumed to have a prospect pattern, as shown
in Figure \ref{D2}. Its derivative $D_2'$ is also drawn.

In our model, the function $D_2'$ is given by
\begin{displaymath}
D_2'(y)=\left\{\begin{aligned}&110e^{-150|x-0.008|^{1.5}},&\
x\leqslant 0.008,\\
&110e^{-40|x-0.008|^{1.3}},&\ x> 0.008,
\end{aligned}\right.
\end{displaymath} and $D_2(y)$ is the numerical integral of
$D_2'(y)$ with $D_2(0.008)=0$.

The change of demand in discrete time is given by
$D_2(Y_{i+1})-D_2(Y_i)$. If the first order approximation is used,
we get $D_2(Y_{i+1})-D_2(Y_i)\approx D_2'(Y_i)(Y_{i+1}-Y_i)$. That
is, their excessive demand of the asset is proportional to the
change of $Y_i$, and the rate equals $D_2'(Y_i)$. When $\Delta
Y_i>0$, their excessive demand increases, and when $\Delta Y_i<0$,
their excessive demand decreases, and the rate equals $D_2'(Y_i)$.
To be more specific, when $Y_i$ is already very high (or very low),
their excessive demand is not very sensitive to $\Delta Y_i$. One
explanation is that $Y_i$ is often thought to be a mean reverting
process, and when $Y_i$ is very high (or very low), the probability
that this return process stays high (or low) is small, and as a
consequence, the excessive demand of this type of traders is
insensitive to $\Delta Y_i$ in this case.

So far we have introduced the concept of prospect agents (traders),
which is novel to our knowledge. In the next section, by assuming
the existence of the usual noisy demand and trend chasing demand as
in \cite{Horst05}, we shall derive an ARCH model for $Y_i$.

\subsection{An ARCH Model for the Yield Process}\label{diff}

Let $\nu W_t$ be the cumulative noisy demand and $\xi \log P_t$ be
the cumulative trend chasing demand as in \cite{Horst05}, and $W_t$
is a Wiener process, $\nu<0$, $\xi>0$ are constants. The market
clearing condition states that
\begin{displaymath}
\nu W_t+\xi\log P_t+\int_0^t D_1(Y_t)dt+D_2(Y_t)=M,
\end{displaymath} where $M$ is a constant. Define $\Delta W_i=W_{t_{i+1}}-W_{t_i}$, $\Delta t =
t_{i+1}-t_i$, $i=0,1,...,n$, and $\Delta W_i$ is assumed to follow
the normal $N(0,\Delta t)$ distribution. Since we are interested in
the daily yield process, we choose $\Delta t=1/252$. The associated
difference model can be easily written as follows
\begin{equation}\label{diffm}
\nu\Delta W_i + \xi Y_{i+1}+  D_1(Y_i)\Delta t +
D_2'(Y_i)(Y_{i+1}-Y_i)=0,\ \ i=0,1,...,n,\ \ t=n\cdot\Delta t.
\end{equation}
Now it is a simple step to derive a recursive equation for $Y_i$ by
rewriting (\ref{diffm}) as
\begin{equation}\label{receqn}\begin{split}
Y_{i+1}&=\frac{D_2'(Y_i)Y_i- D_1(Y_i)\Delta t}{\xi+D_2'(Y_i)}+\frac{-\nu}{\xi+D_2'(Y_i)}\Delta W_i\\
&= \frac{D_2'(Y_i)Y_i- D_1(Y_i)\Delta t}{\xi+D_2'(Y_i)}+\frac{-\nu}{\sqrt{252}(\xi+D_2'(Y_i))}\sqrt{252}\Delta W_i\\
&=f(Y_i)+g(Y_i)\epsilon_i,
\end{split}
\end{equation} where
\begin{displaymath}f(Y_i)=\frac{D_2'(Y_i)Y_i- D_1(Y_i)\Delta t}{\xi+D_2'(Y_i)}, \ \ \  \ \ \
g(Y_i)=\frac{-\nu}{\sqrt{252}(\xi+D_2'(Y_i))},\end{displaymath} and
$\epsilon_i=\sqrt{252}\Delta W_i,\ i=1,2,...$ are i.i.d $N(0,1)$
random variables.

\begin{figure}[h!] \centering
\subfigure[$f$]{\includegraphics[width=2.5in,height=2in]{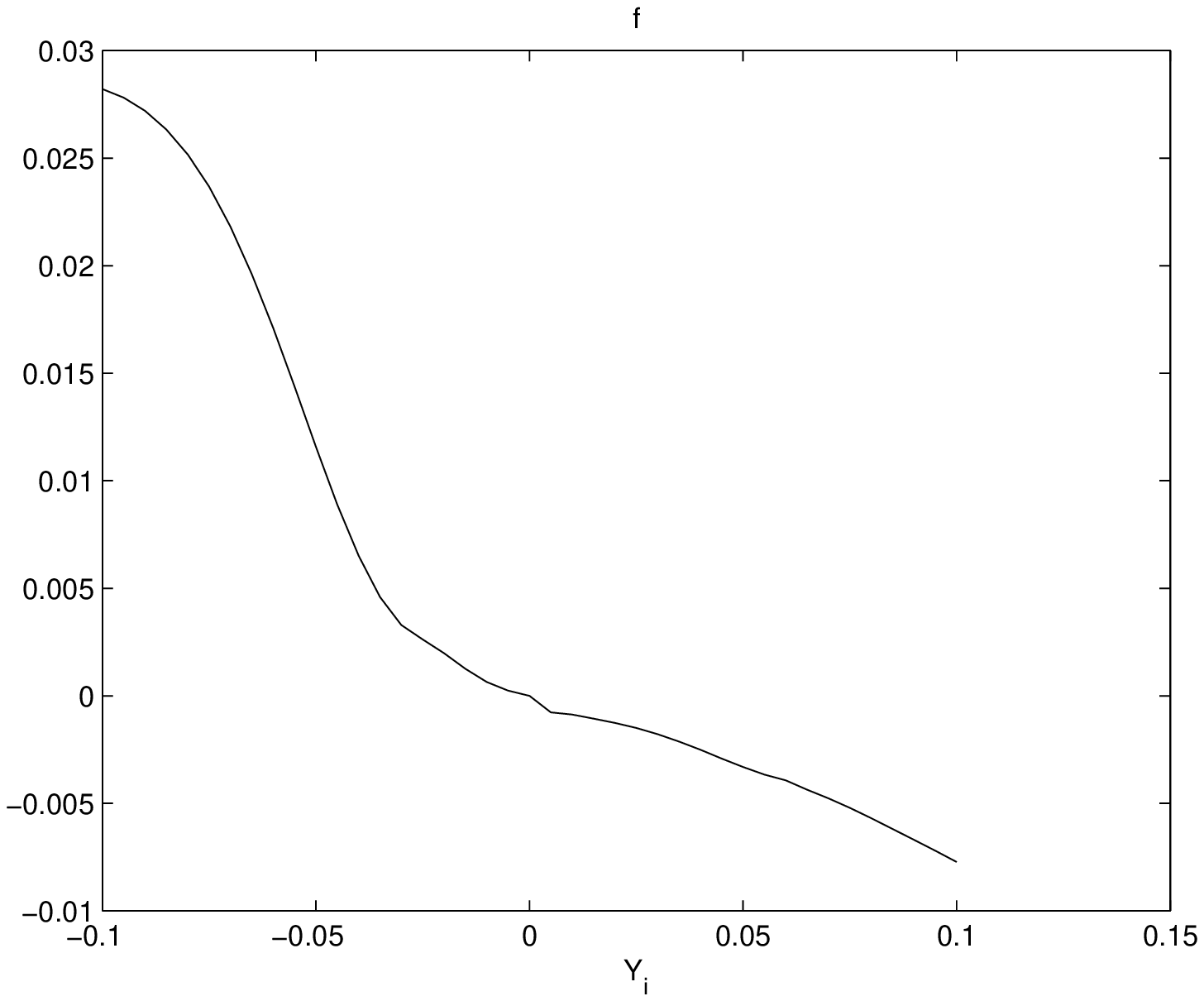}}
\subfigure[$g$]{\includegraphics[width=2.5in,height=2in]{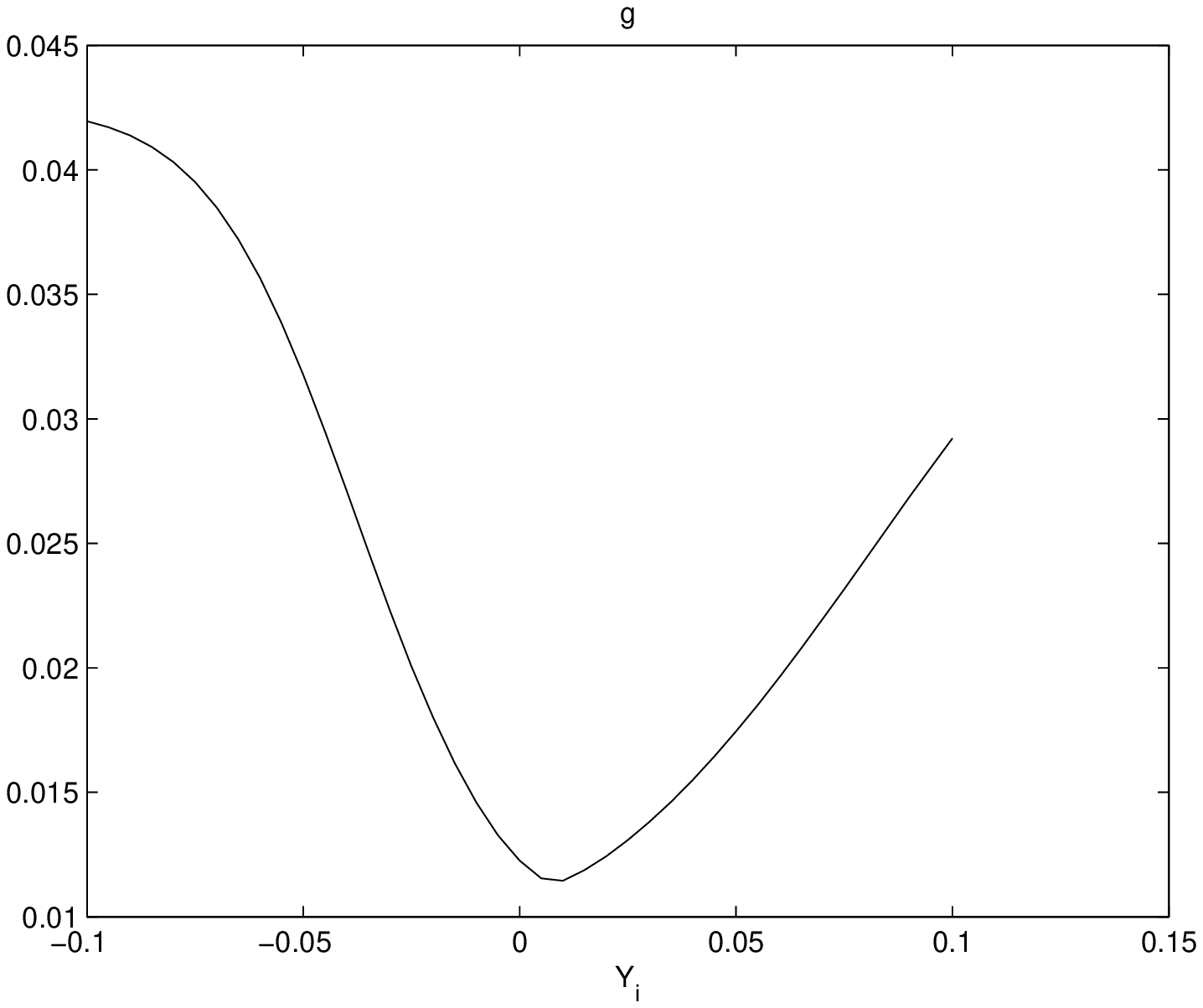}}
\subfigure[$g^2$]{\includegraphics[width=2.5in,height=2in]{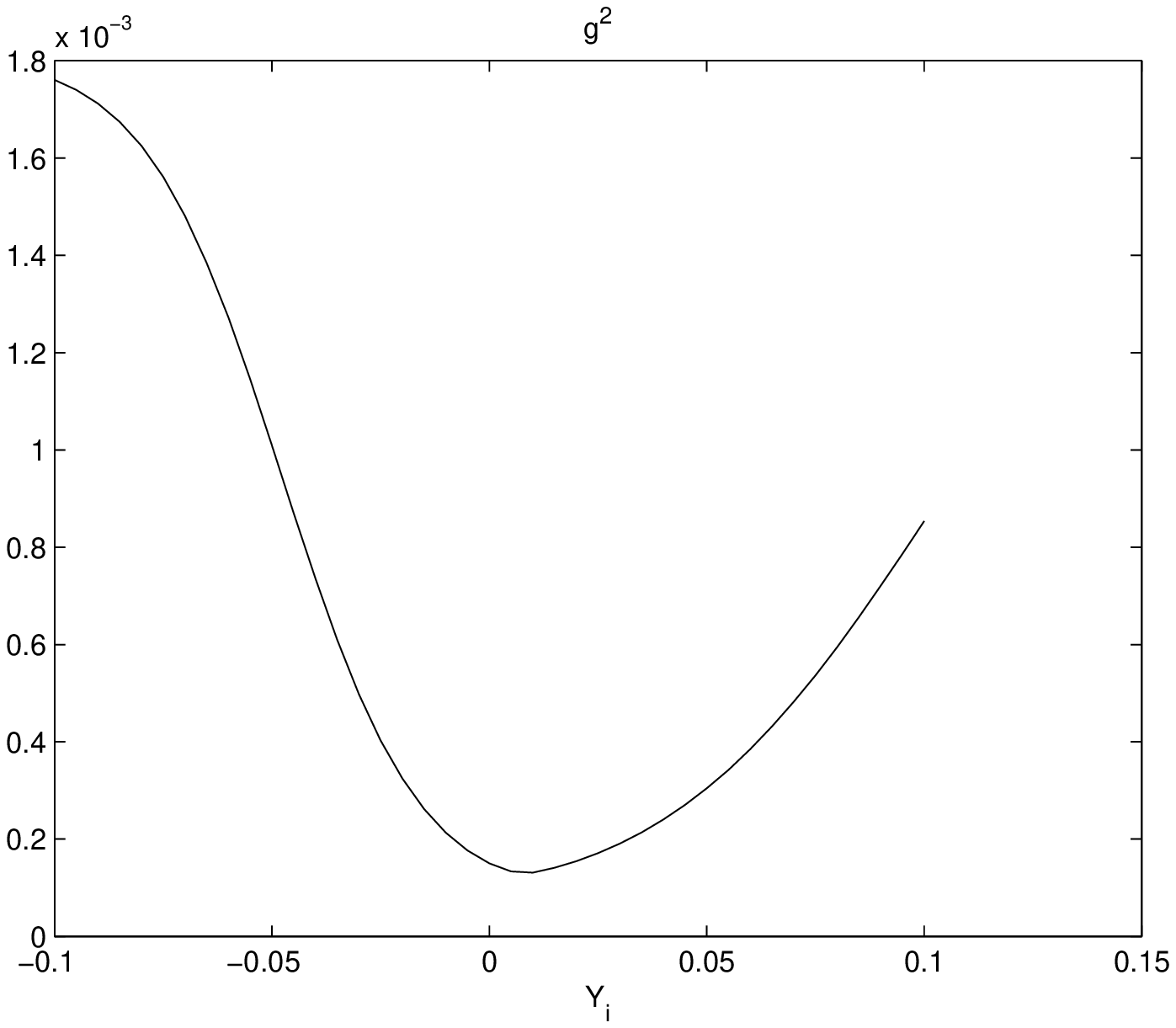}}
\caption{Drift and volatility functions}\label{dfg}
\end{figure}

For the given functions $D_1$ and $D_2'$ and the parameter settings
$\xi=40,\nu=-27$, we can plot the graphs of
 $f(\cdot)$, $g(\cdot)$ and $g^2(\cdot)$ as shown in Figure \ref{dfg}.

It can be easily seen from Figure \ref{dfg} that there are patterns.
For the function $f$, when $y>0$, $f(y)$ tends to be close to zero,
and when $y$ is bigger, $f(y)$ is slightly negative. When $y<0$ is
small, $f(y)$ tends to be positive, and if $y$ is way below zero,
$f(y)$ tends to be bigger. In other words, if $Y_i$ is positive, on
expectation, $Y_{i+1}$ will be negative, and if $Y_i$ is negative,
on expectation, $Y_{i+1}$ will be positive. For the volatility
$g^2$, its graph shows a `U-shaped' smiling face.

\section{Data Calibration}\label{locreg}
Local polynomial regression \cite{Tsybakov97} \cite{Fan97} is used
to estimate the drift (f) and volatility (g) functions. Let
$\{Y_t\}$ be a time series, we want to fit an ARCH model so that
\begin{displaymath}
Y_i=f(Y_{i-1})+g(Y_{i-1})\epsilon_i,
\end{displaymath} where $\epsilon_i$ are i.i.d. $N(0,1)$ random
variables. The procedure is as follows: we seek a function $g(x)$
which satisfies
\begin{displaymath}
g^2(x)=E(Y_i^2|Y_{i-1}=x)-E^2(Y_i|Y_{i-1}=x).
\end{displaymath}
Let $n$ be the size of $\{Y_t\}$, then for each $x$, consider the
following two minimization problems:
\begin{equation}\begin{split}
&[\alpha_1(x),\alpha_2(x),\alpha_3(x)]\\
&=arg\min_{\alpha_1,\alpha_2,\alpha_3}\sum_{i=1}^n\left(Y_i^2-\alpha_1-\alpha_2\left(\frac{Y_{i-1}-x}{h}\right)-\frac{\alpha_3}{2}\left(\frac{Y_{i-1}-x}{h}\right)^2\right)K\left(\frac{Y_{i-1}-x}{h}\right),
\end{split}
\end{equation}
\begin{equation}\begin{split}
&[\beta_1(x),\beta_2(x),\beta_3(x)]\\
&=arg\min_{\beta_1,\beta_2,\beta_3}\sum_{i=1}^n\left(Y_i-\beta_1-\beta_2\left(\frac{Y_{i-1}-x}{h}\right)-\frac{\beta_3}{2}\left(\frac{Y_{i-1}-x}{h}\right)^2\right)^2K\left(\frac{Y_{i-1}-x}{h}\right),
\end{split}
\end{equation}
where $K(\cdot)$ denotes a nonnegative weight function and $h$ is a
positive number called the bandwidth. We choose $K$ to be the
standard normal pdf. Then $g^2(x)$ is estimated by
$\hat{g}^2(x)=\alpha_1(x)-\beta_1^2(x)$, and the estimation of $f$
is  given by $\hat{f}(x)=\beta_1(x)$.

Let $\{P_i\}$ be the daily data of S\&P500 index where the last date
is 5/2/2013, and let $\{Y_i\}$ be the yield process, i.e., $Y_i=\log
P_i-\log P_{i-1}$. For each data  size $S$, we choose the bandwidth
$h$ to be $h=\{\max(Y_i)-\min(Y_i)\}/{\gamma}$, where all $Y_i$'s
belong to this set and $\gamma$ is a chosen positive constant. In
what follows we shall pick different data  sizes and different
values of $\gamma$, and plot the graphs of $\hat{f}$ and
$\hat{g}^2$.

\begin{figure}[h!] \centering
\subfigure[$\hat{f}$]{\includegraphics[width=2.5in,
height=2in]{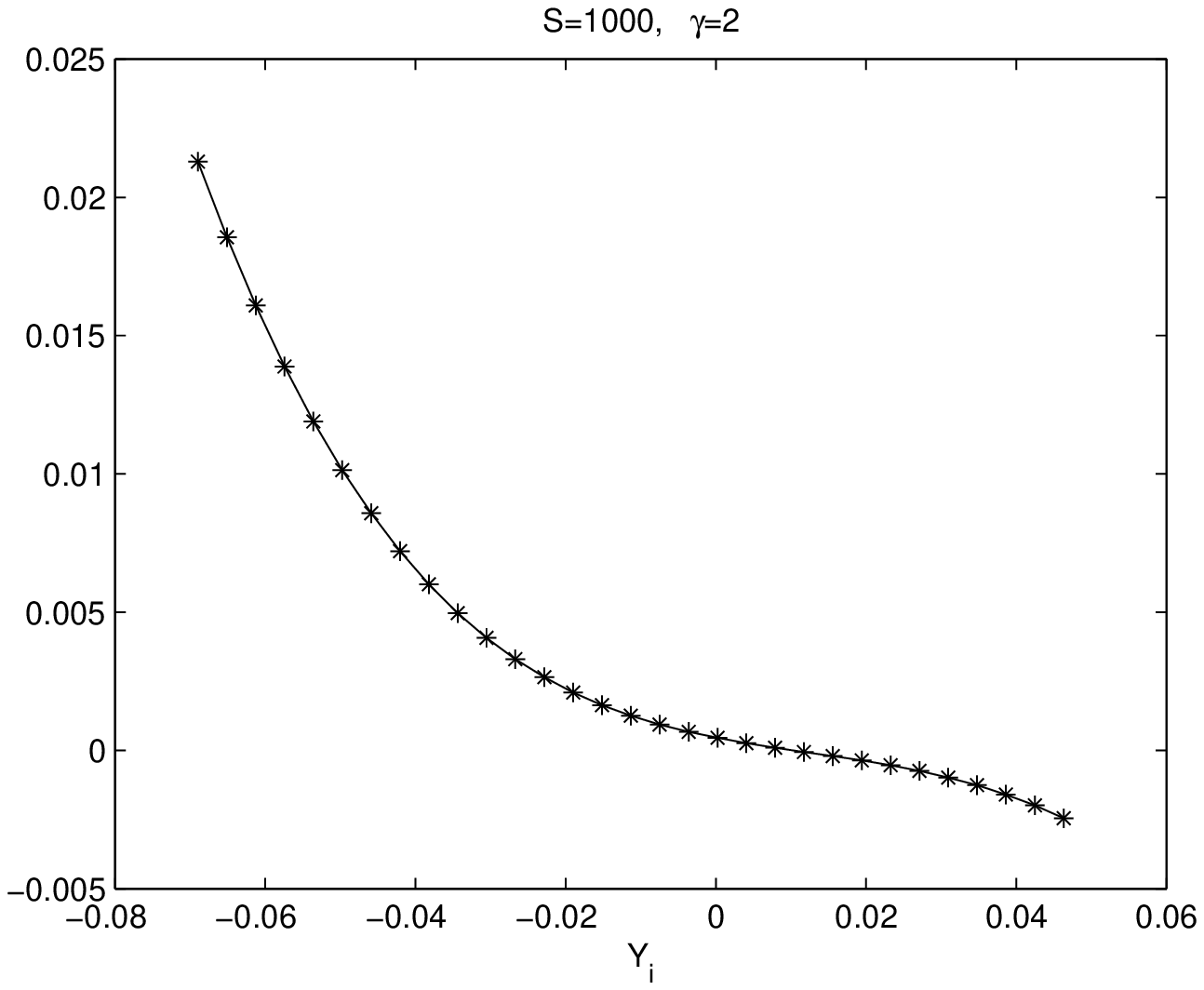}}
\subfigure[$\hat{g}^2$]{\includegraphics[width=2.5in,height=2in]{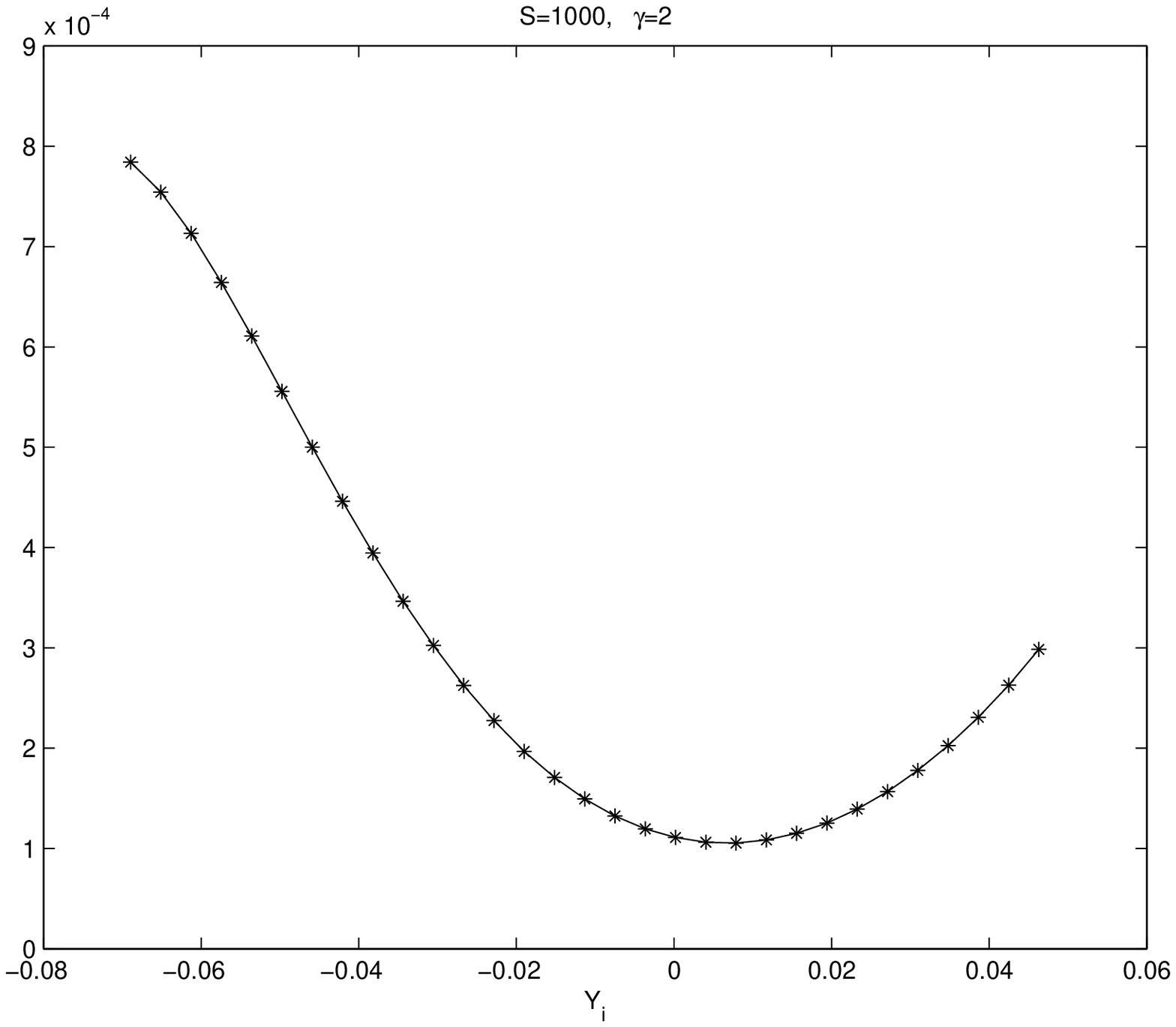}}
\caption{$S= 1000, \gamma=2$}\label{2-1000}
\end{figure}

\begin{figure}[h!] \centering
\subfigure[$\hat{f}$]{\includegraphics[width=2.5in,
height=2in]{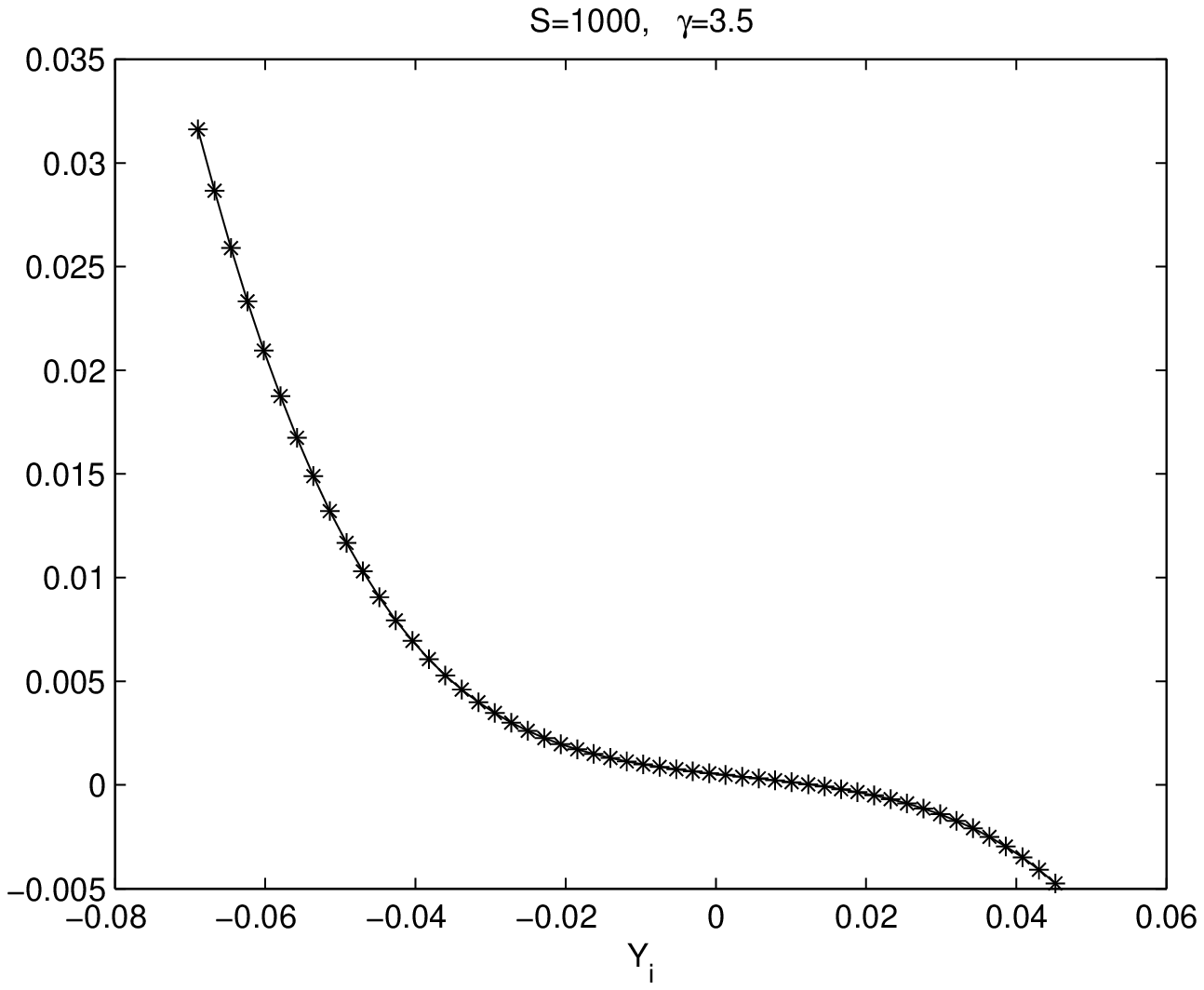}}
\subfigure[$\hat{g}^2$]{\includegraphics[width=2.5in,height=2in]{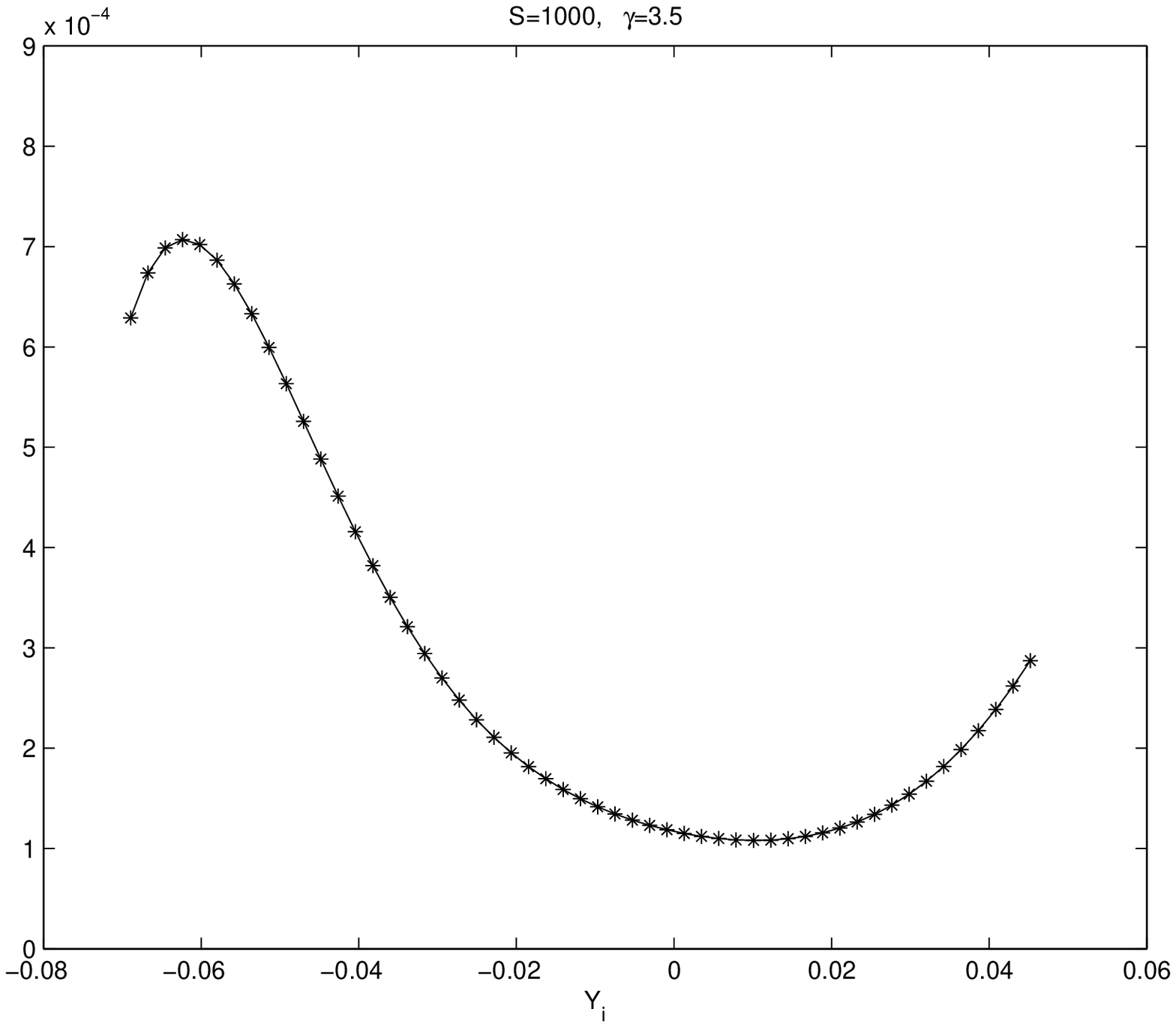}}
\caption{$S= 1000, \gamma=3.5$}\label{35-1000}
\end{figure}

\begin{figure}[h!] \centering
\subfigure[$\hat{f}$]{\includegraphics[width=2.5in,
height=2in]{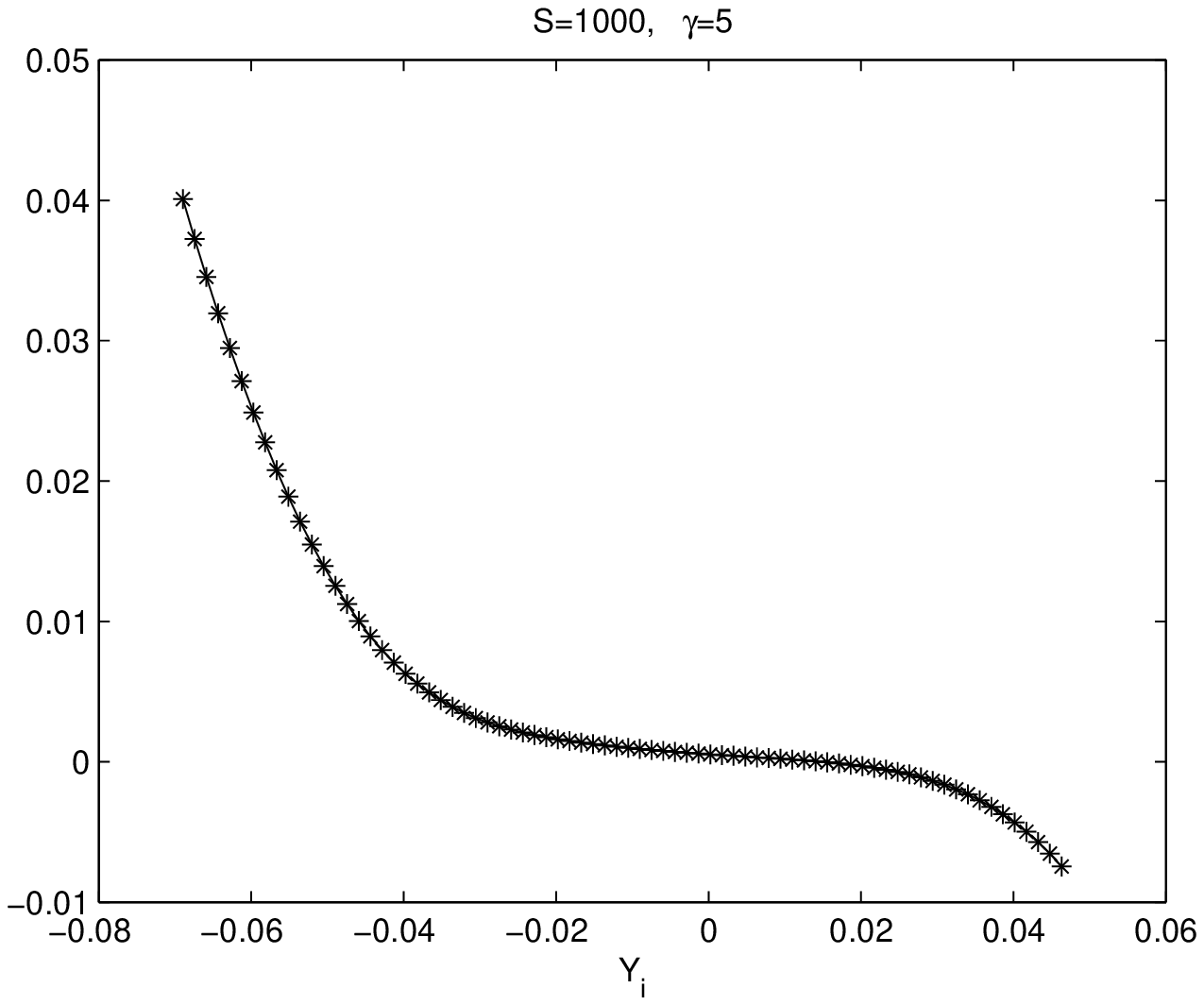}}
\subfigure[$\hat{g}^2$]{\includegraphics[width=2.5in,height=2in]{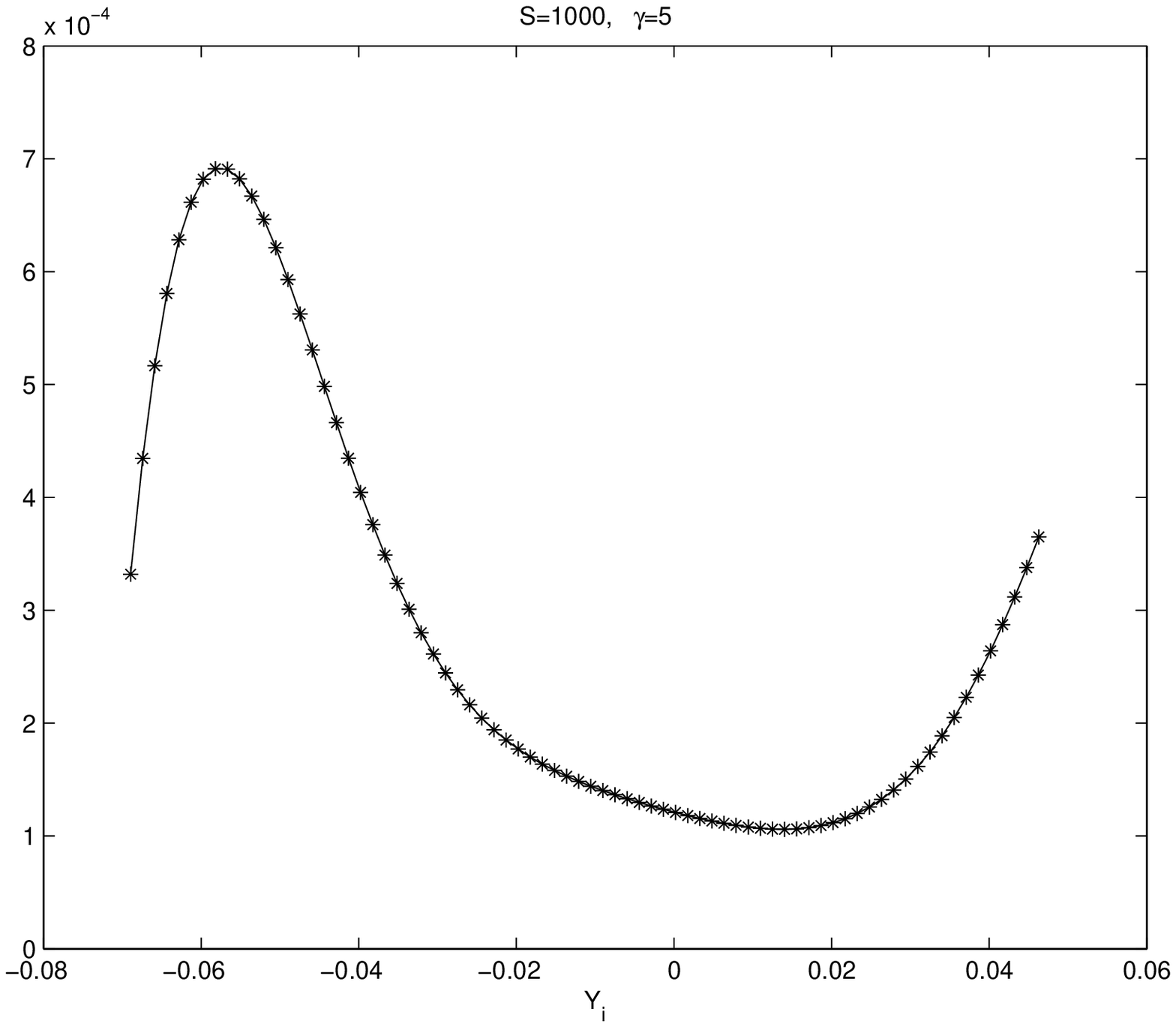}}
\caption{$S= 1000, \gamma=5$}\label{5-1000}
\end{figure}

\begin{figure}[h!] \centering
\subfigure[$\hat{f}$]{\includegraphics[width=2.5in,
height=2in]{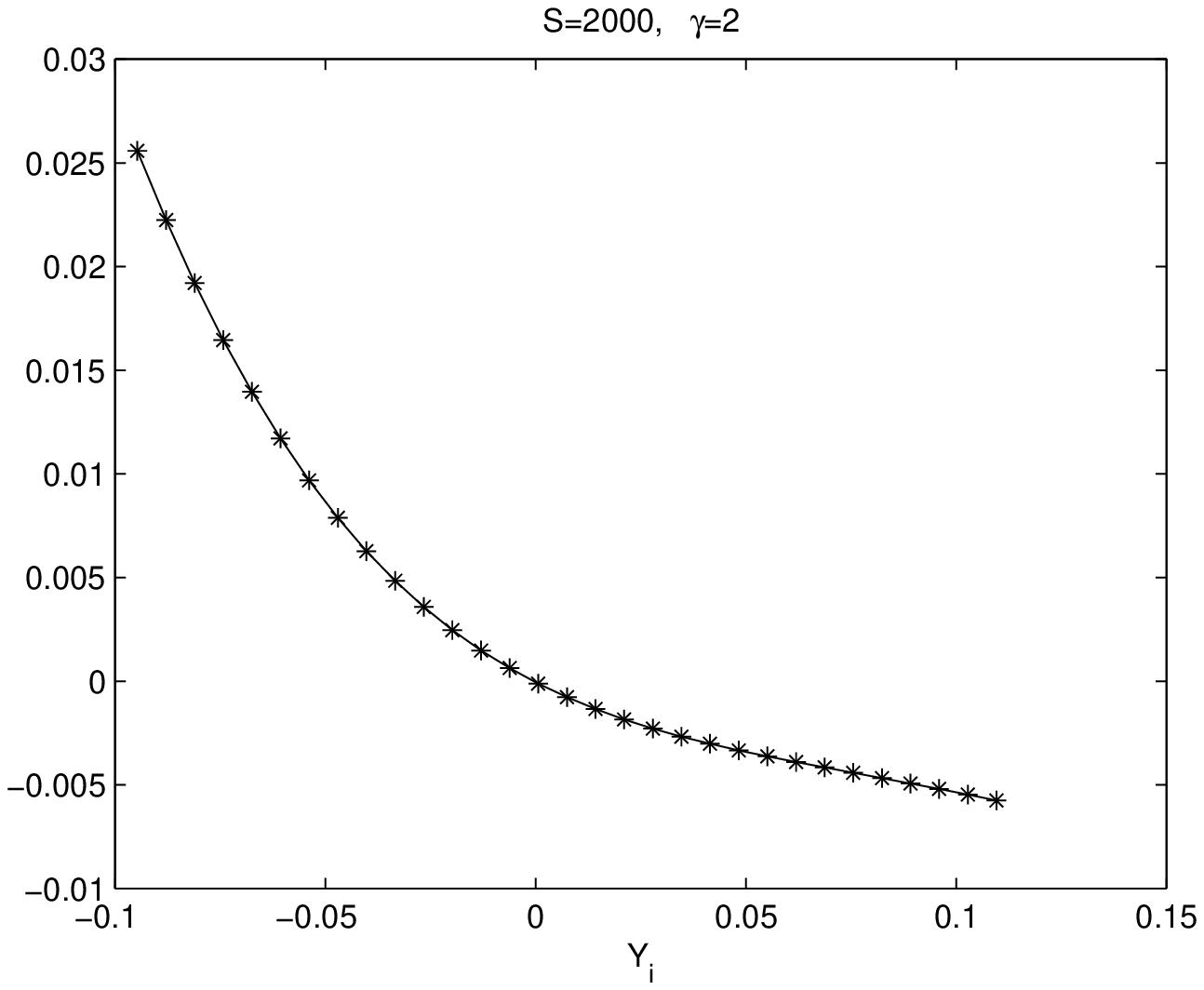}}
\subfigure[$\hat{g}^2$]{\includegraphics[width=2.5in,height=2in]{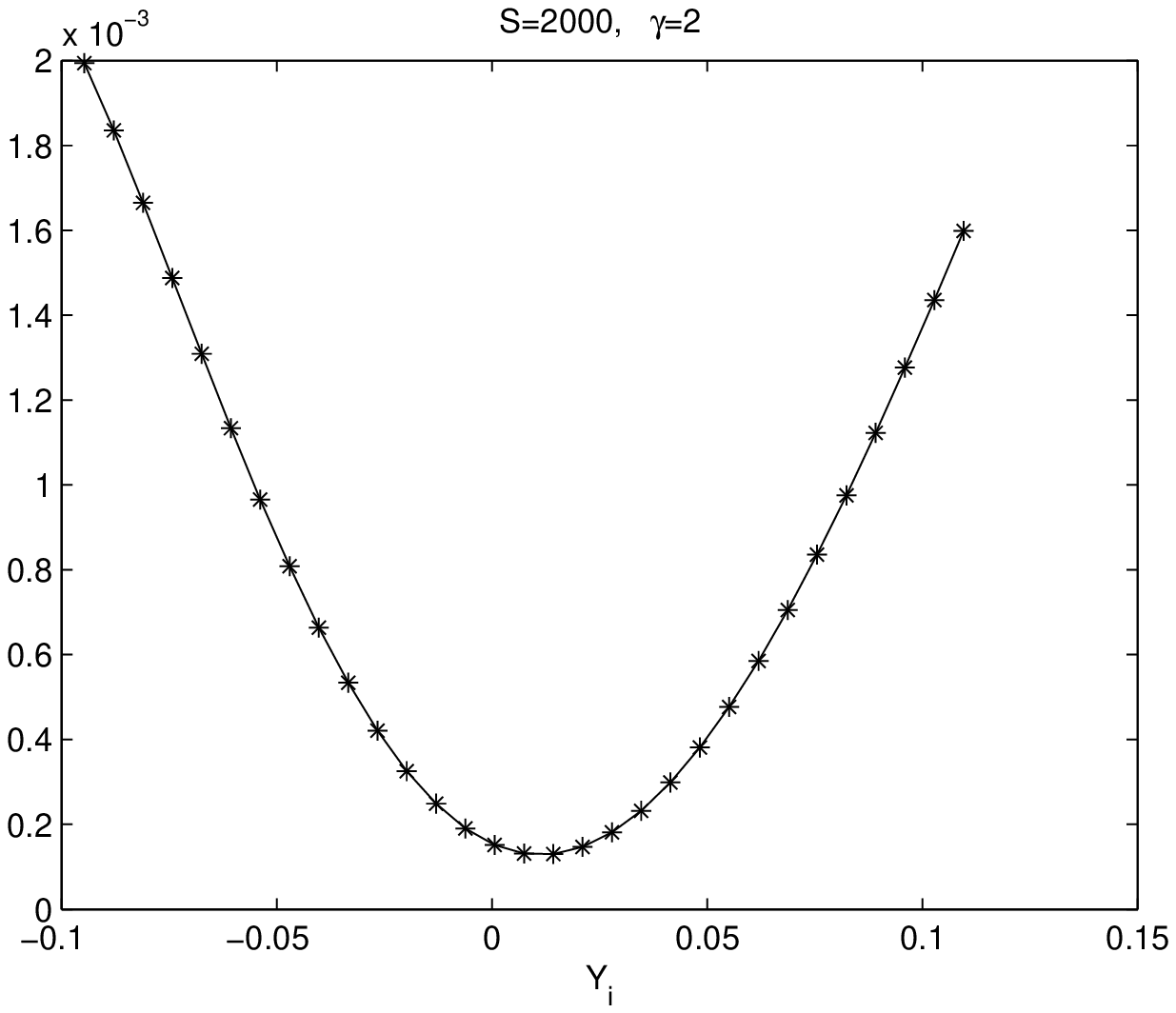}}
\caption{$S= 2000, \gamma=2$}\label{2-2000}
\end{figure}

\begin{figure}[h!] \centering
\subfigure[$\hat{f}$]{\includegraphics[width=2.5in,
height=2in]{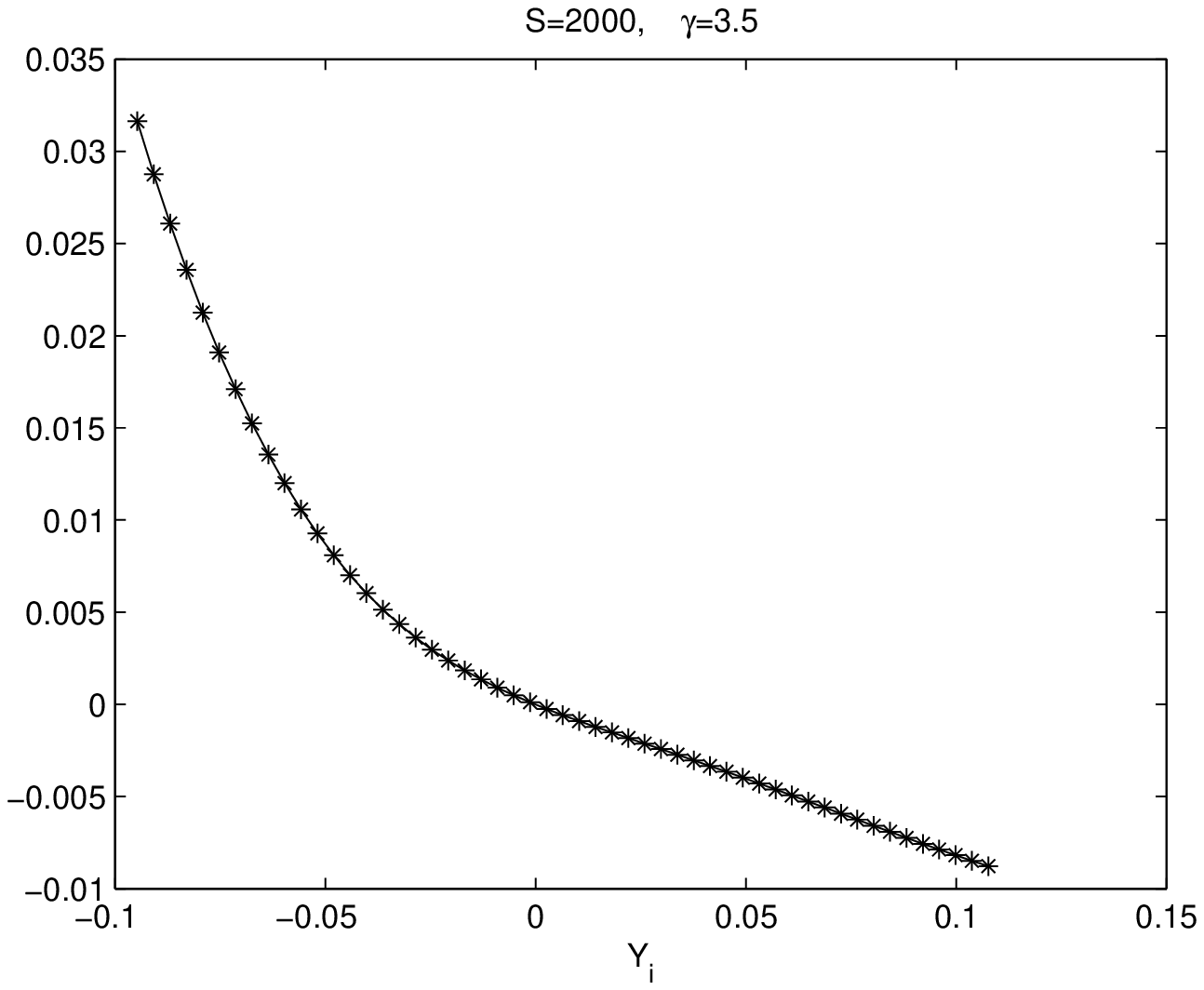}}
\subfigure[$\hat{g}^2$]{\includegraphics[width=2.5in,height=2in]{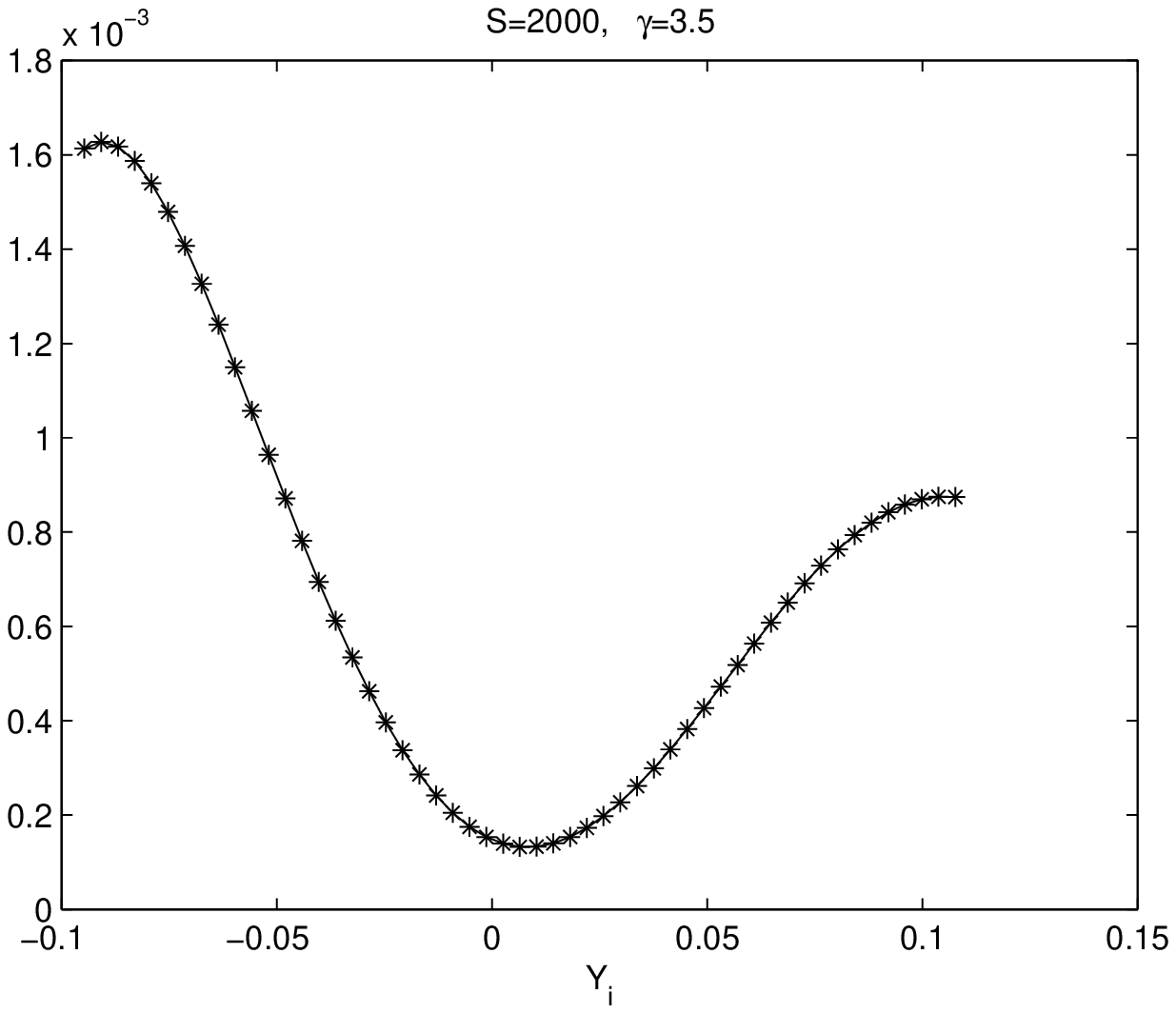}}
\caption{$S= 2000, \gamma=3.5$}\label{35-2000}
\end{figure}

\begin{figure}[h!] \centering
\subfigure[$\hat{f}$]{\includegraphics[width=2.5in,
height=2in]{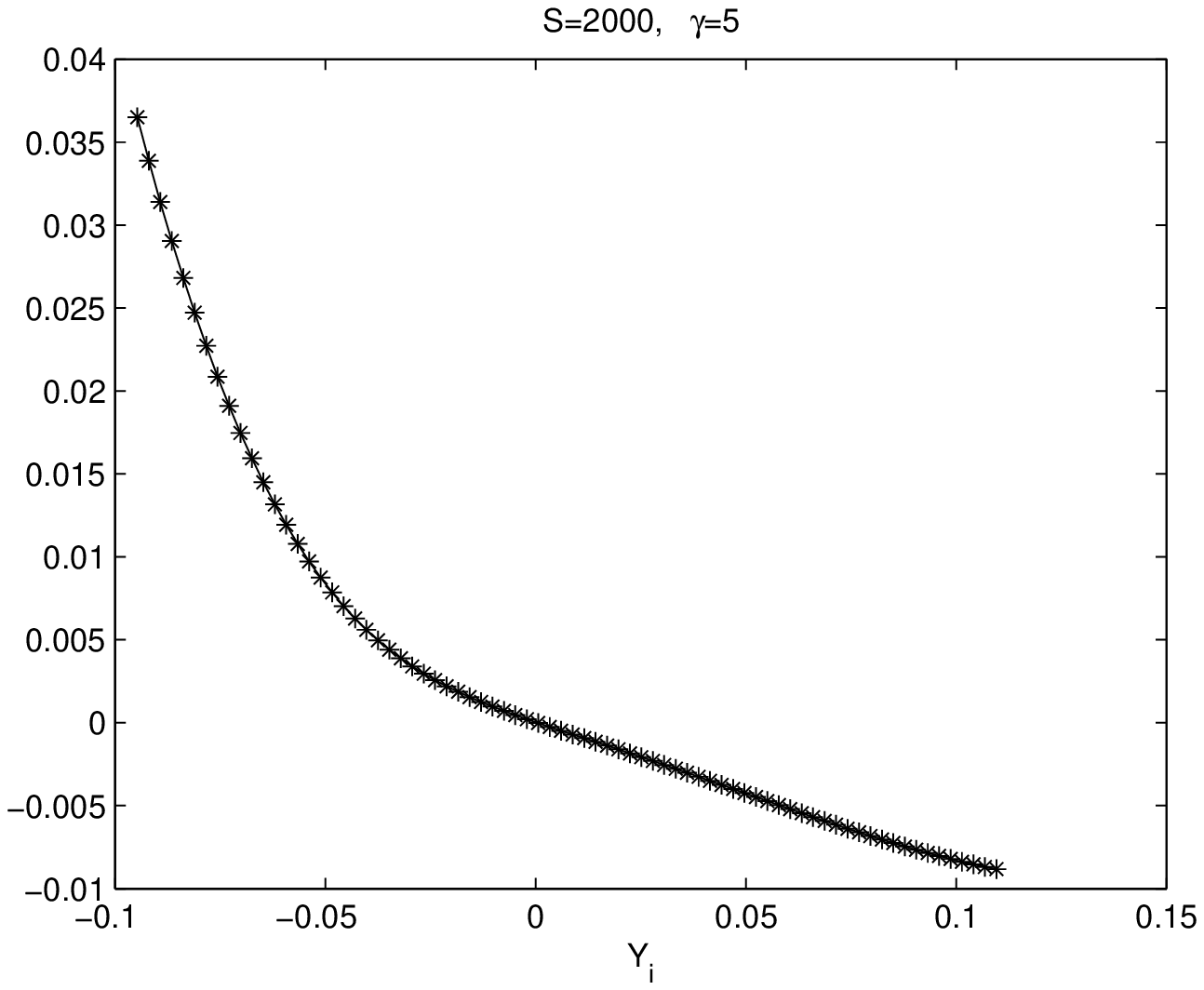}}
\subfigure[$\hat{g}^2$]{\includegraphics[width=2.5in,height=2in]{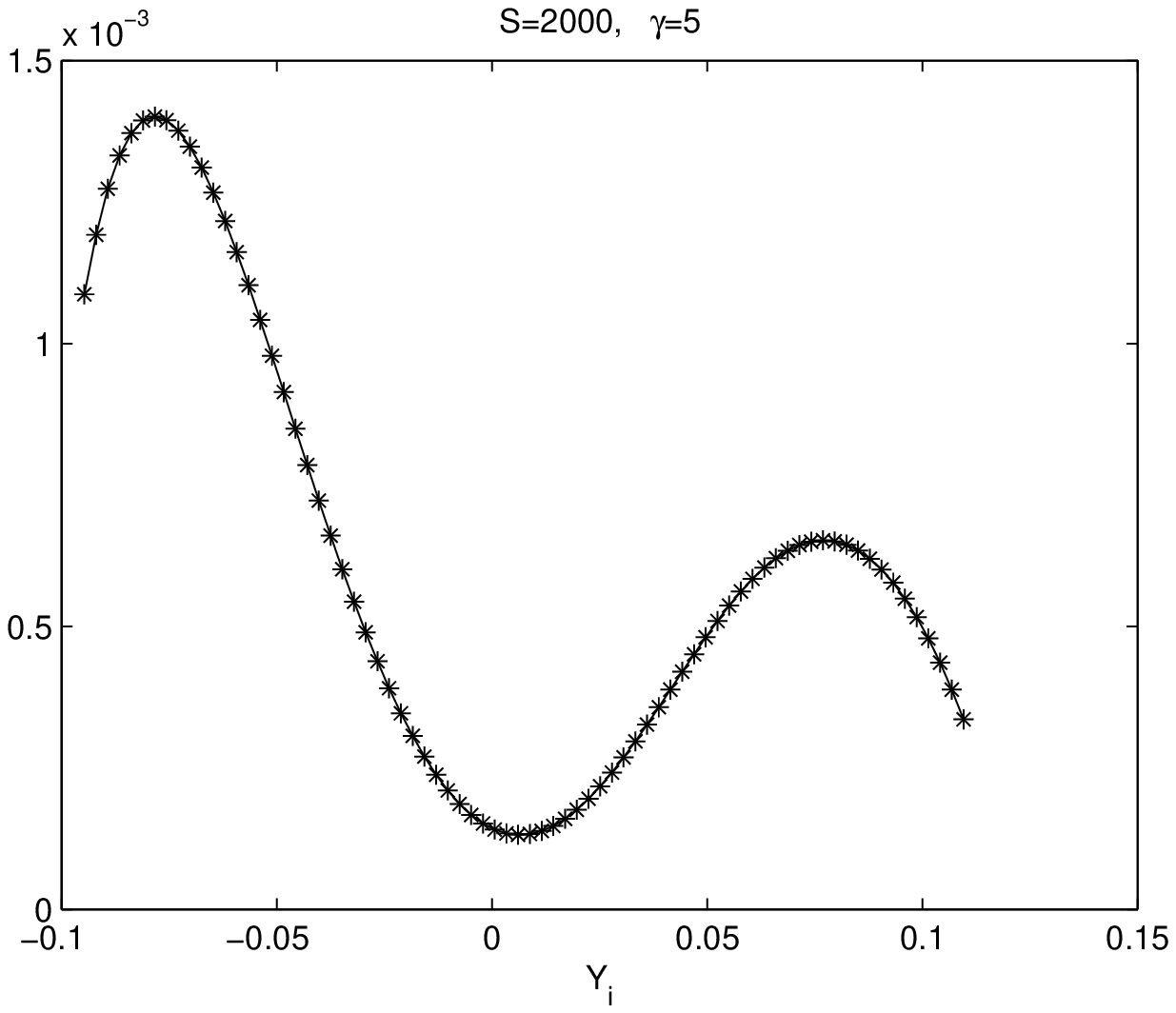}}
\caption{$S= 2000, \gamma=5$}\label{5-2000}
\end{figure}

\begin{figure}[h!] \centering
\subfigure[$\hat{f}$]{\includegraphics[width=2.5in,
height=2in]{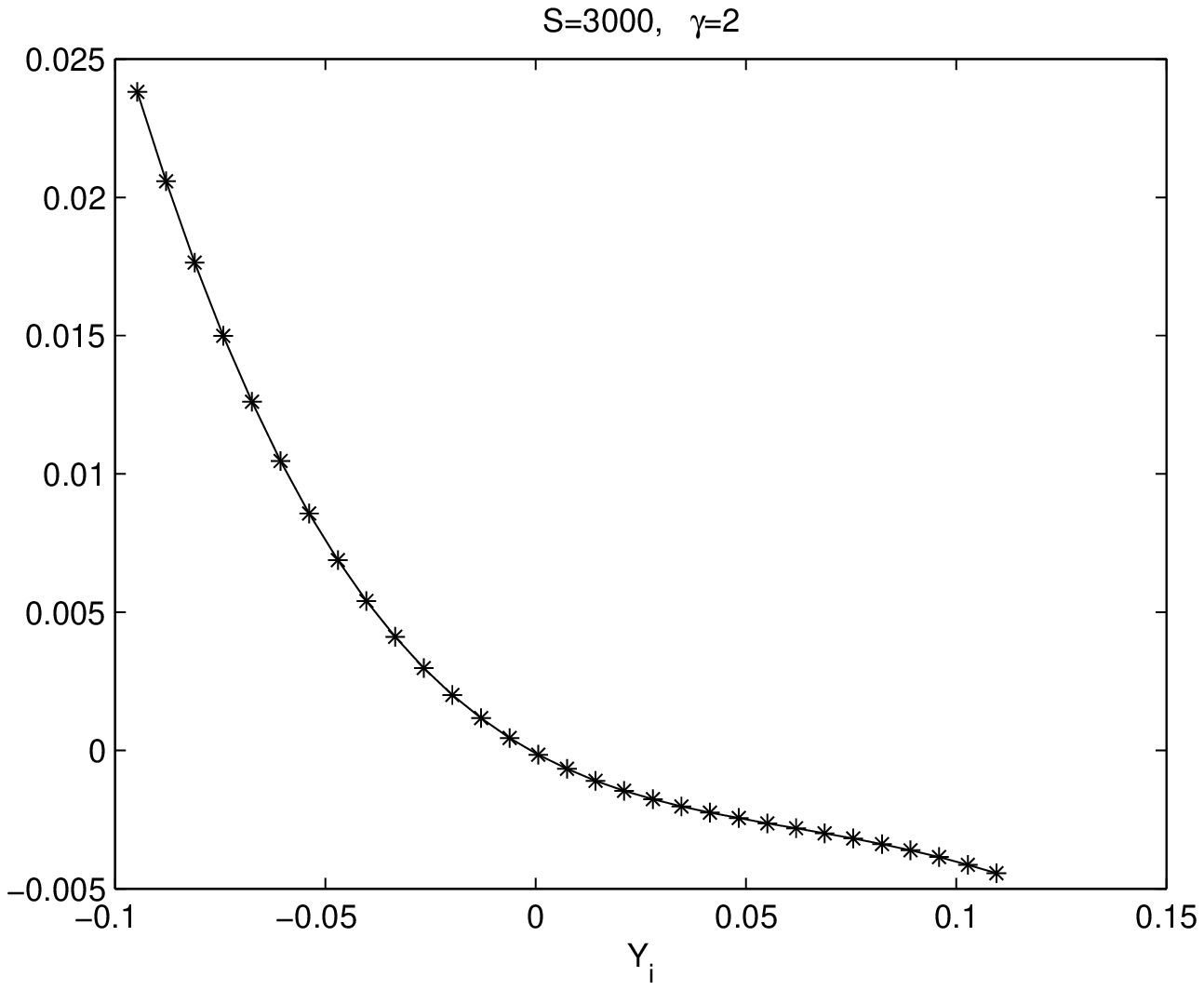}}
\subfigure[$\hat{g}^2$]{\includegraphics[width=2.5in,height=2in]{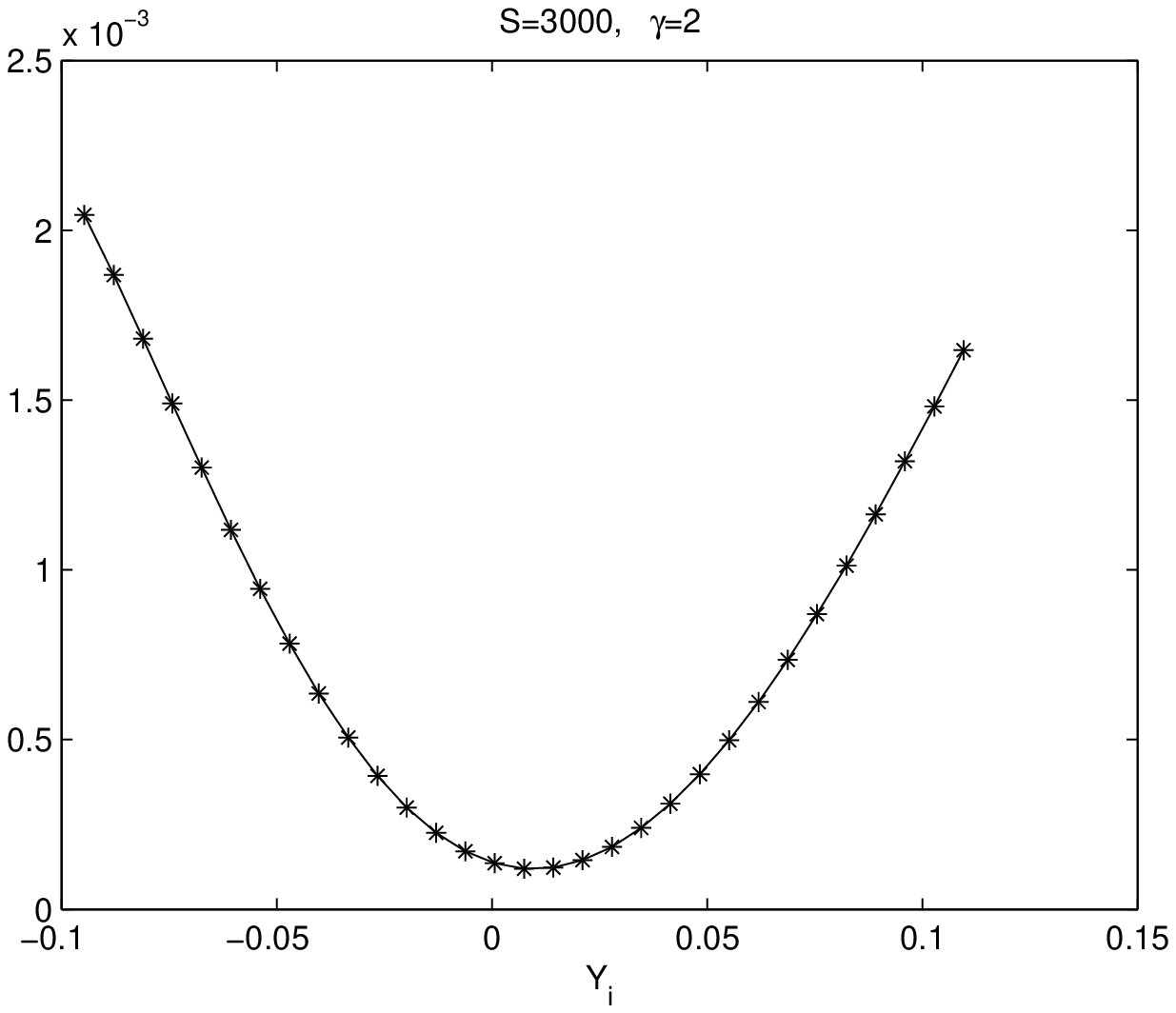}}
\caption{$S= 3000, \gamma=2$}\label{2-3000}
\end{figure}

\begin{figure}[h!] \centering
\subfigure[$\hat{f}$]{\includegraphics[width=2.5in,
height=2in]{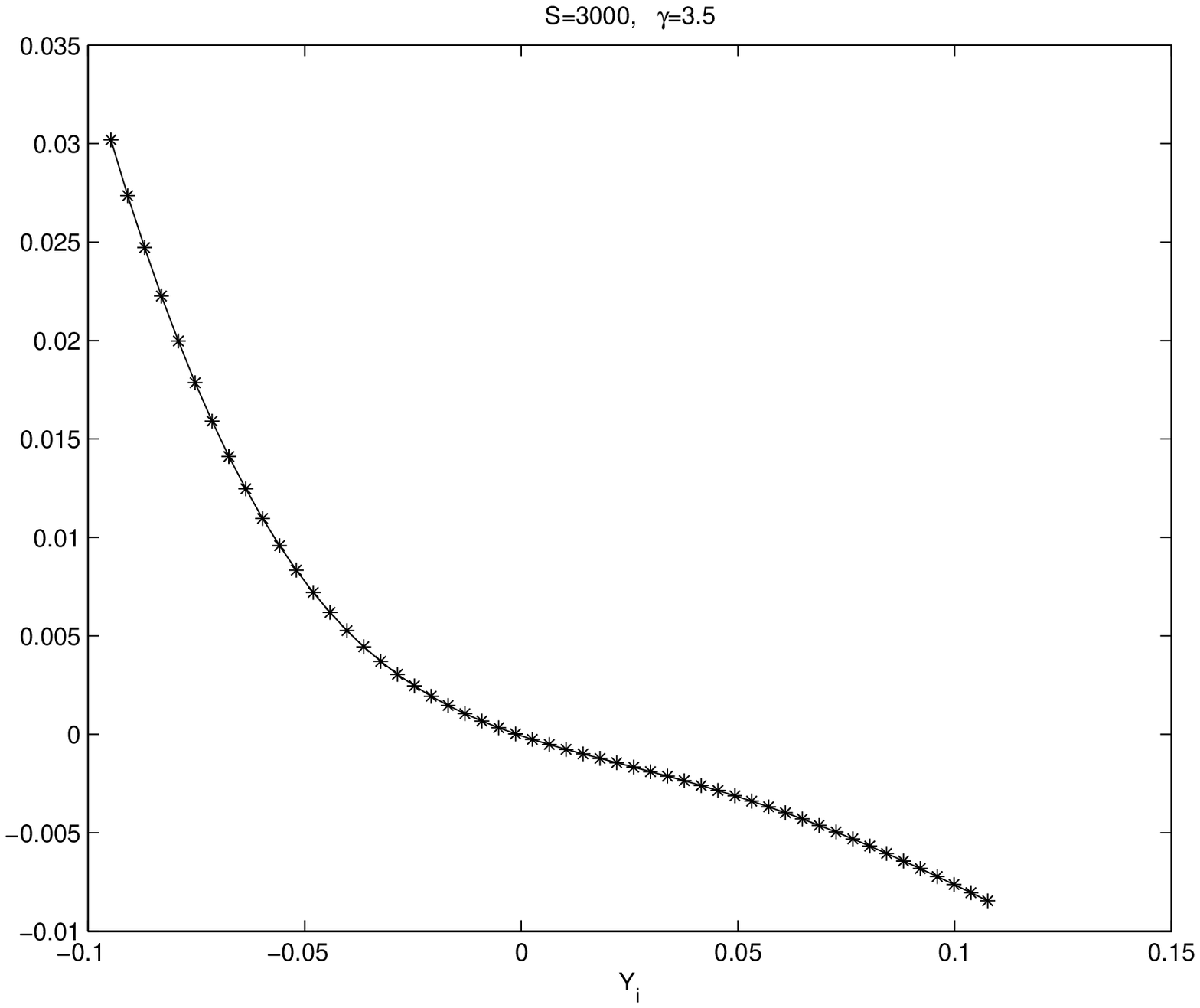}}
\subfigure[$\hat{g}^2$]{\includegraphics[width=2.5in,height=2in]{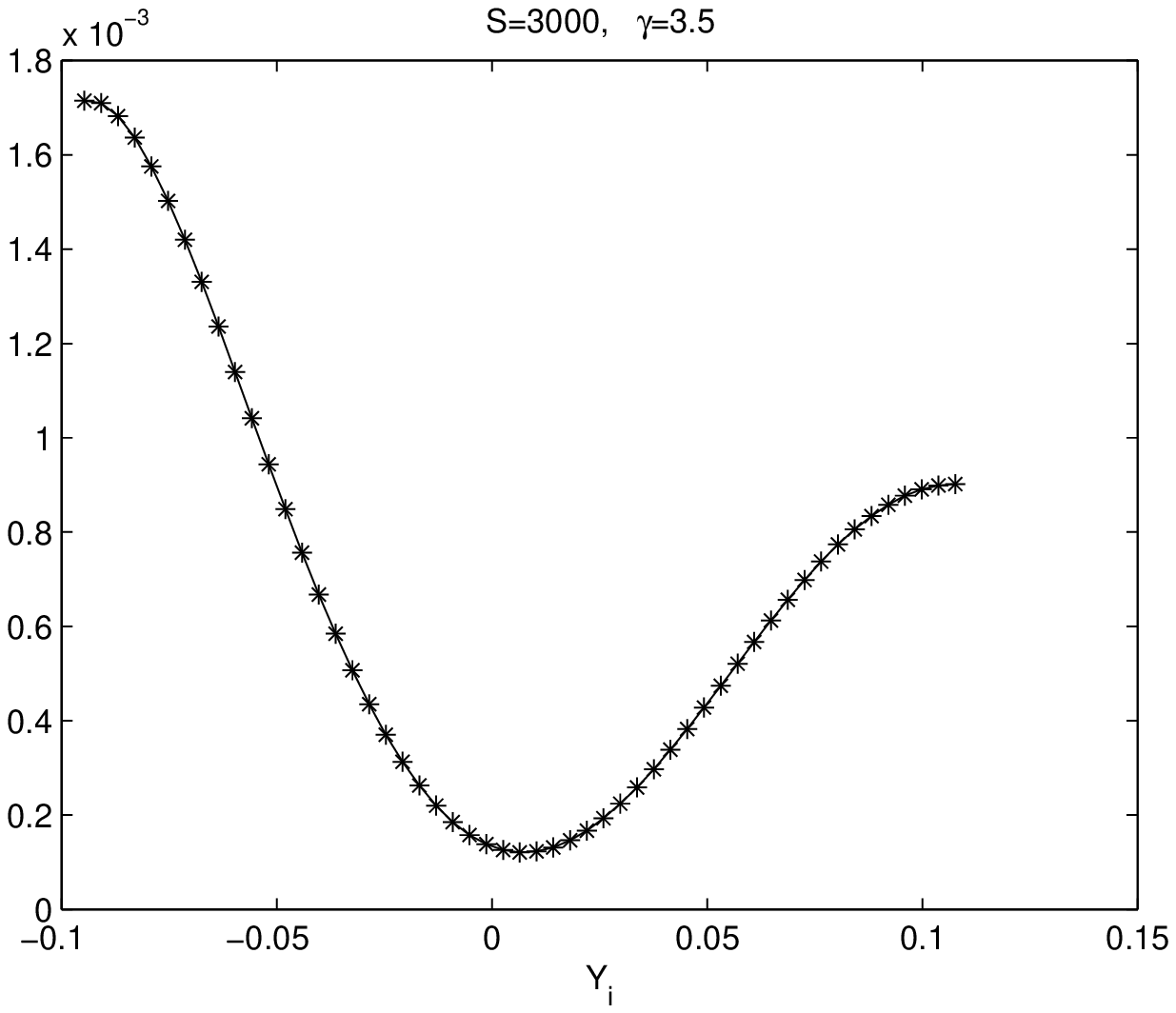}}
\caption{$S= 3000, \gamma=3.5$}\label{35-3000}
\end{figure}

\begin{figure}[h!] \centering
\subfigure[$\hat{f}$]{\includegraphics[width=2.5in,
height=2in]{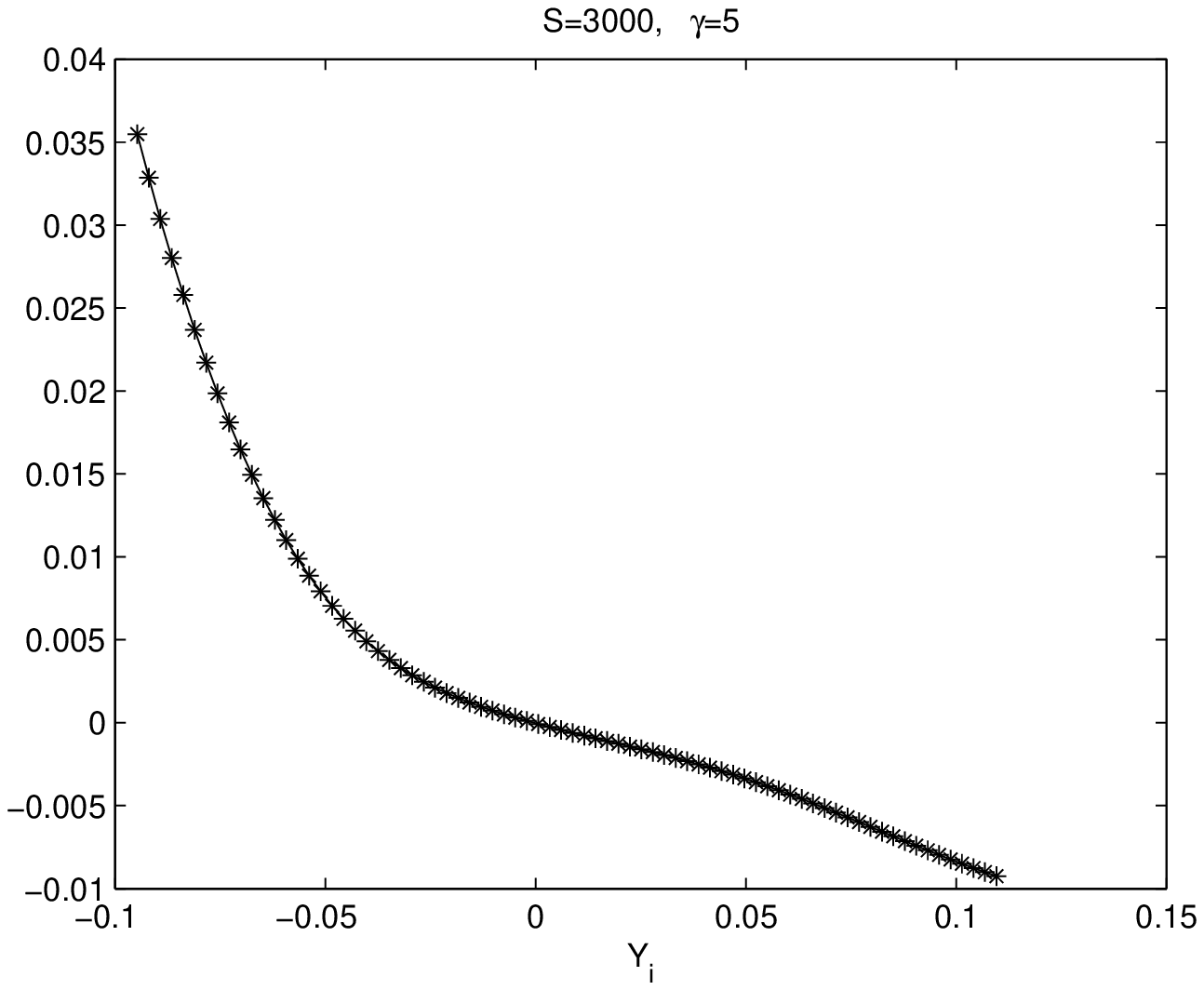}}
\subfigure[$\hat{g}^2$]{\includegraphics[width=2.5in,height=2in]{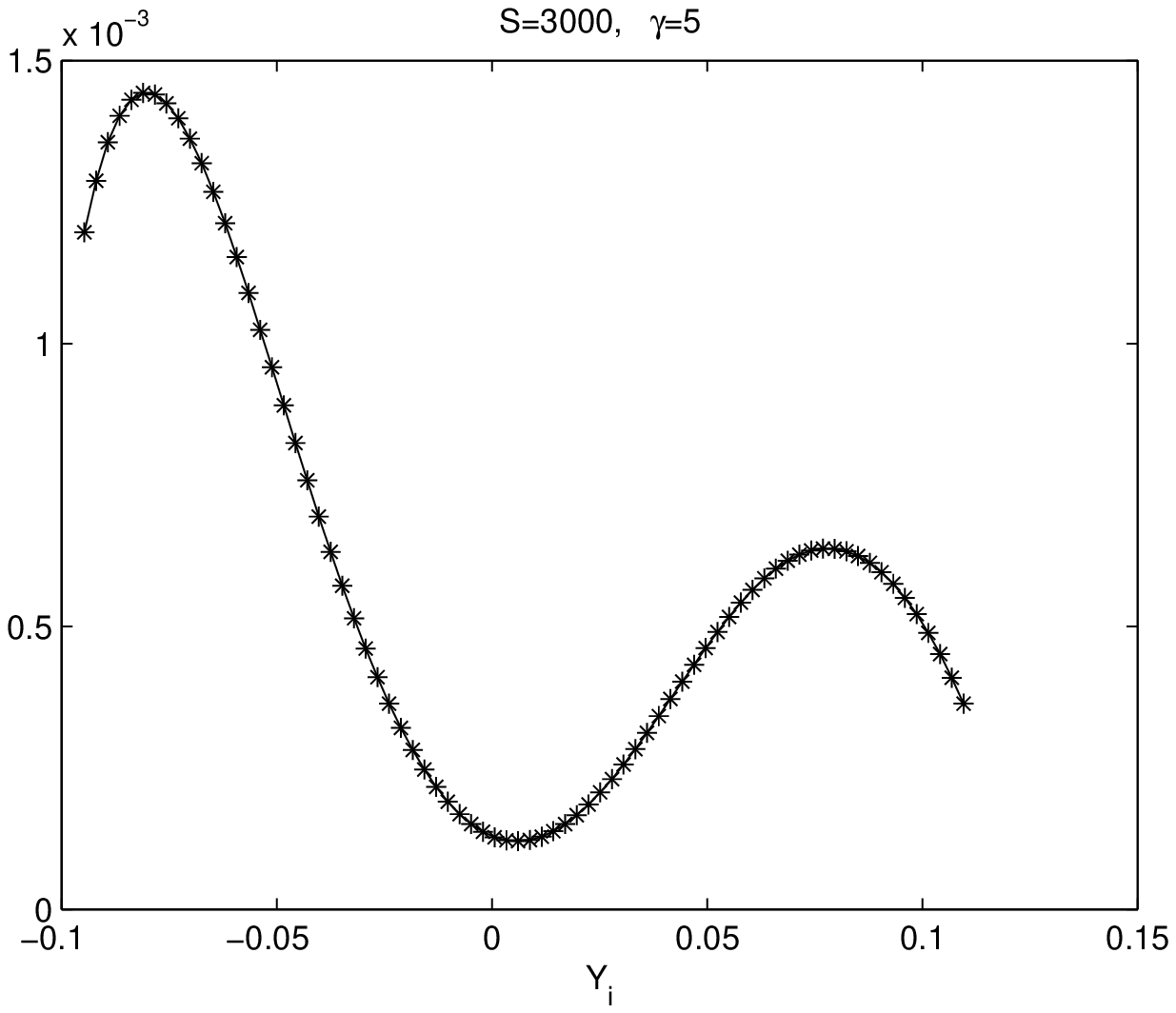}}
\caption{$S= 3000, \gamma=5$}\label{5-3000}
\end{figure}

From Figures \ref{2-1000} through \ref{5-3000}, we can observe the
following phenomena.

\noindent {\bf Observations}:
\begin{enumerate}
\item There is a pattern of $\hat{f}$. If $Y_i=y>0$, $f(Y_i)$ is
slightly negative. If $Y_i<0$, $f(Y_i)$ is positive. That means, if
$Y_i>0$, i.e., there is a gain at time $t_i$, then on expectation,
$Y_{i+1}$ tends to be slightly negative, i.e., there will be a
slight loss at time $t_{i+1}$. However, if $Y_i<0$, i.e., there is a
loss at time $t_i$, then on expectation, $Y_{i+1}$ tends to be big,
i.e., there will be a big gain at time $t_{i+1}$. This represents a
mean reverting pattern.
\item All the graphs of $\hat{g}^2$ show U-shaped `smiling faces', and
the minimum is achieved at a point of $Y_i$ close to zero. In fact,
a point that is slightly to the right of zero. What is more, on each
`smiling face', the left side of the curve is higher than the right
side of the curve. So these are tilted `smiling faces', or skew.
\item When the data size is small, e.g., $S=1000$, we
observe the boundary effect on the edges of the interval, see, e.g.,
the graphs of $\hat{g}^2$. This phenomenon was also observed in
\cite{Tsybakov97} where an explanation was also provided.
\end{enumerate}

If we compare Figures \ref{2-1000} through \ref{5-3000} with Figure
\ref{dfg}, we can see that model (\ref{receqn}) provides a good fit.

\section{The Implied Volatility}\label{smile}

Since $Y_i = \log P_i-\log P_{i-1}$, we can rewrite model
(\ref{receqn}) as
\begin{displaymath}
\log P_{i+1}-\log P_{i}=f(\log P_i-\log P_{i-1})+g(\log P_i-\log
P_{i-1})\epsilon_i, i=1,2,...
\end{displaymath} which provides us an ARCH model for the process $\log
P_i$ as follows
\begin{equation}
\log P_{i+1}=\log P_{i}+f(\log P_i-\log P_{i-1})+g(\log P_i-\log
P_{i-1})\epsilon_i, i=1,2,...
\end{equation}

\begin{figure}[h!] \centering
\subfigure[$\tilde{f}$]{\includegraphics[width=2.5in,
height=2in]{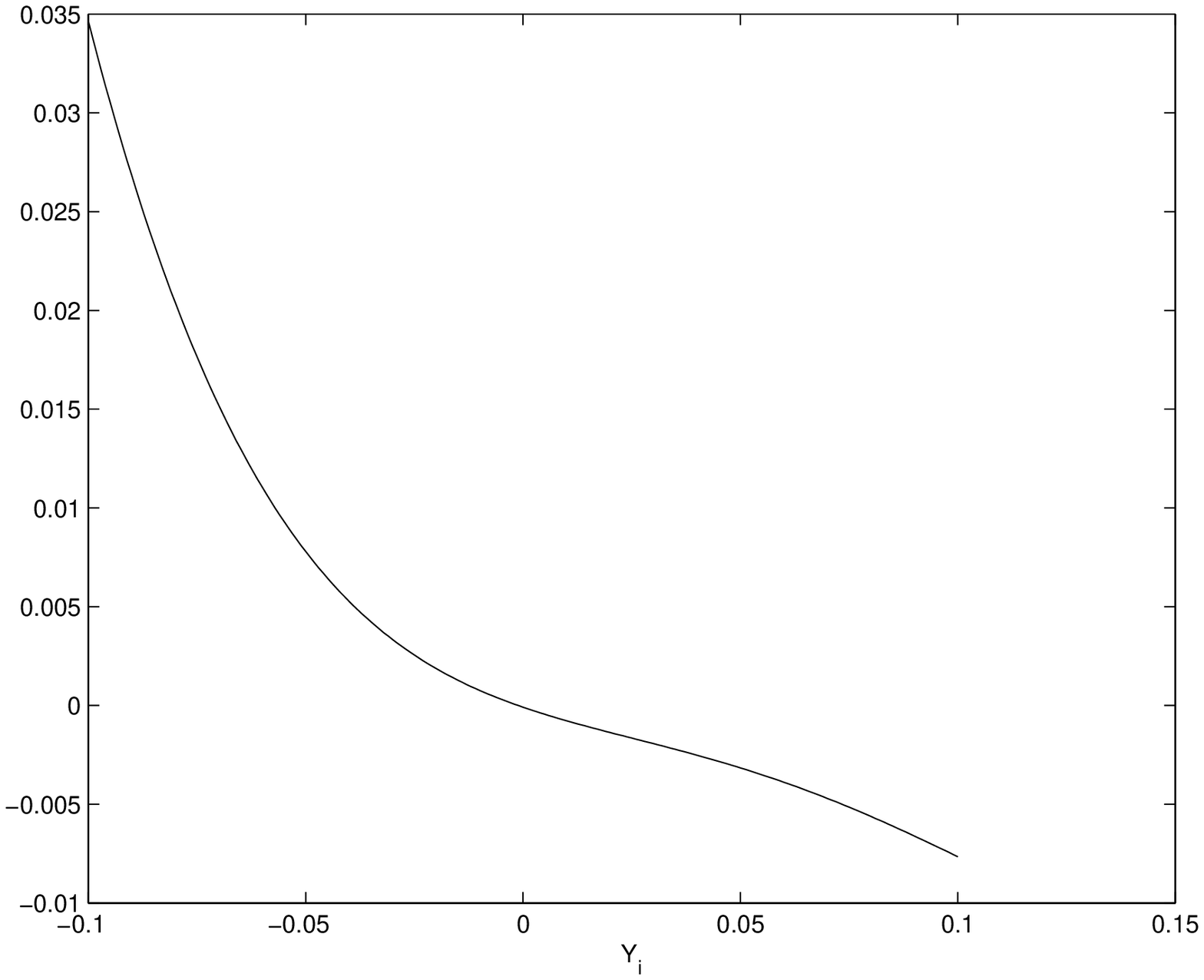}}
\subfigure[$\tilde{g}$]{\includegraphics[width=2.5in,height=2in]{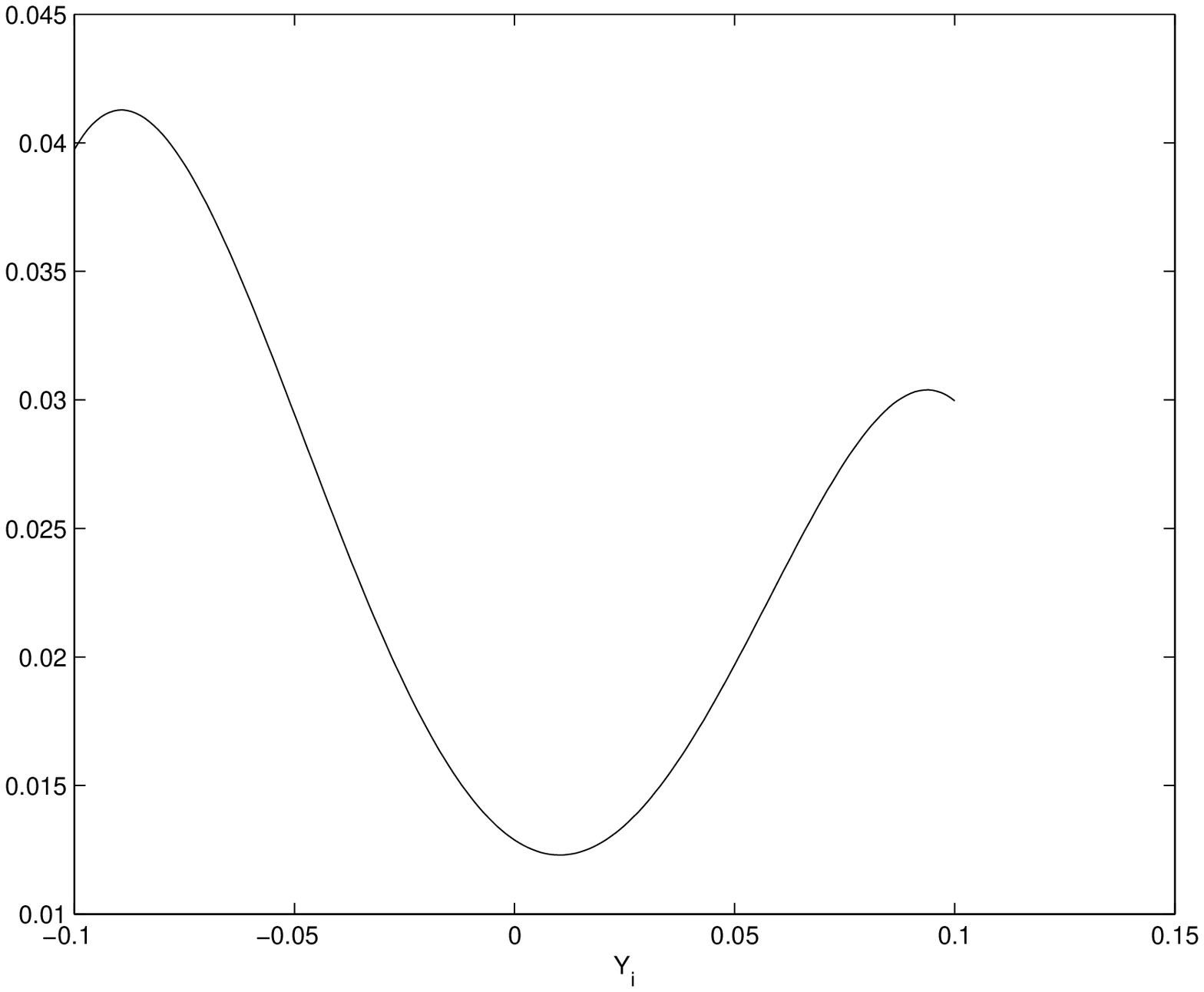}}
\subfigure[$\tilde{g}^2$]{\includegraphics[width=2.5in,height=2in]{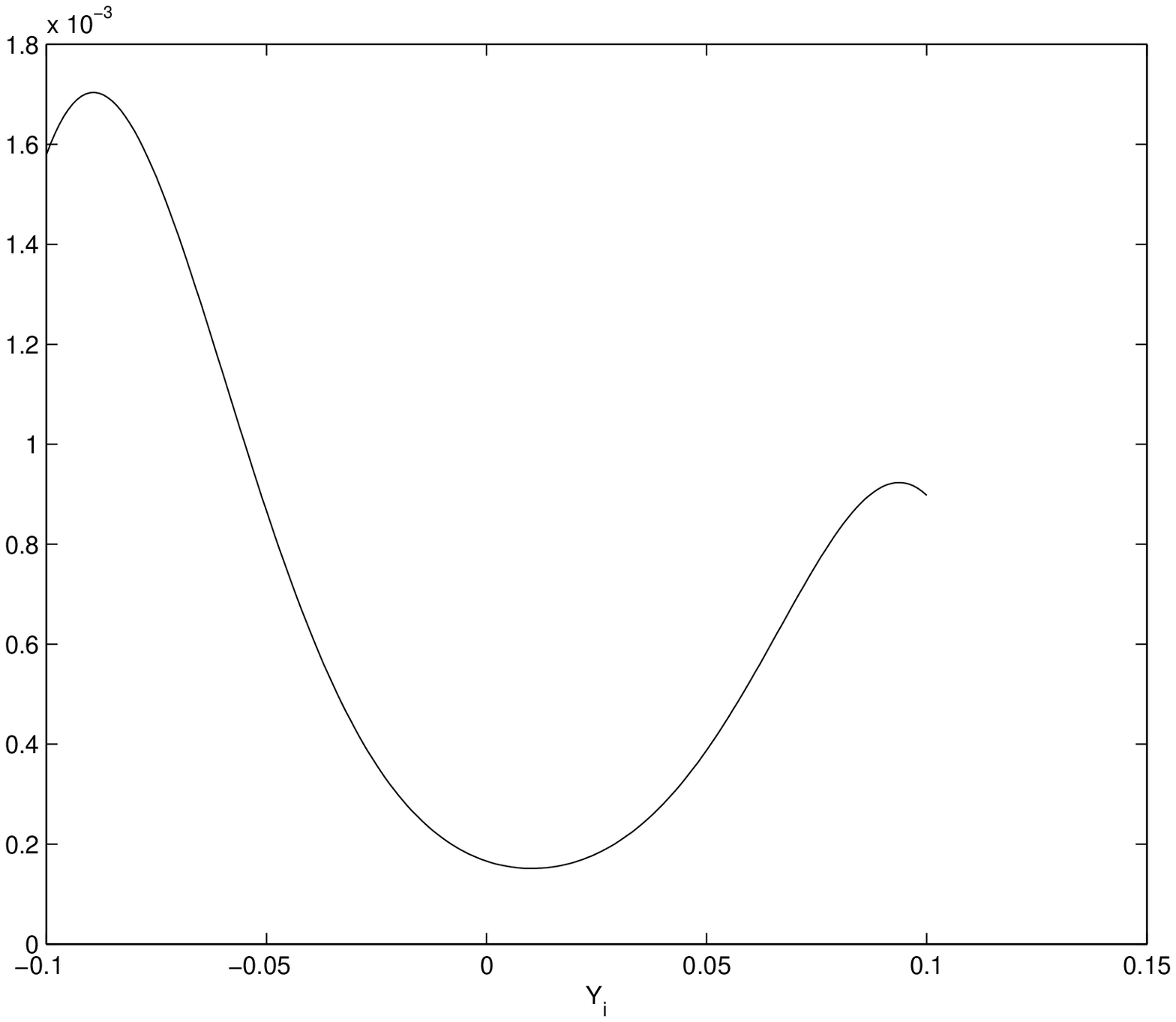}}\caption{Polynomial
Approximations of $f,g,g^2$}\label{fgtilde}
\end{figure}

We shall use this model as the fundamental to study the option
pricing problem. In doing this, it might be easier to use a simple
polynomial regression to approximate $f$ and $g$, and we get the
following ARCH model
\begin{equation}\label{archP}\begin{split}
\log P_{i+1}=&\log P_{i} \\
&- 8.948\times 10^{-5} -7.557\times 10^{-2} Y_i+0.8305Y_i^2-13.60Y_i^3+52.84Y_i^4\\
&+\left(1.288\times 10^{-2} -0.1138
Y_i+5.503Y_i^2+6.492Y_i^3-3.306\times 10^2Y_i^4\right)\epsilon_i,
\end{split}
\end{equation} where $Y_i = \log P_i-\log P_{i-1}$. That is,
\begin{displaymath}\begin{split}
\tilde{f}(Y_i)&=- 8.948\times 10^{-5} -7.557\times 10^{-2}
Y_i+0.8305Y_i^2-13.60Y_i^3+52.84Y_i^4, \\
\tilde{g}(Y_i)&=1.288\times 10^{-2} -0.1138
Y_i+5.503Y_i^2+6.492Y_i^3-3.306\times 10^2Y_i^4,
\end{split}
\end{displaymath} where $\tilde{f},\tilde{g}$ represent the
polynomial approximations of $f,g$, respectively. Figure
(\ref{fgtilde}) shows the graphs of $\tilde{f},\tilde{g}$ and
$\tilde{g}^2$.

It can be seen that this model is a local volatility model, but
different from most well known local volatility models which aimed
to replicate the implied volatility surface only, our model
replicates both the drift term and volatility term, through real
data calibration using the technique of local polynomial regression.

Now the model (\ref{archP}) is easy to use in a Monte Carlo
simulation. In order to get the fair price of the European call
options, we rewrite this model under the risk neutral measure and
get
\begin{equation}\label{rnmodel}
\log P_{i+1}-\log P_{i} = (r-\frac{1}{2}\sigma^2(Y_i))\Delta
t+\sigma(Y_i)\sqrt{\Delta t}\tilde{\epsilon}_i,
\end{equation} where $r$ is the risk free interest rate, $Y_i = \log P_i-\log
P_{i-1}$, $\tilde{\epsilon}_i$ is the standard normal random
variable under the risk neutral measure, and
\begin{displaymath}
\sigma(Y_i)=\tilde{g}(Y_i)/\sqrt{\Delta t}.
\end{displaymath}

\begin{figure}[h!]\centering
\includegraphics[width=2.5in,
height=2in]{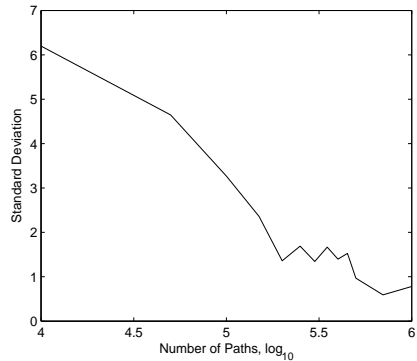}\caption{Standard
Deviation}\label{sample}
\end{figure}

In order to find a proper number of paths, we use Monte Carlo
simulation to price a European call option with the following
parameter settings: annual interest rate $r=0.03$, time to maturity
$T=60$ months, strike price $K=800$ and underlying price
$P_0=1462.42,P_1=1459.37$. We chose this option because it has the
largest variance in our model. The number of paths ranges from
10,000 to 1,000,000, and for each setting, 20 trials are run and the
standard deviation of the option prices are calculated. The result
is shown in Figure \ref{sample}.

\begin{figure}[h!]\centering
\includegraphics[width=4in,
height=3in]{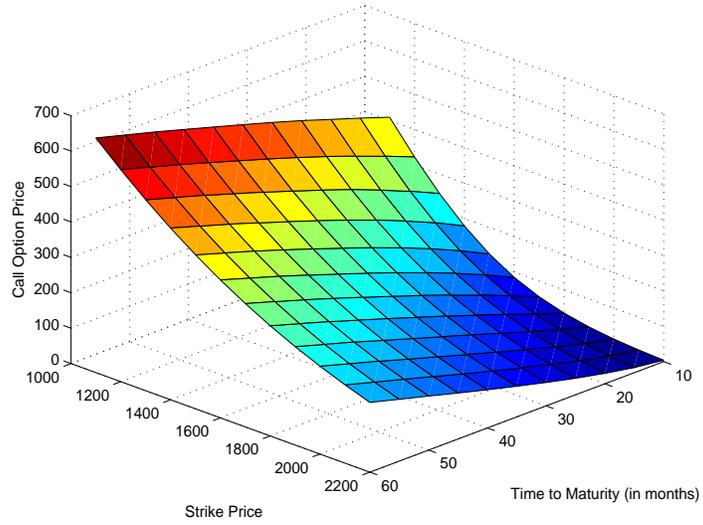}\caption{Call Options Prices}\label{op}
\end{figure}

We can now set the number of paths to be 200,000. Figure \ref{op}
shows a Monte Carlo simulation on call option prices with 200,000
sample paths.

We pick the region $1100\leq K\leq 2000$ and $6\leq T\leq 60$ to
recover the implied volatility surface from the Black-Scholes
formula. The reason is that, on this region the computation is
considered to be stable because $Vega$ is not too close to zero. A
detailed explanation on this issue is provided in the Appendix.
Figure \ref{ivsurf} shows the implied volatility surface through the
Black-Scholes formula.

\begin{figure}[h!]\centering
\includegraphics[width=4in,
height=3in]{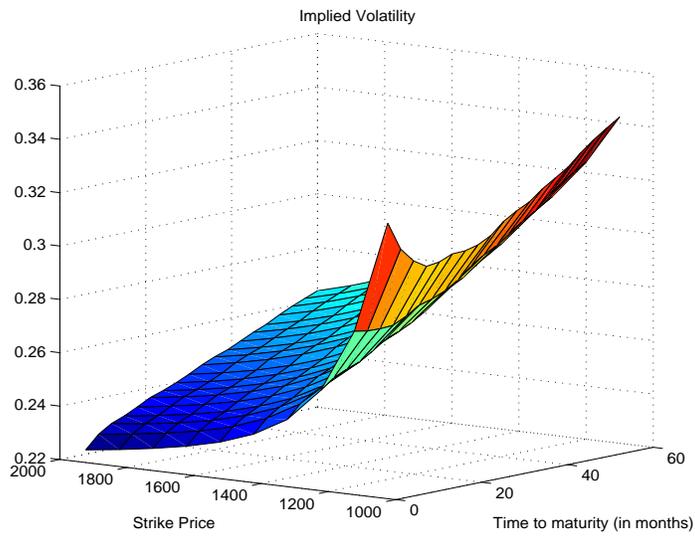}\caption{Implied Volatility
Surface}\label{ivsurf}
\end{figure}

\section{Foreign Currency Market}\label{fcm}
In this section, we shall construct an ARCH model for risky assets
on currency market. Let $P_i$ be the USD/GBP exchange rate, and we
again use the same model:
\begin{equation}\label{usdrate}
Y_{i+1}=f(Y_i)+g(Y_i)\epsilon_i,
\end{equation} where
\begin{displaymath}f(Y_i)=\frac{D_2'(Y_i)Y_i- D_1(Y_i)\Delta t}{\xi+D_2'(Y_i)}, \ \ \  \ \ \
g(Y_i)=\frac{-\nu}{\sqrt{252}(\xi+D_2'(Y_i))},\end{displaymath}
$Y_i=\log P_i-\log P_{i-1}$, and $\epsilon_i=\sqrt{252}\Delta W_i,\
i=1,2,...$ are i.i.d $N(0,1)$ random variables.

\begin{figure}[h!] \centering
{\includegraphics[width=2.5in, height=2in]{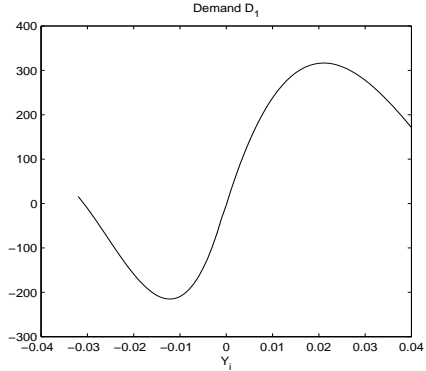}}
\caption{Excessive demand as a function of yield}\label{dem1-p}
\end{figure}

There are obviously some differences in the currency market.
Realistically, unlike the stock market, one can not expect that
$P_i$ (USD/GBP) increases exponentially at a certain speed. It is
commonly believed that $P_i$ shows a mean reverting dynamics.
Therefore, if there is a big jump of $P_i$ in one day, due to the
prospect theory, more holders tend to sell the share in order to
realize the profit, showing an extreme risk averse behavior, and
vice versa. Therefore, the demand $D_1$ in the currency market is
believed to show a stronger prospect phenomenon than that on the
stock market. Here the function $D_1(y)$ is assumed to take the
following form:
\begin{displaymath}
D_1(y)=\left\{\begin{aligned} &5.381+4.039\times 10^4y+2.169\times 10^6y^2+2.802\times 10^7 y^3, &\  y\leqslant 0 \\
&-3.27+3.391\times 10^4y-1.056\times 10^6y^2+7.935\times 10^6y^3, &\
y
> 0
\end{aligned}\right.
\end{displaymath} and its graph is shown in Figure \ref{dem1-p}.

It can be seen that if $Y_i$ is much bigger than zero, $D_1$
decreases, showing an extreme risk averse behavior; and if $Y_i$ is
far below zero, $D_1$ increases, showing an extreme risk seeking
behavior.

\begin{figure}[h!] \centering
\subfigure[$D_2$]{\includegraphics[width=2.5in,height=2in]{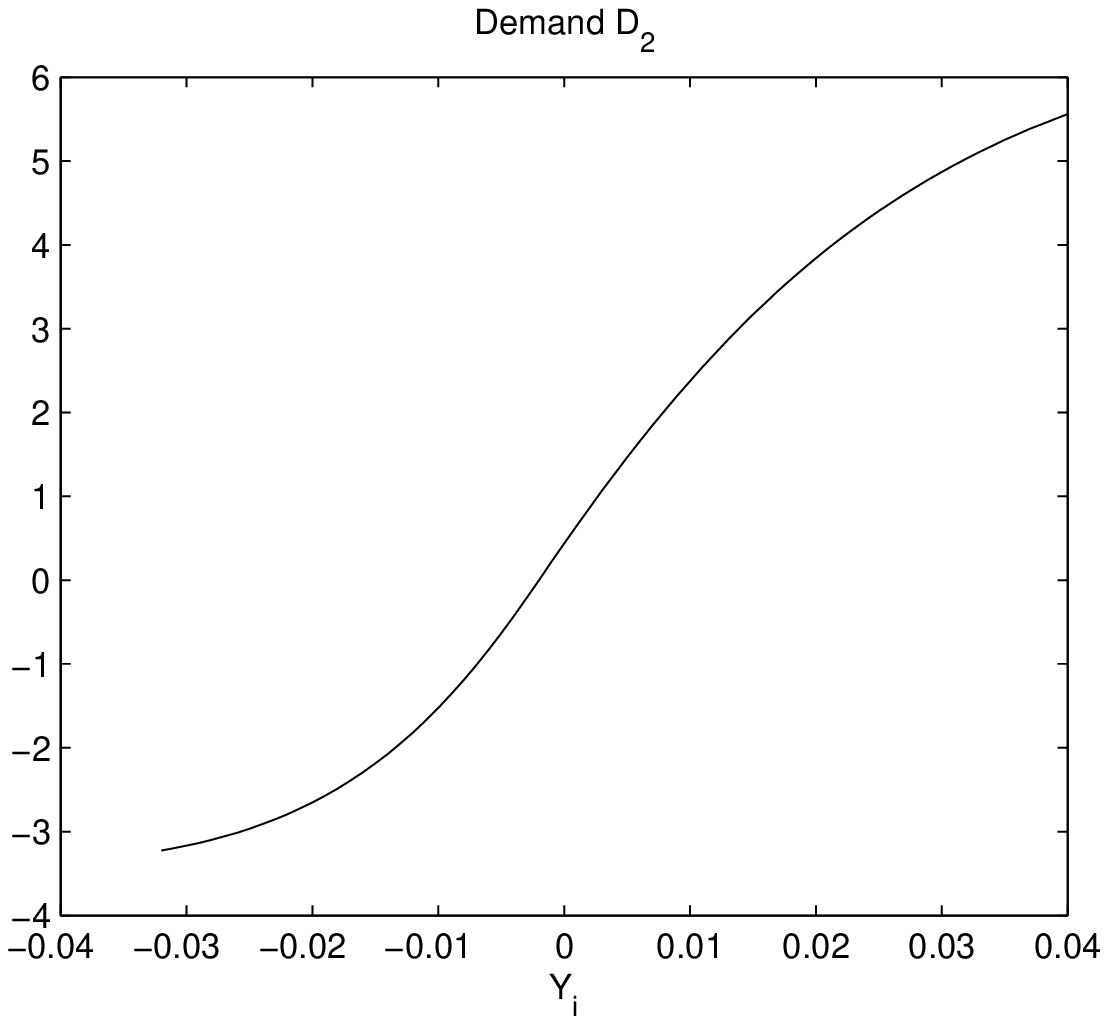}}
\subfigure[$D_2'$]{\includegraphics[width=2.5in,height=2in]{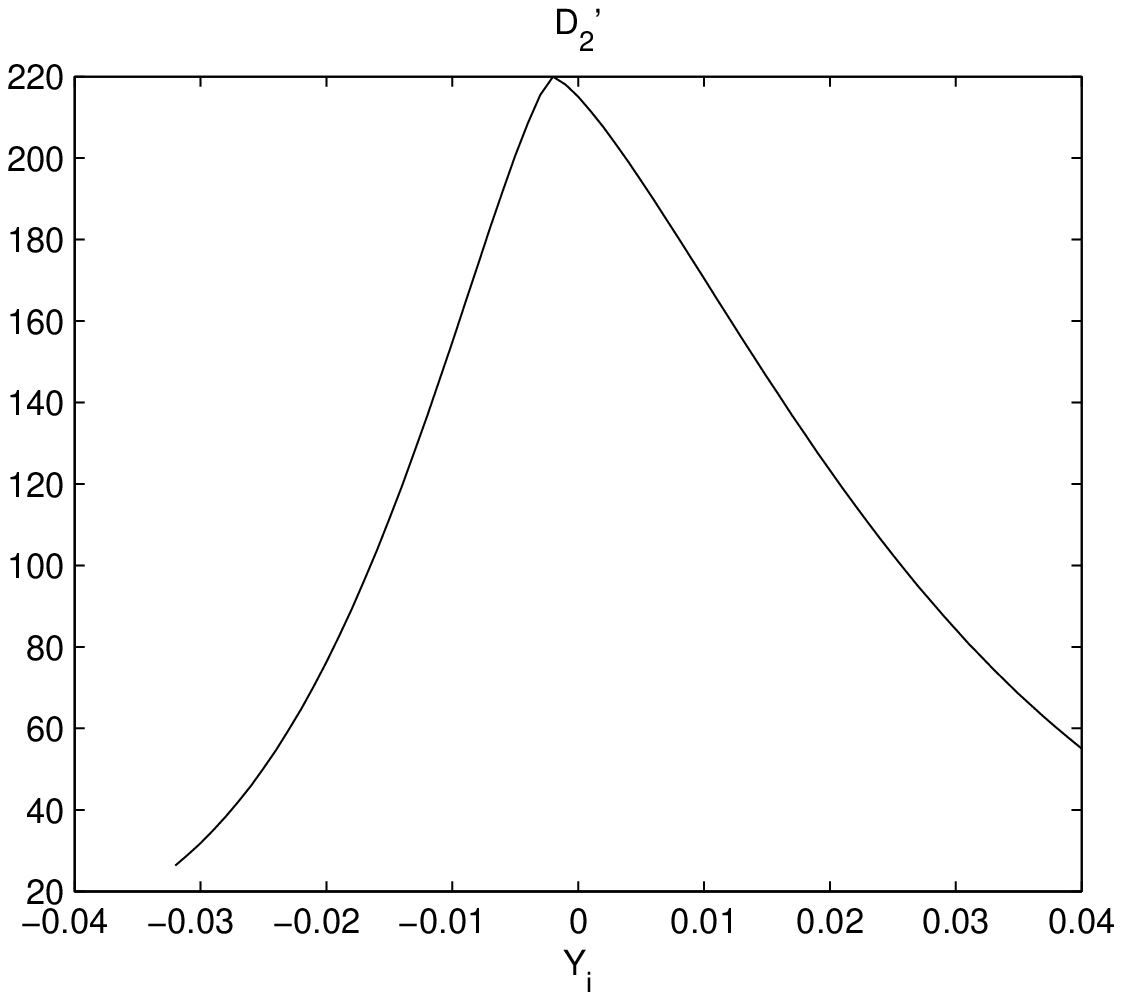}}
\caption{Cumulative demand as a function of yield}\label{D2_cur}
\end{figure}

The function $D_2'$ in this model is given by
\begin{displaymath}
D_2'(y)=\left\{\begin{aligned}&220e^{-250|x+0.002|^{1.36}},&\
x\leqslant -0.002,\\
&220e^{-100|x+0.002|^{1.35}},&\ x> -0.002,
\end{aligned}\right.
\end{displaymath} and $D_2(y)$ is the numerical integral of
$D_2'(y)$ with $D_2(-0.002)=0$. The center point $-0.002$ is very
close to and even slightly less than zero, showing that $P_i$ is not
expected to increase exponentially at a certain speed, but rather,
$Y_i=\log P_i-\log P_{i-1}$ is expected to stay around zero. The
graphs of $D_2,D_2'$ are shown in Figure \ref{D2_cur}.

\begin{figure}[h!] \centering
\subfigure[$f$]{\includegraphics[width=2.5in,height=2in]{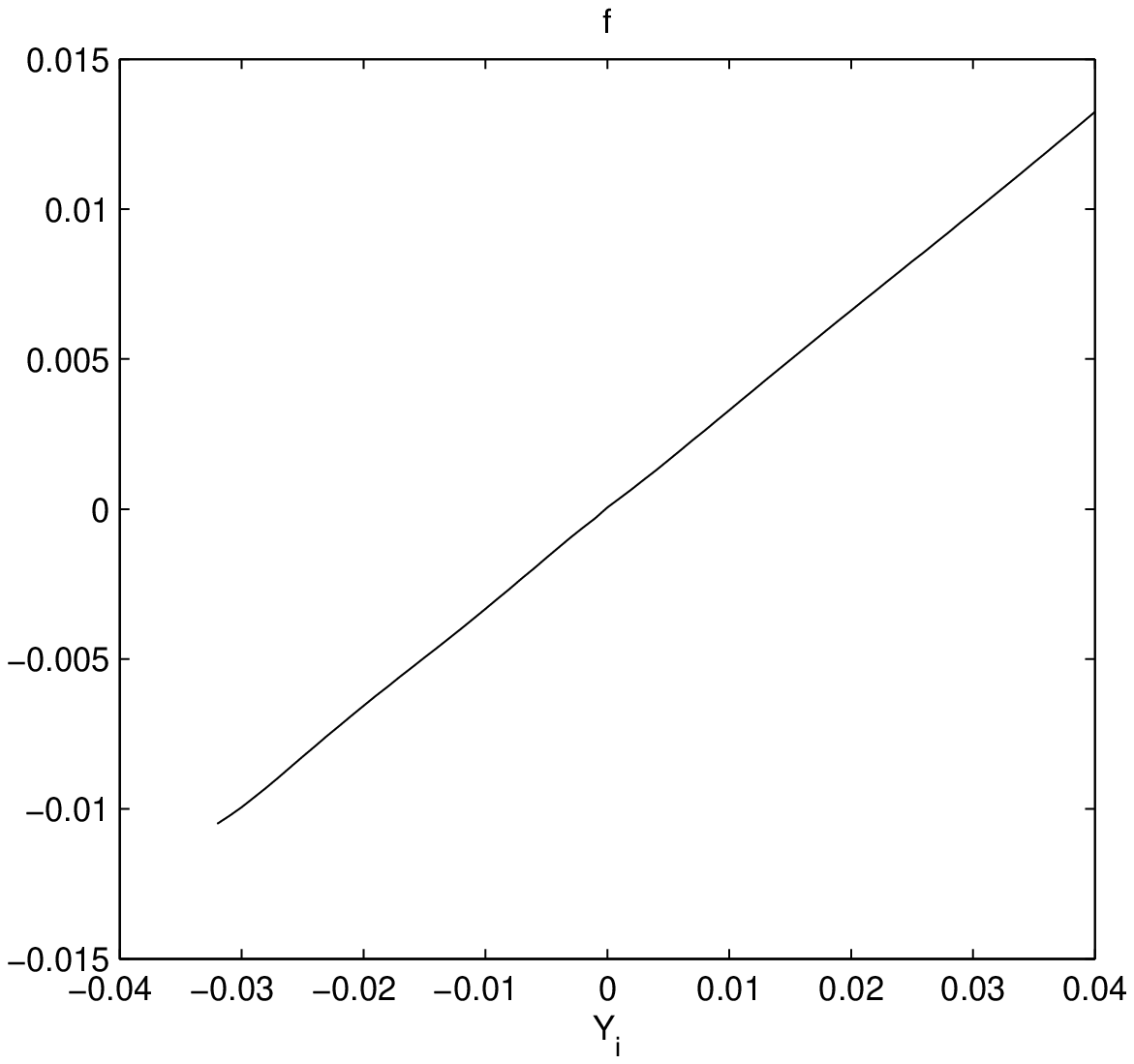}}
\subfigure[$g$]{\includegraphics[width=2.5in,height=2in]{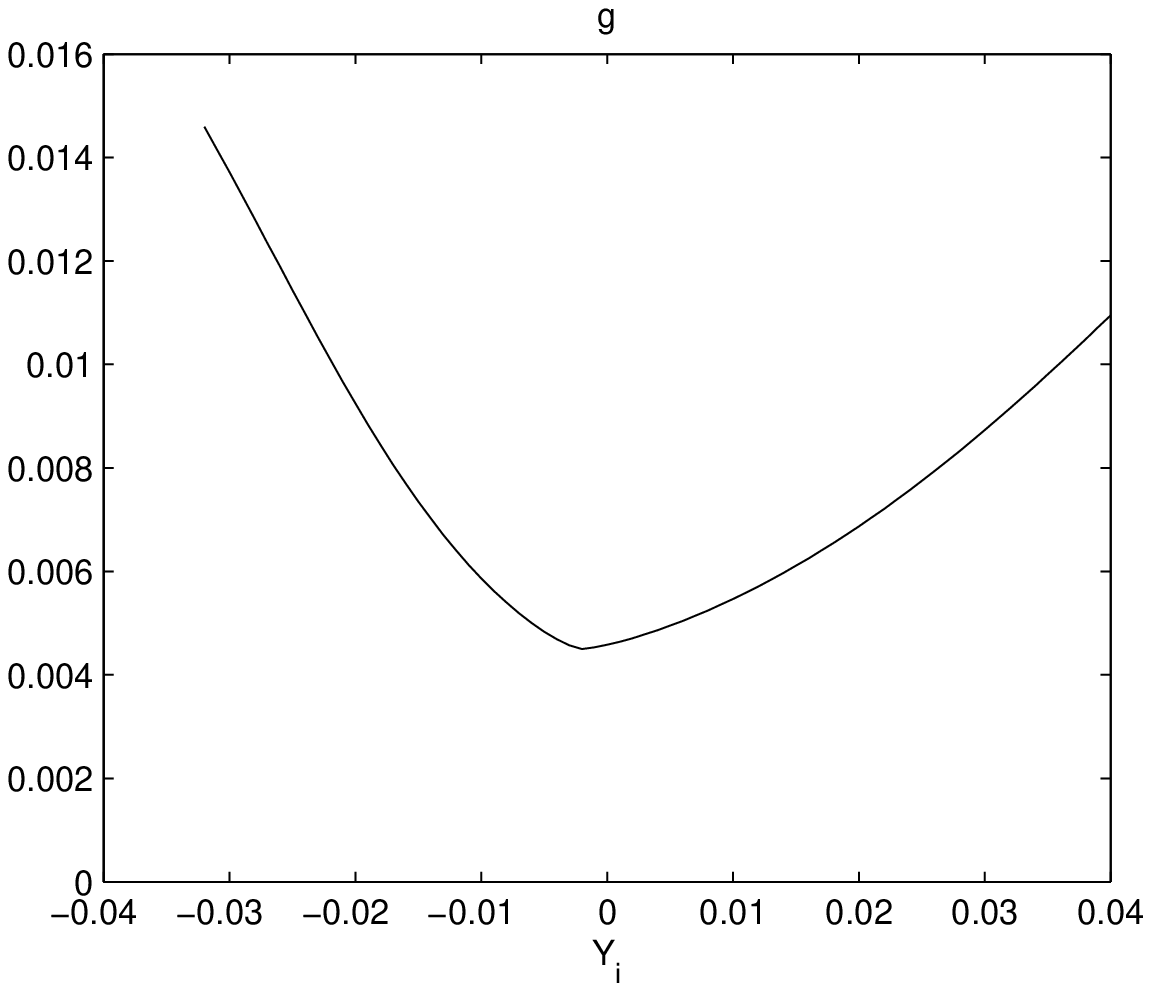}}
\subfigure[$g^2$]{\includegraphics[width=2.5in,height=2in]{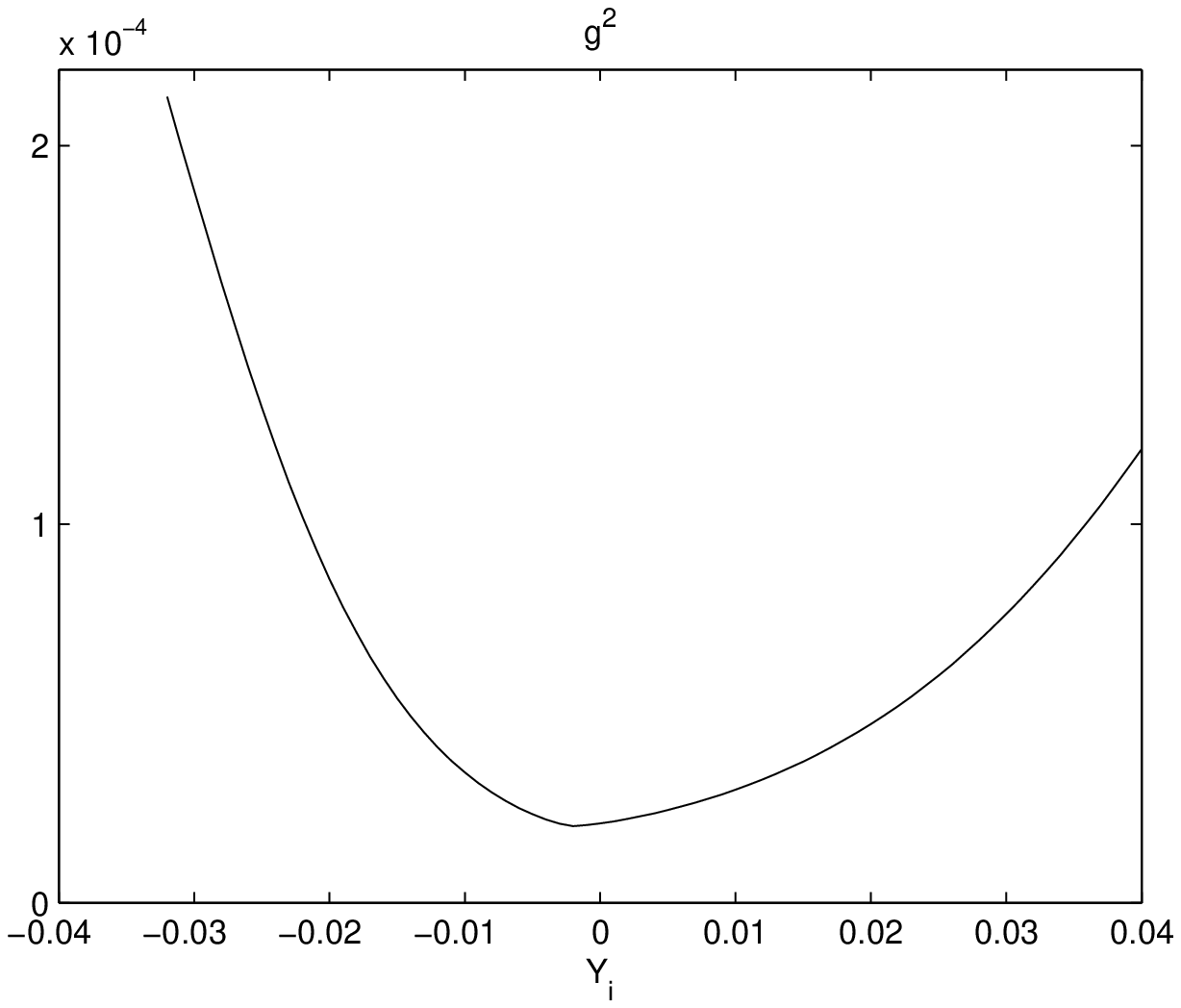}}
\caption{Drift and volatility functions}\label{dfg_cur}
\end{figure}

With the parameter settings $\xi=60$, $\nu=-20$, we plot the
functions $f,g,g^2$ in Figure \ref{dfg_cur}.

Once again the graphs of $g,g^2$ show `smiling faces', which are
comparatively more symmetric than that in Figure \ref{dfg}. However,
the graph of $f$ shows a very different pattern. It can be seen that
$f(Y_i)$ is an increasing function in $Y_i$, which means, if $Y_i$
is positive, then on expectation, $Y_{i+1}$ tends to be positive,
and this tendency is expected to last for several days. On the
contrary, if $Y_i$ is negative, $Y_{i+1}$ tends to be negative and
this pattern may last for some time, expectedly. Also we  notice
that the slope of $f(Y_i)$ is less than $1$, which indicates that
 $Y_{i+1}=f(Y_i)$($i\to\infty$) converges to zero if only the drift term is considered.

\begin{figure}[h!] \centering
\subfigure[$\hat{f}$]{\includegraphics[width=2.5in,
height=2in]{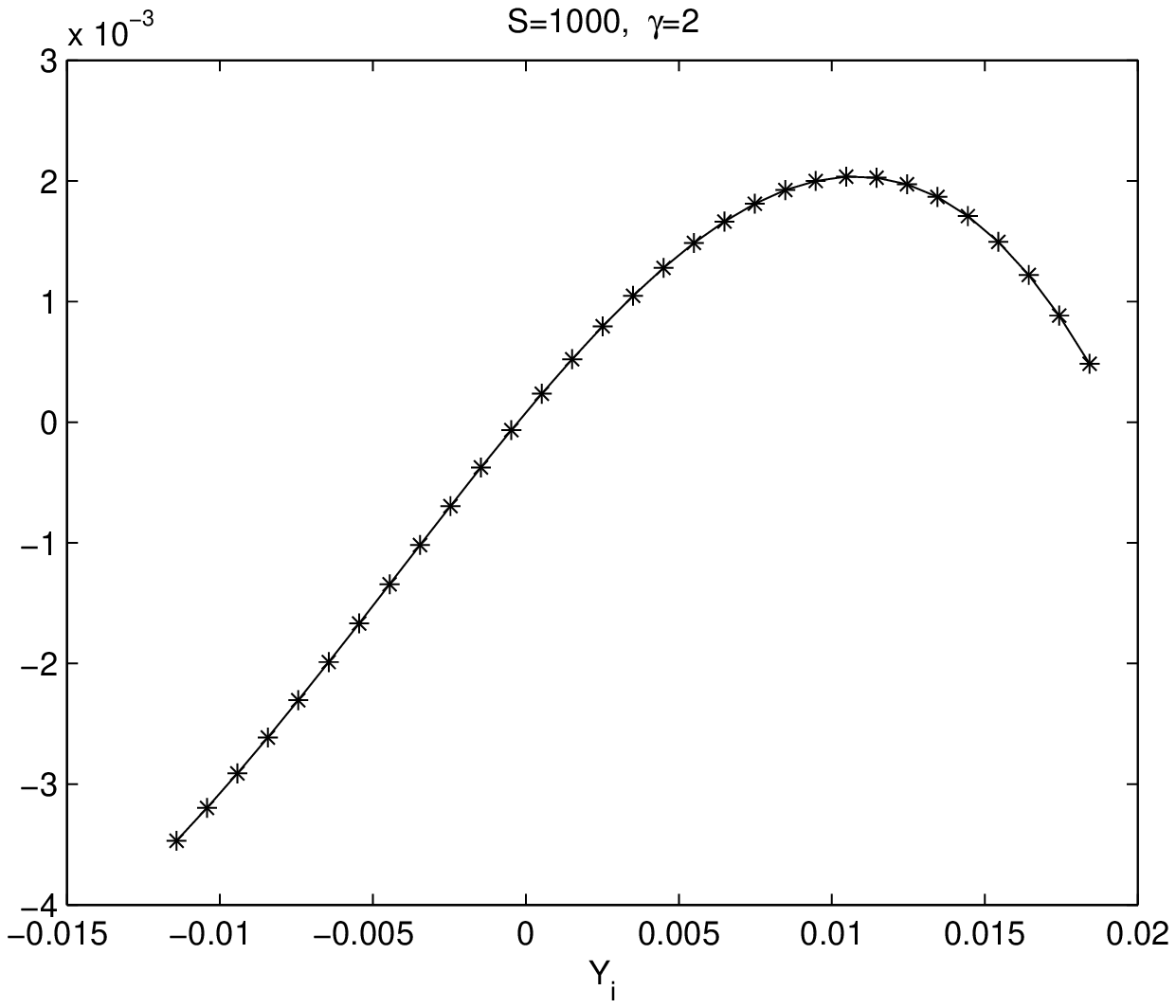}}
\subfigure[$\hat{g}^2$]{\includegraphics[width=2.5in,height=2in]{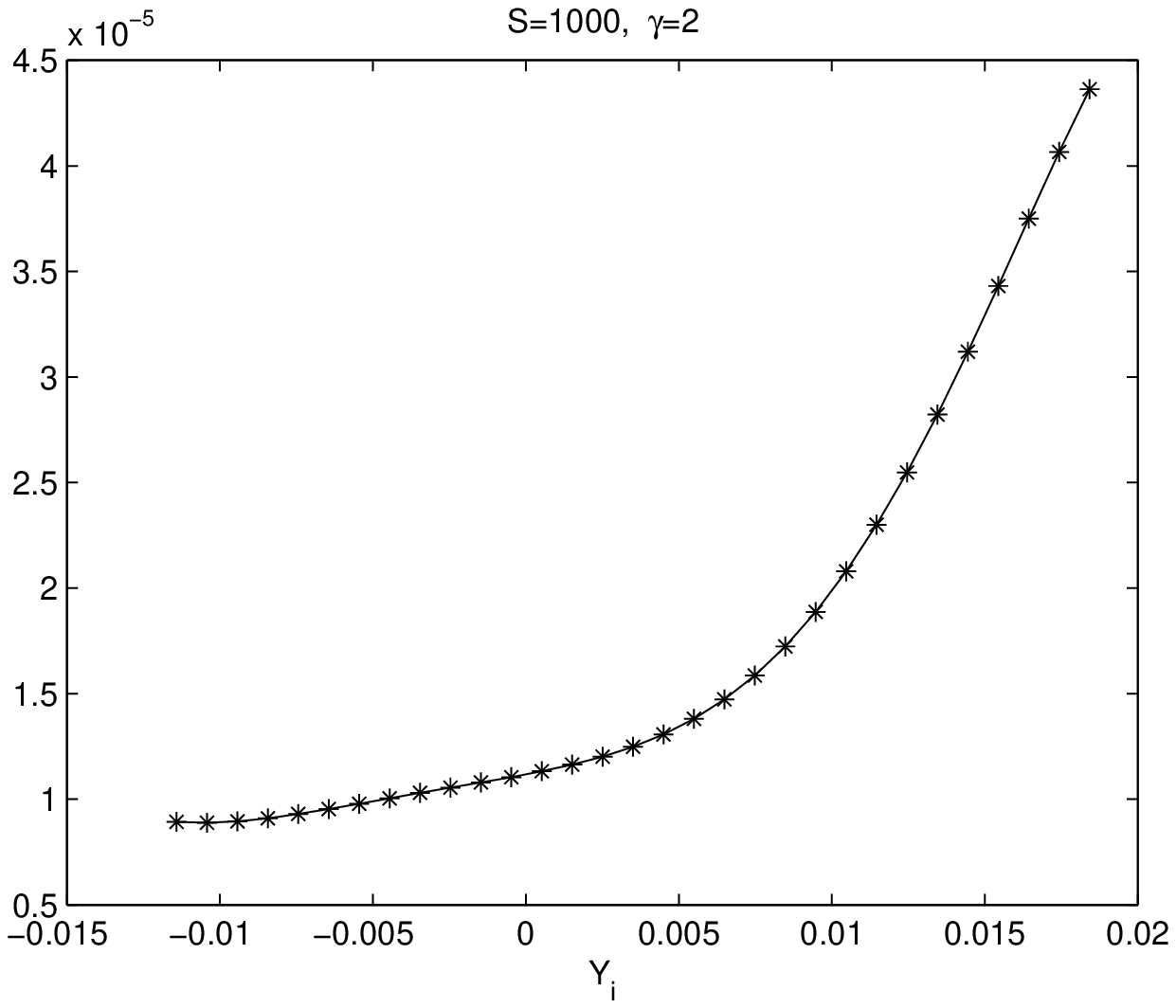}}
\caption{$S= 1000, \gamma=2$}\label{c2_cur-1000}
\end{figure}

\begin{figure}[h!] \centering
\subfigure[$\hat{f}$]{\includegraphics[width=2.5in,
height=2in]{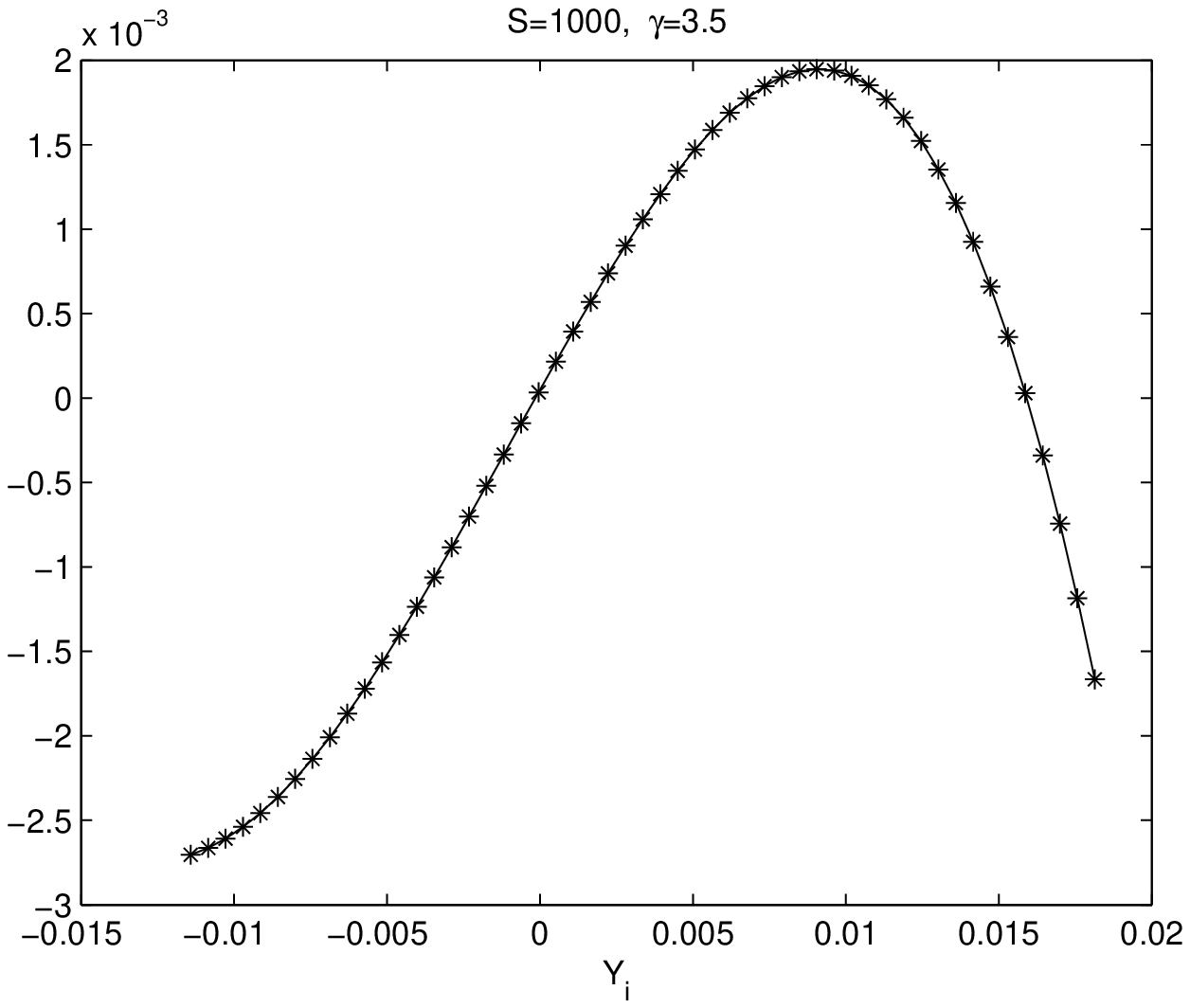}}
\subfigure[$\hat{g}^2$]{\includegraphics[width=2.5in,height=2in]{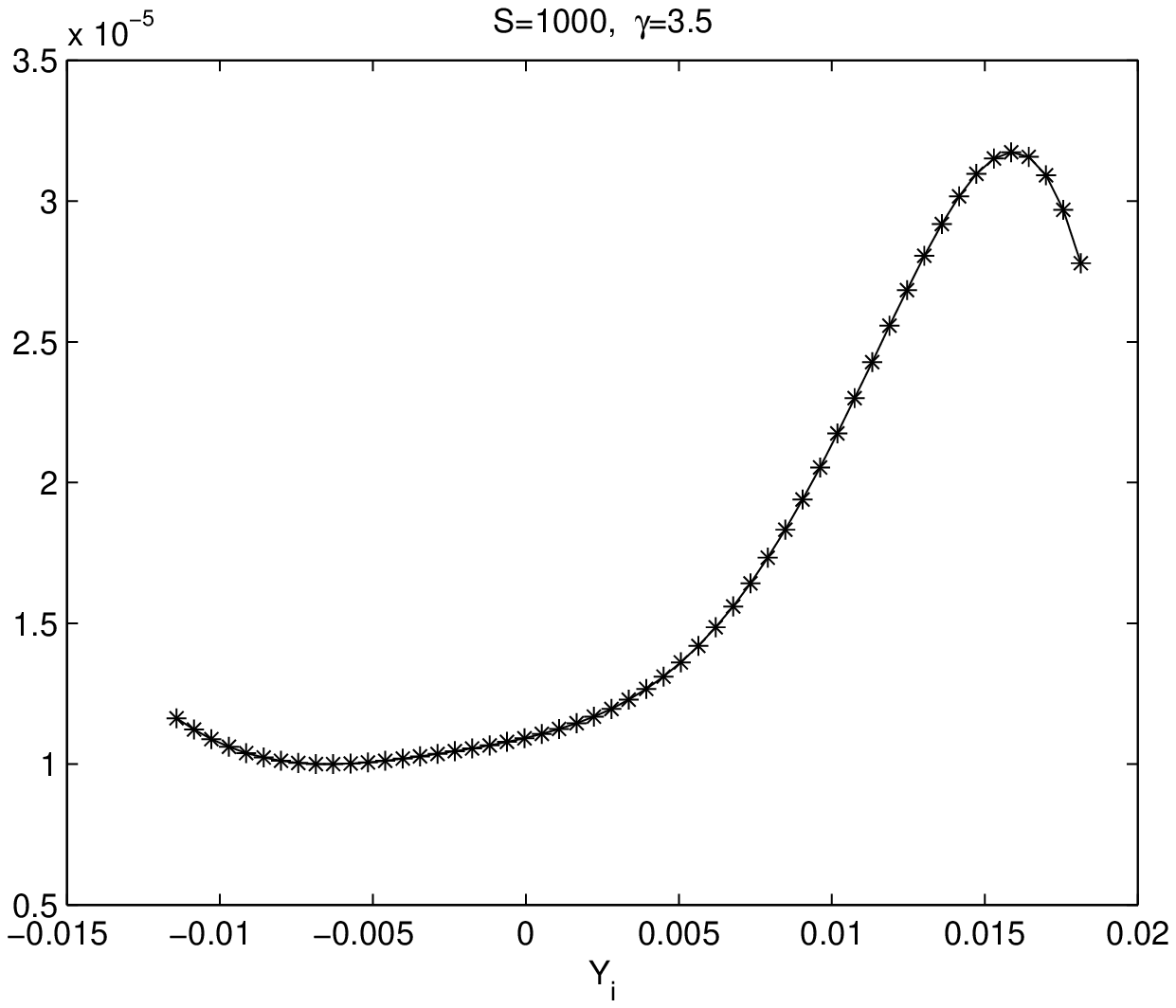}}
\caption{$S= 1000, \gamma=3.5$}\label{c35_cur-1000}
\end{figure}

\begin{figure}[h!] \centering
\subfigure[$\hat{f}$]{\includegraphics[width=2.5in,
height=2in]{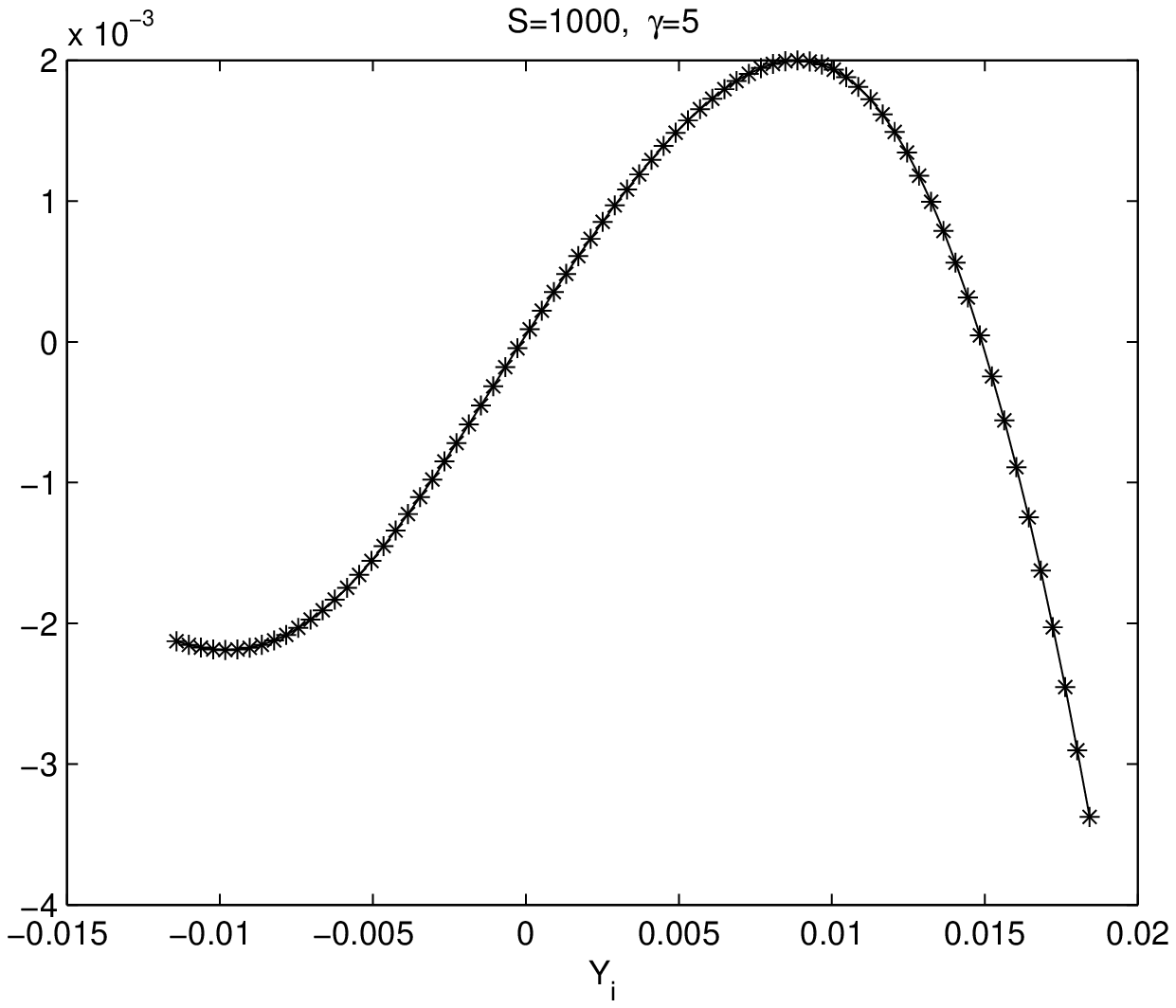}}
\subfigure[$\hat{g}^2$]{\includegraphics[width=2.5in,height=2in]{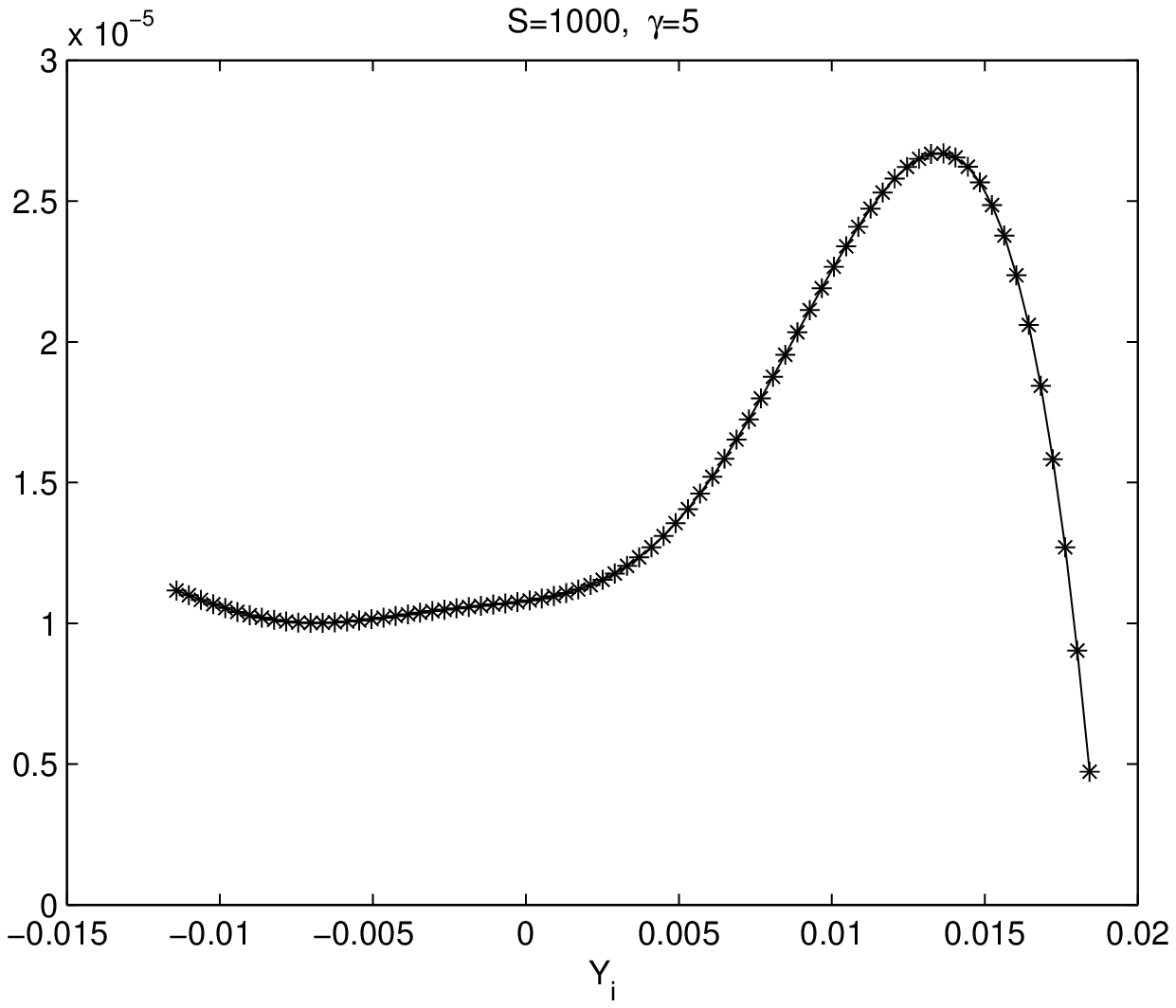}}
\caption{$S= 1000, \gamma=5$}\label{c5_cur-1000}
\end{figure}

\begin{figure}[h!] \centering
\subfigure[$\hat{f}$]{\includegraphics[width=2.5in,
height=2in]{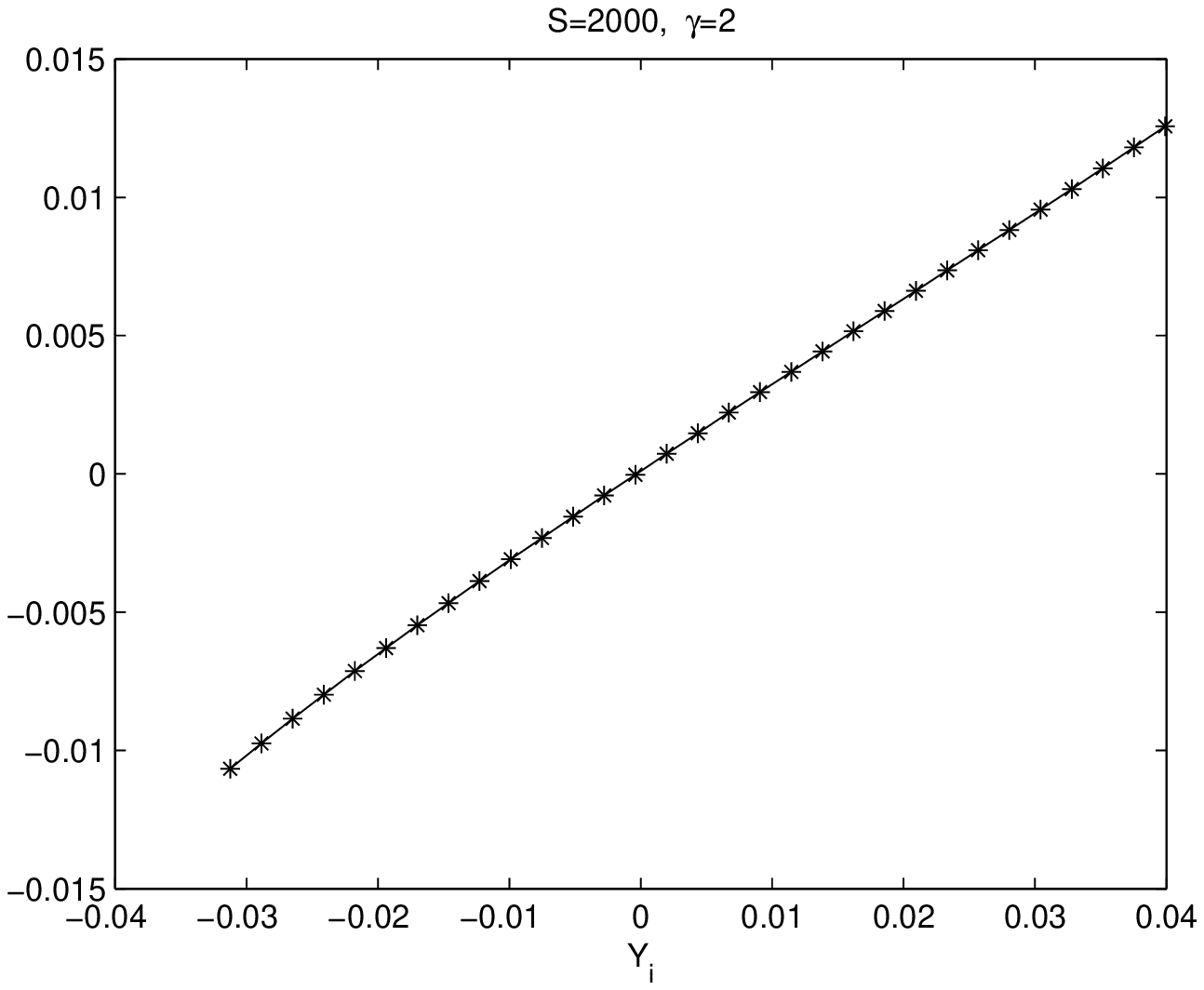}}
\subfigure[$\hat{g}^2$]{\includegraphics[width=2.5in,height=2in]{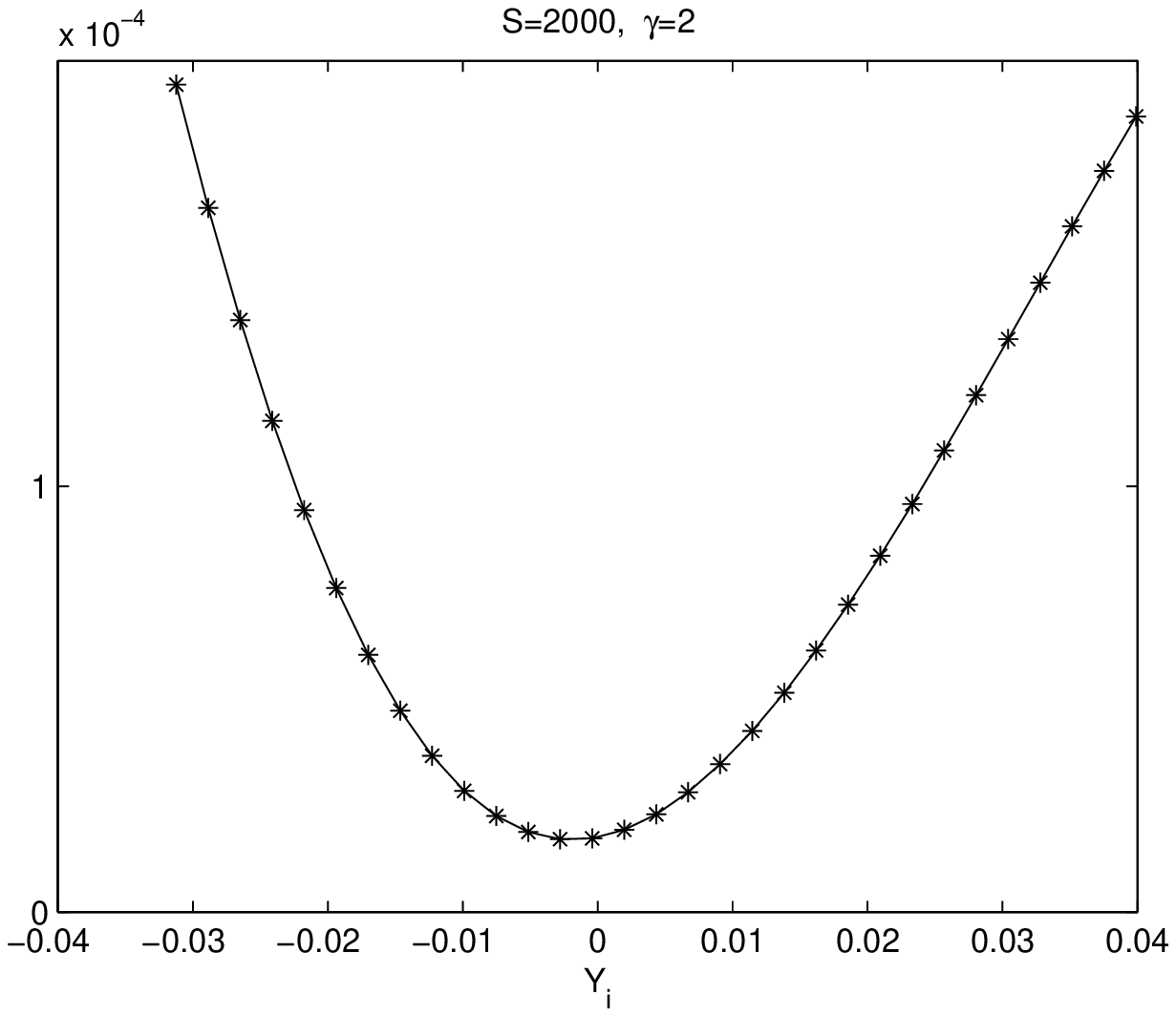}}
\caption{$S= 2000, \gamma=2$}\label{c2_cur-2000}
\end{figure}

\begin{figure}[h!] \centering
\subfigure[$\hat{f}$]{\includegraphics[width=2.5in,
height=2in]{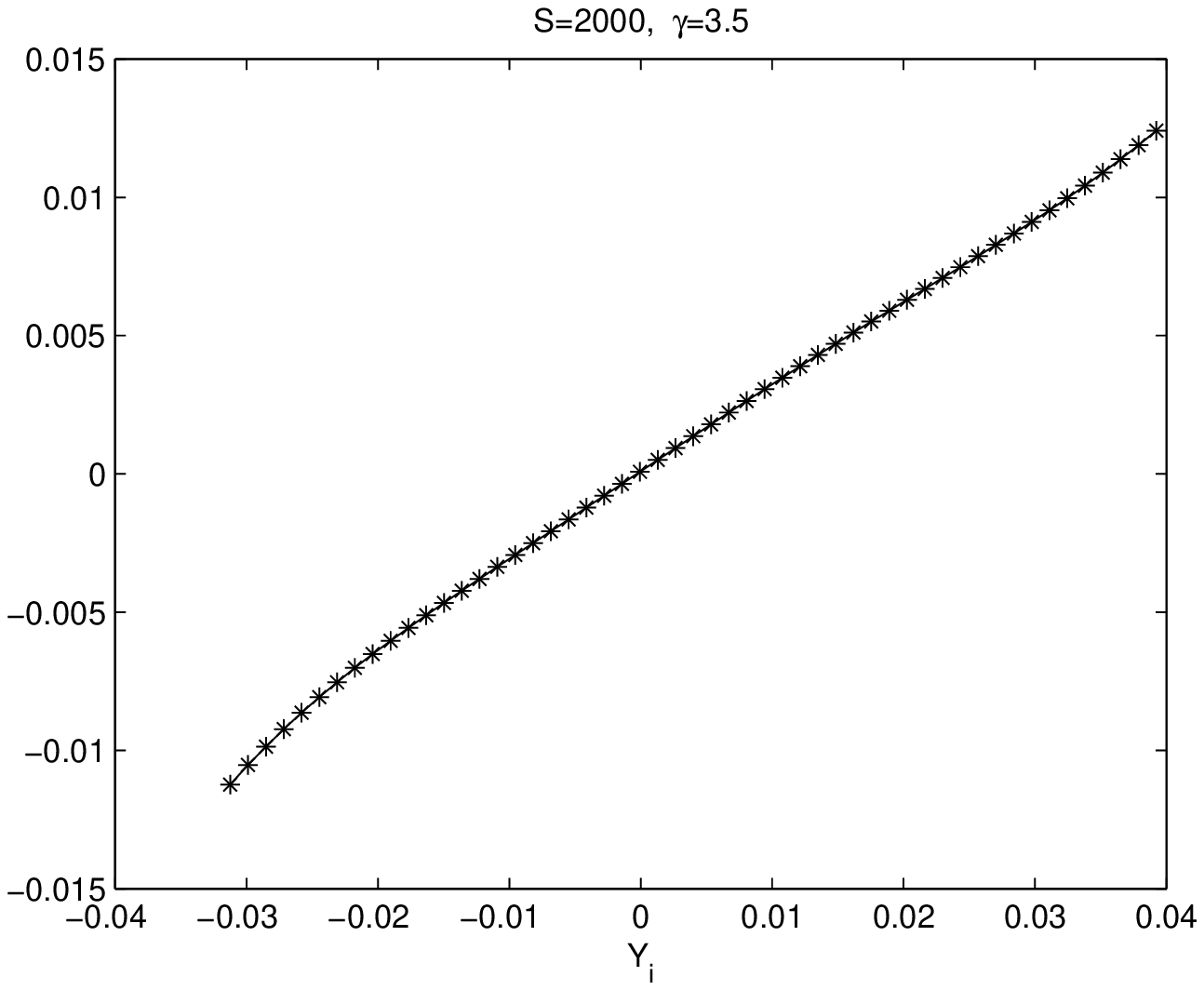}}
\subfigure[$\hat{g}^2$]{\includegraphics[width=2.5in,height=2in]{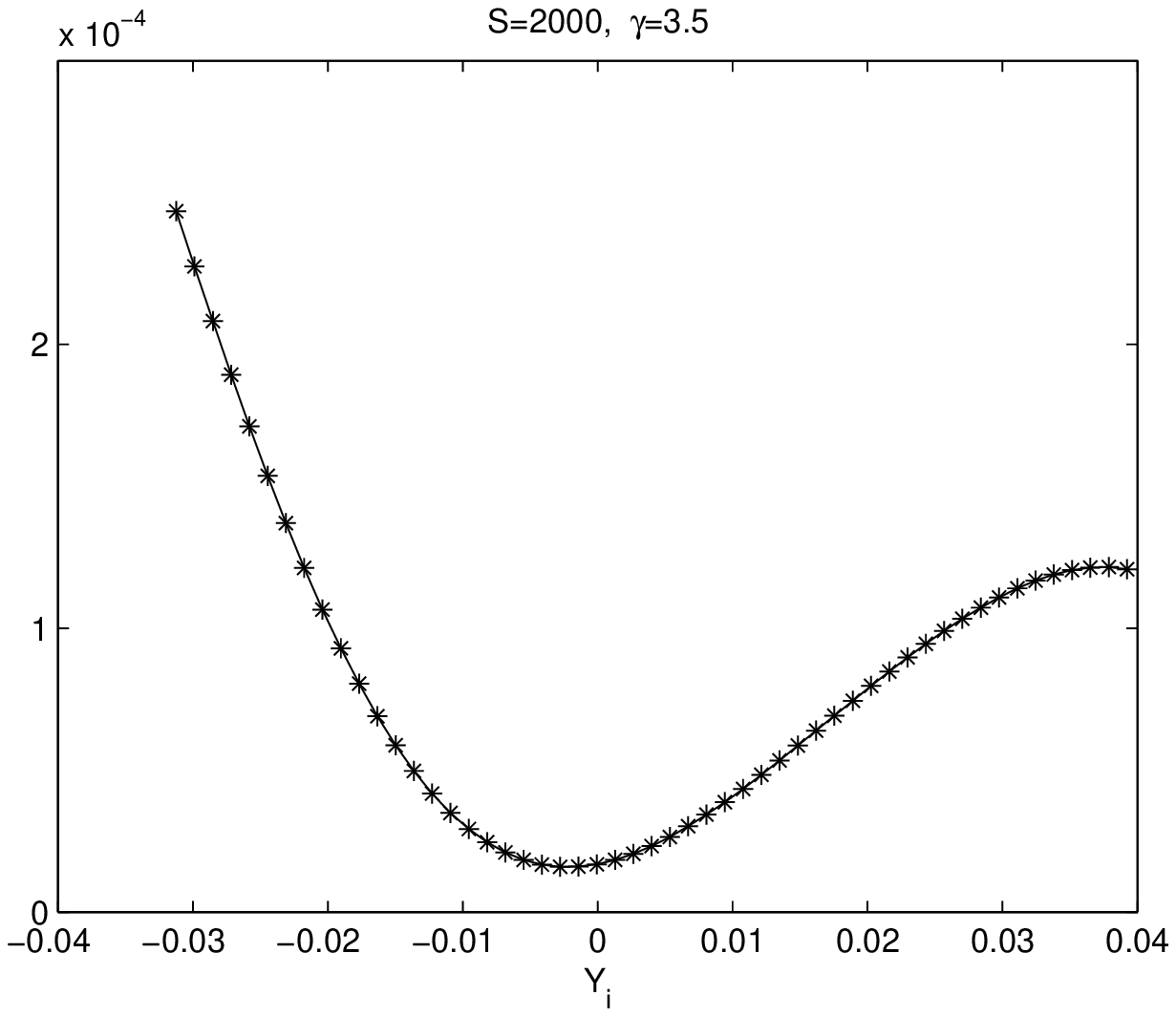}}
\caption{$S= 2000, \gamma=3.5$}\label{c35_cur-2000}
\end{figure}

\begin{figure}[h!] \centering
\subfigure[$\hat{f}$]{\includegraphics[width=2.5in,
height=2in]{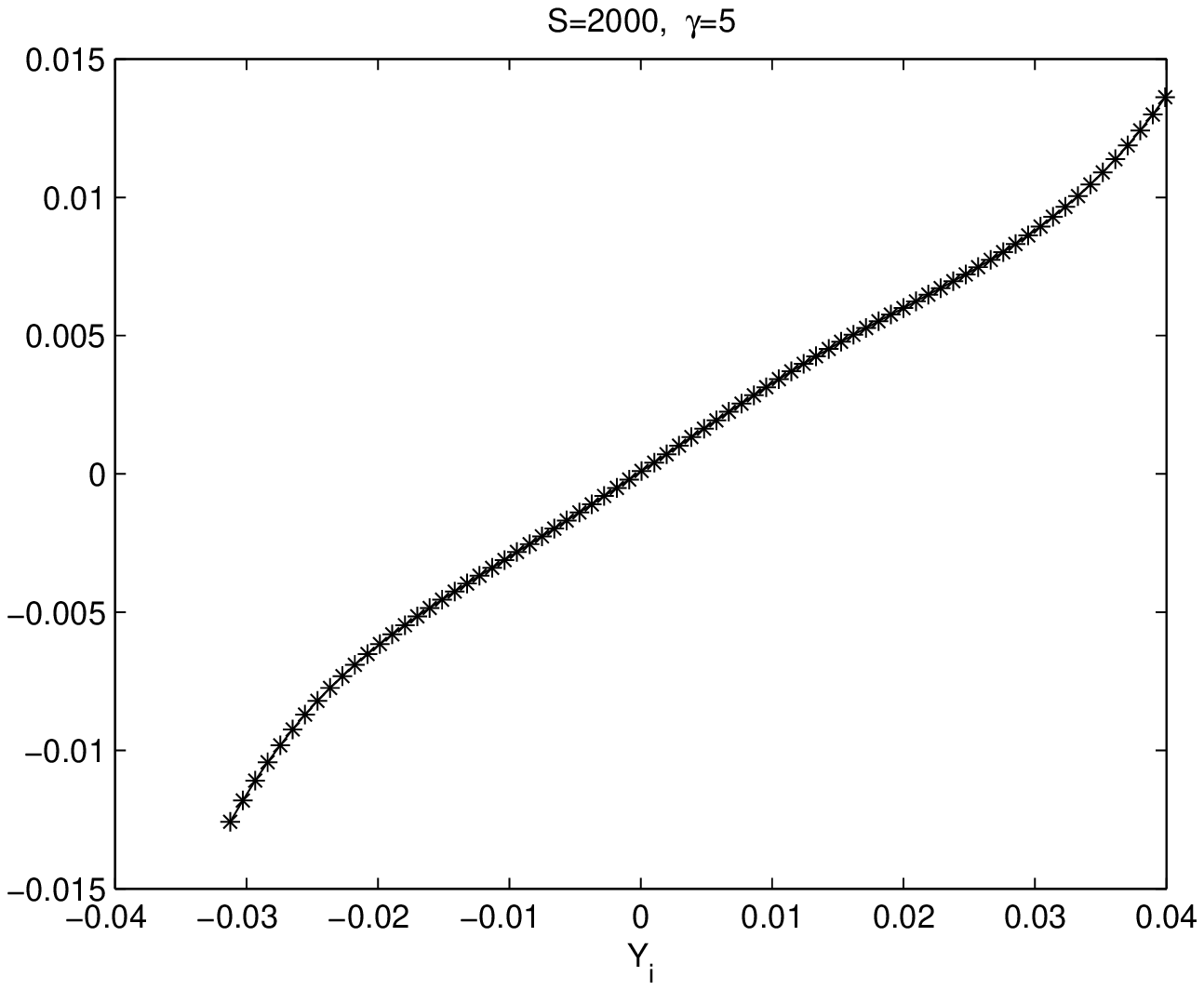}}
\subfigure[$\hat{g}^2$]{\includegraphics[width=2.5in,height=2in]{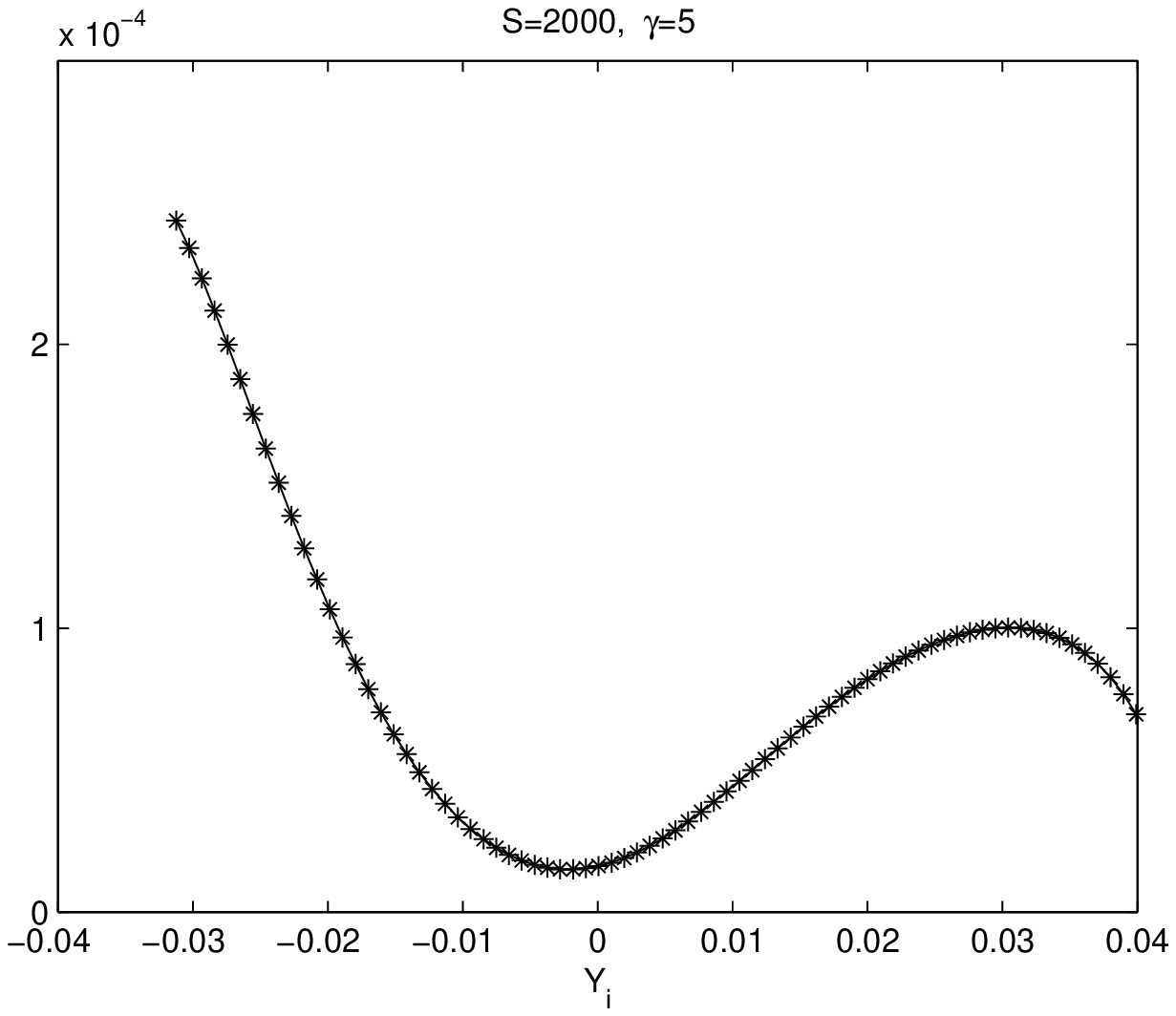}}
\caption{$S= 2000, \gamma=5$}\label{c5_cur-2000}
\end{figure}

\begin{figure}[h!] \centering
\subfigure[$\hat{f}$]{\includegraphics[width=2.5in,
height=2in]{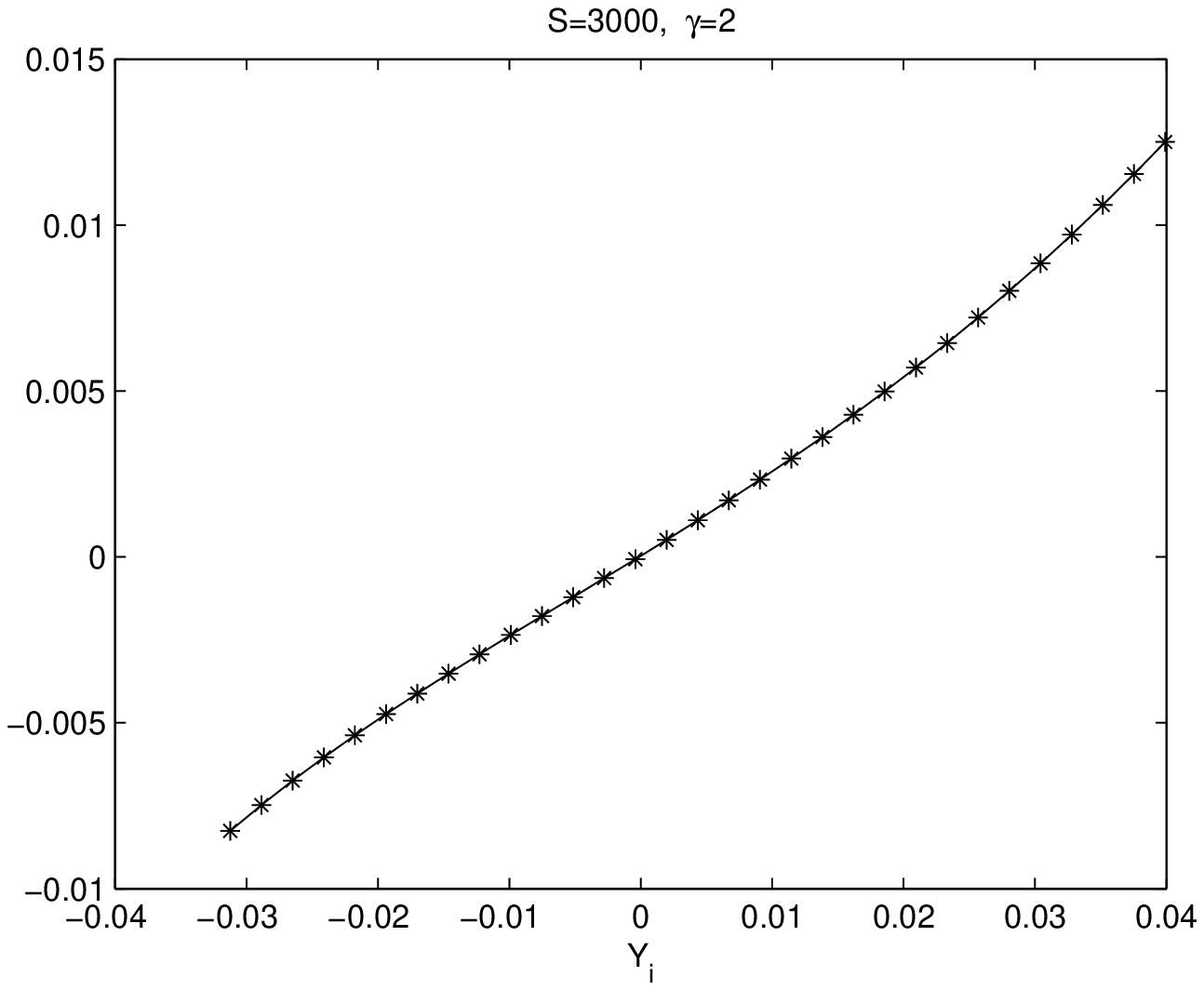}}
\subfigure[$\hat{g}^2$]{\includegraphics[width=2.5in,height=2in]{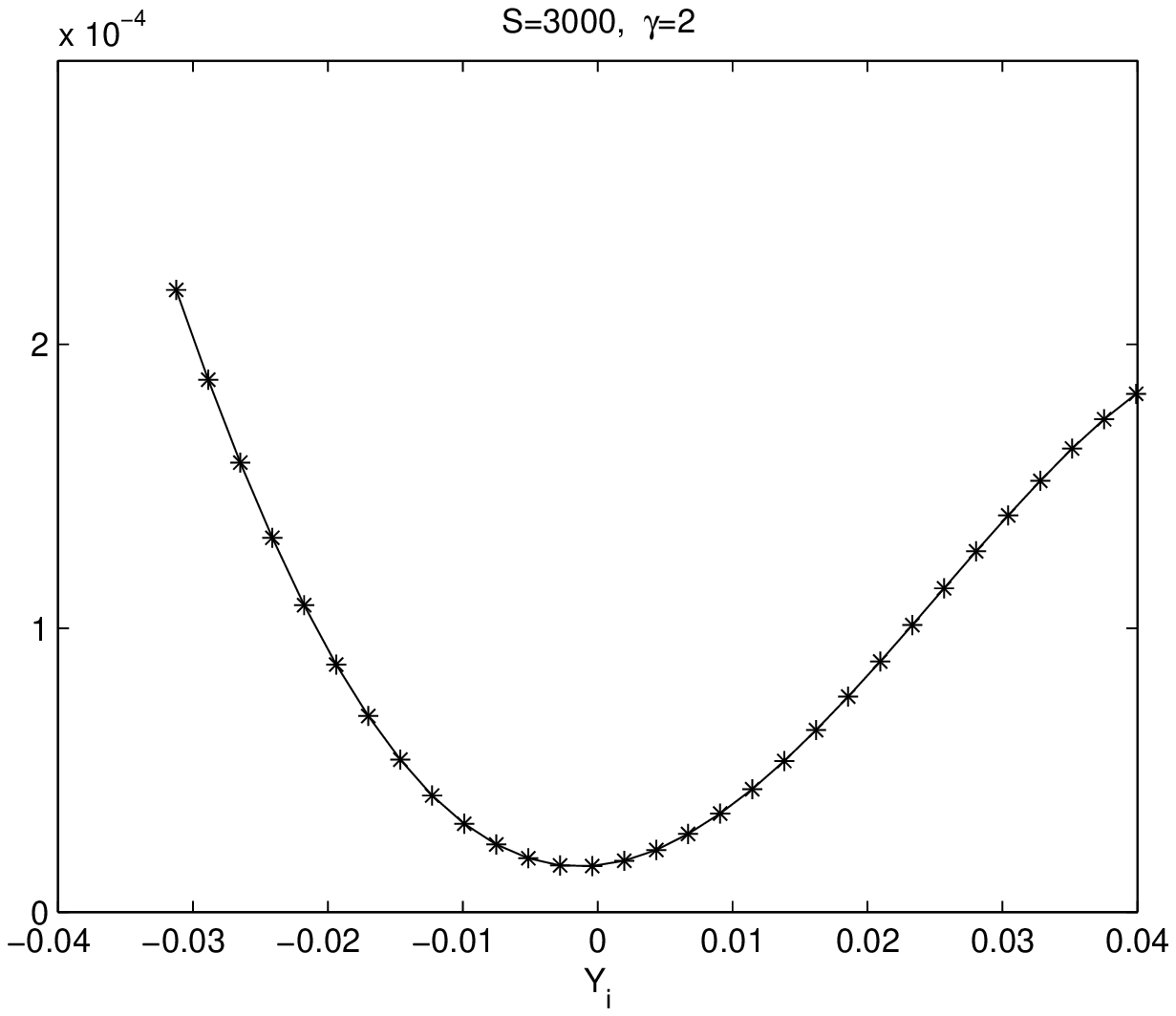}}
\caption{$S= 3000, \gamma=2$}\label{c2_cur-3000}
\end{figure}

\begin{figure}[h!] \centering
\subfigure[$\hat{f}$]{\includegraphics[width=2.5in,
height=2in]{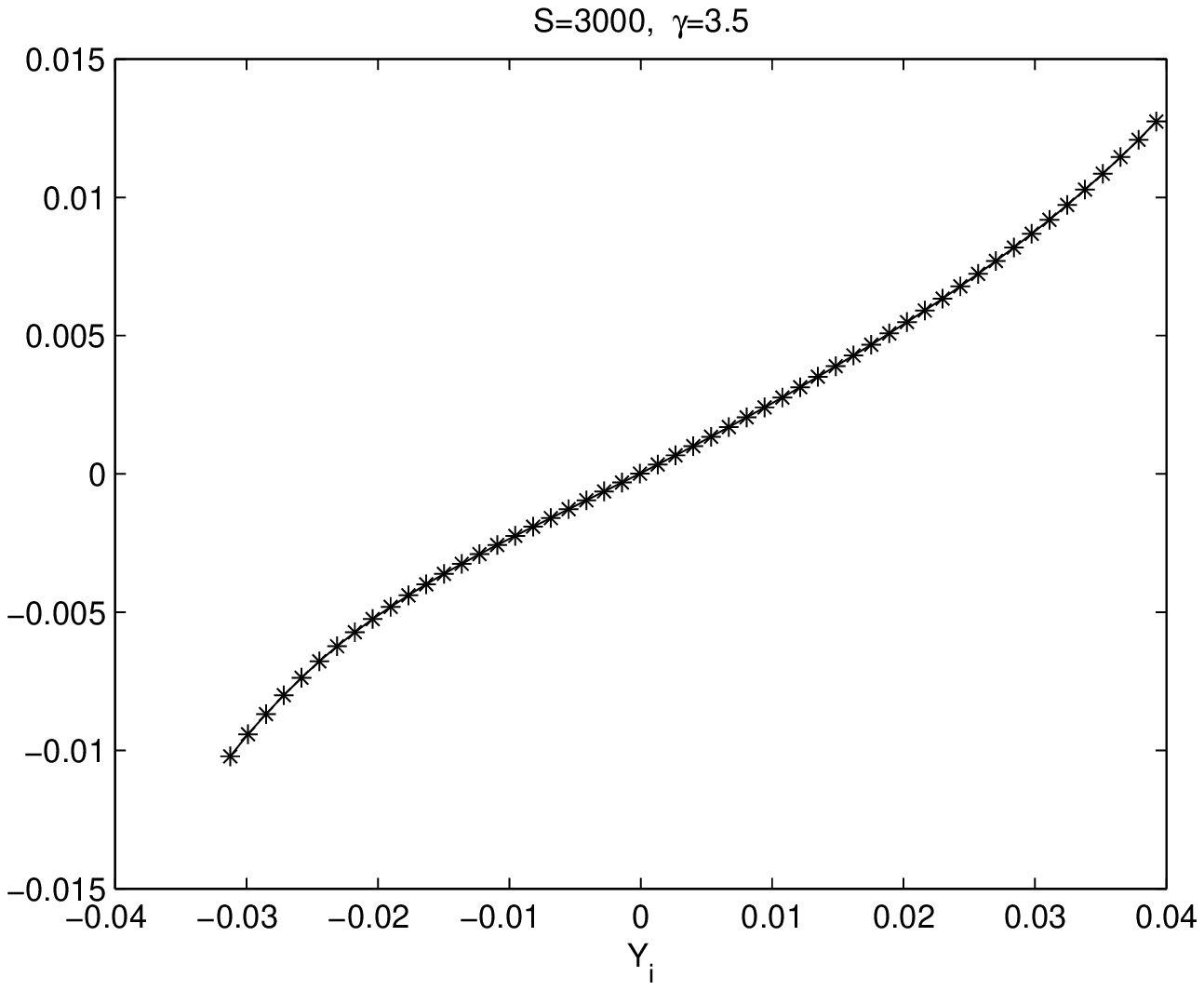}}
\subfigure[$\hat{g}^2$]{\includegraphics[width=2.5in,height=2in]{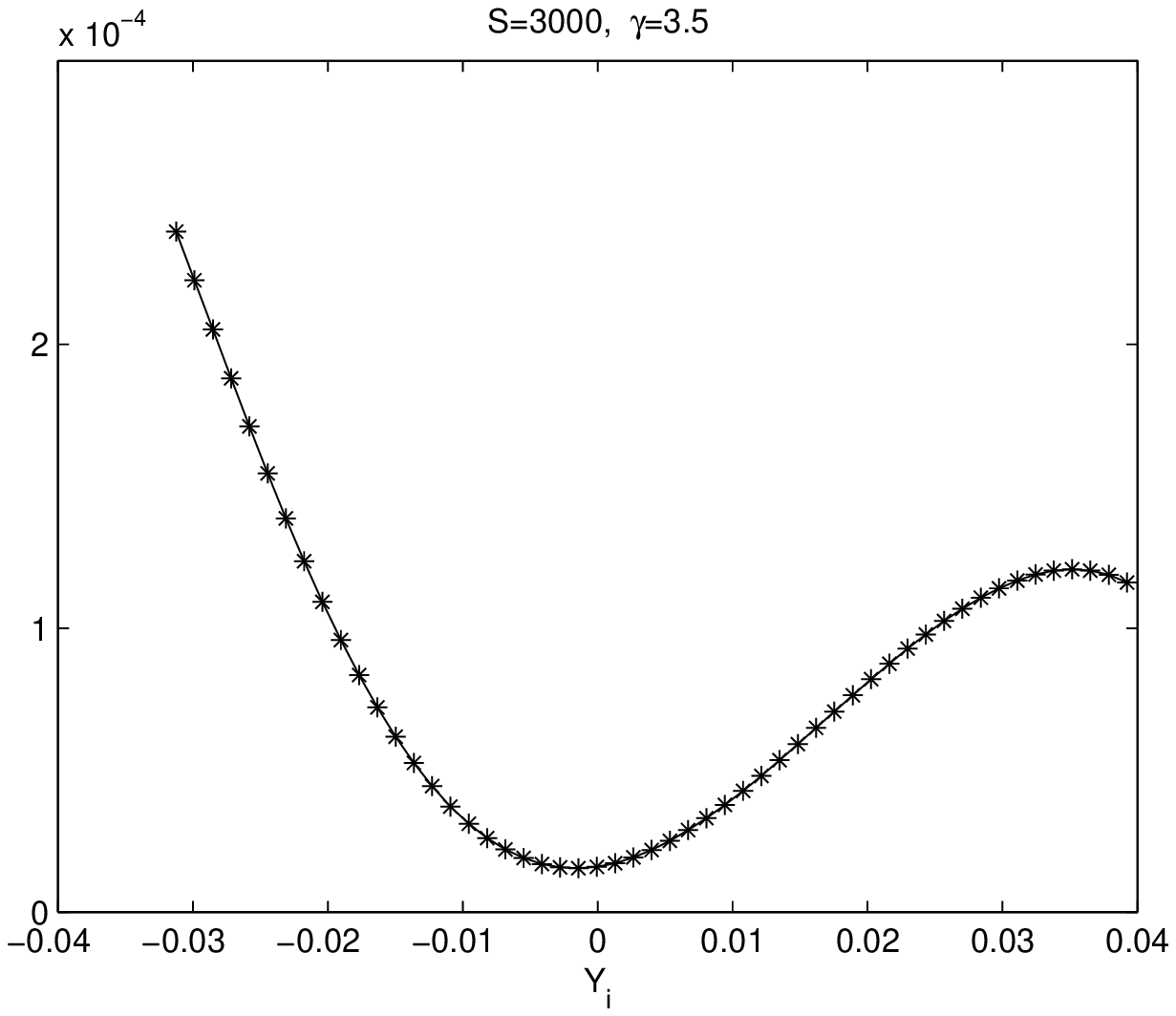}}
\caption{$S= 3000, \gamma=3.5$}\label{c35_cur-3000}
\end{figure}

\begin{figure}[h!] \centering
\subfigure[$\hat{f}$]{\includegraphics[width=2.5in,
height=2in]{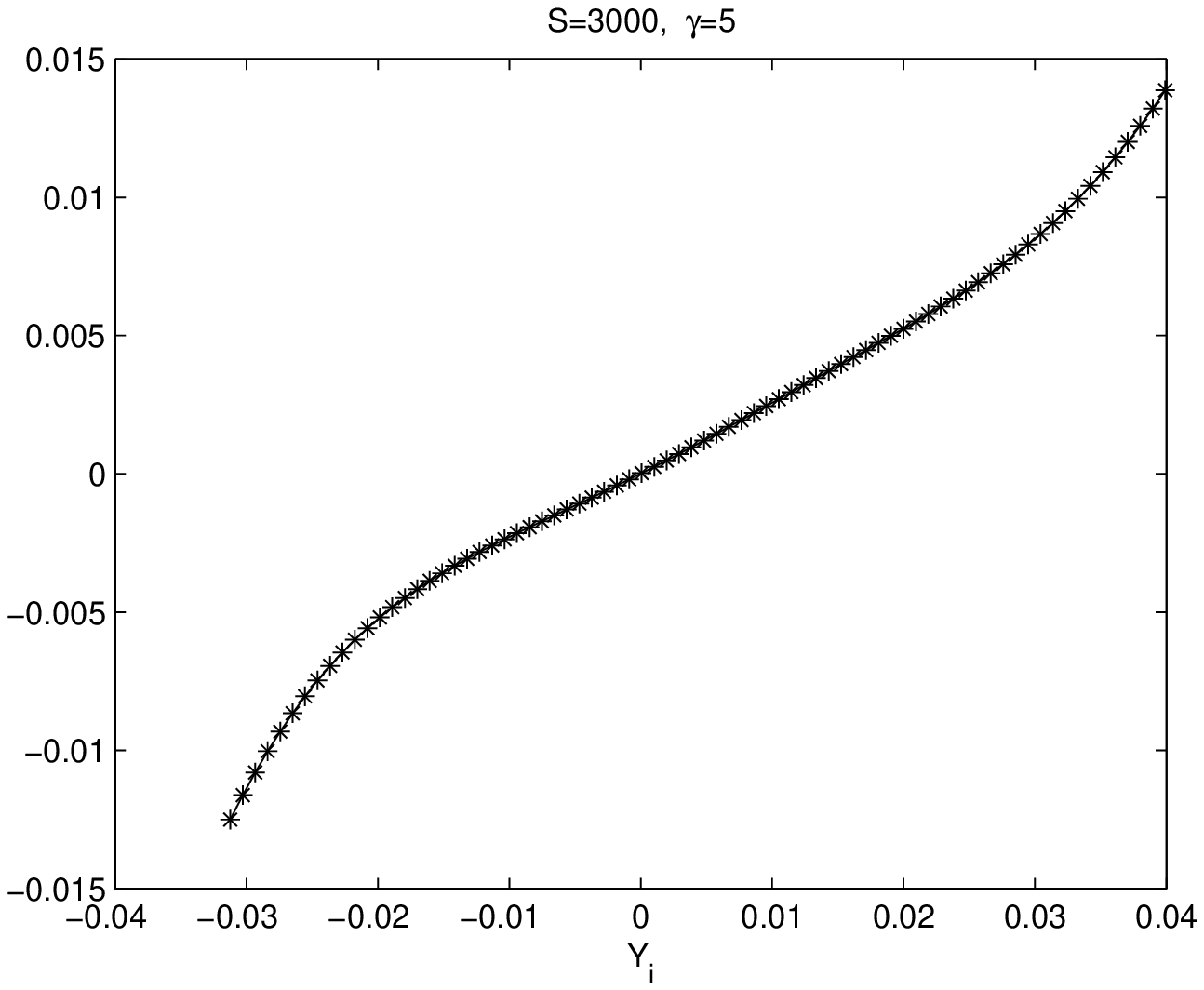}}
\subfigure[$\hat{g}^2$]{\includegraphics[width=2.5in,height=2in]{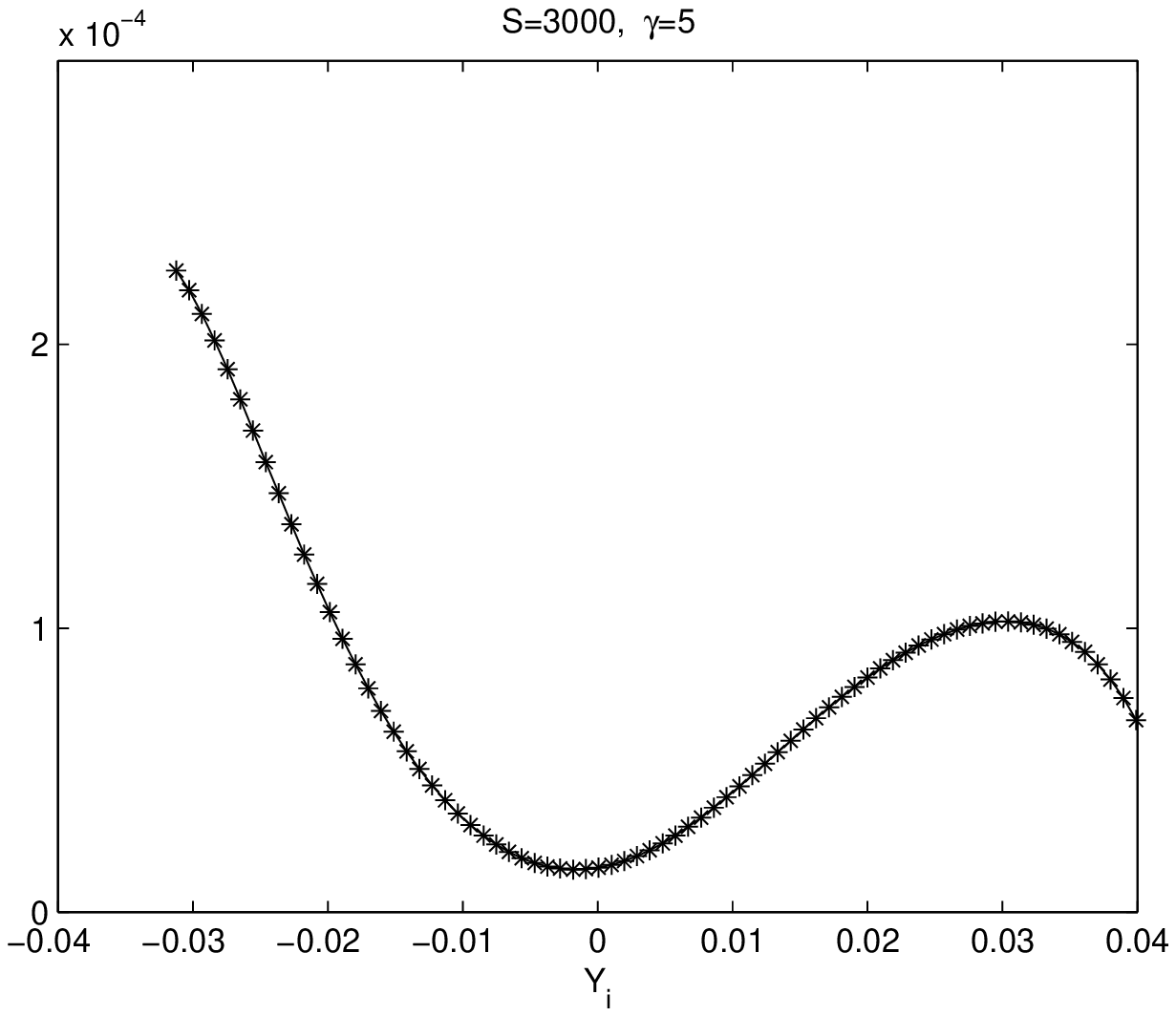}}
\caption{$S= 3000, \gamma=5$}\label{c5_cur-3000}
\end{figure}

We use real market data and local polynomial regression to estimate
the functions $f$ and $g^2$. Let $P_i$ be the USD/GBP exchange rate,
where the last date in the data set is 5/21/2013. Different data
sizes and bandwidths are chosen and the result is shown in Figures
\ref{c2_cur-1000}-\ref{c5_cur-3000}.

\noindent {\bf Observations}:
\begin{enumerate}
\item When the data size is small, e.g., $S=1000$, the boundary
effect is very significant. Similar graphs are found in
\cite{Tsybakov97}.
\item When $S=2000$ or $S=3000$, $\hat{f}(Y_i)$ is increasing and almost linear in
$Y_i$, but the slope of the graph is less than $1$.
\item Graphs of $\hat{g}^2$ all show `smiling faces'.
\end{enumerate}

\begin{figure}[h!] \centering
\subfigure[$\tilde{g}$]{\includegraphics[width=2.5in,
height=2in]{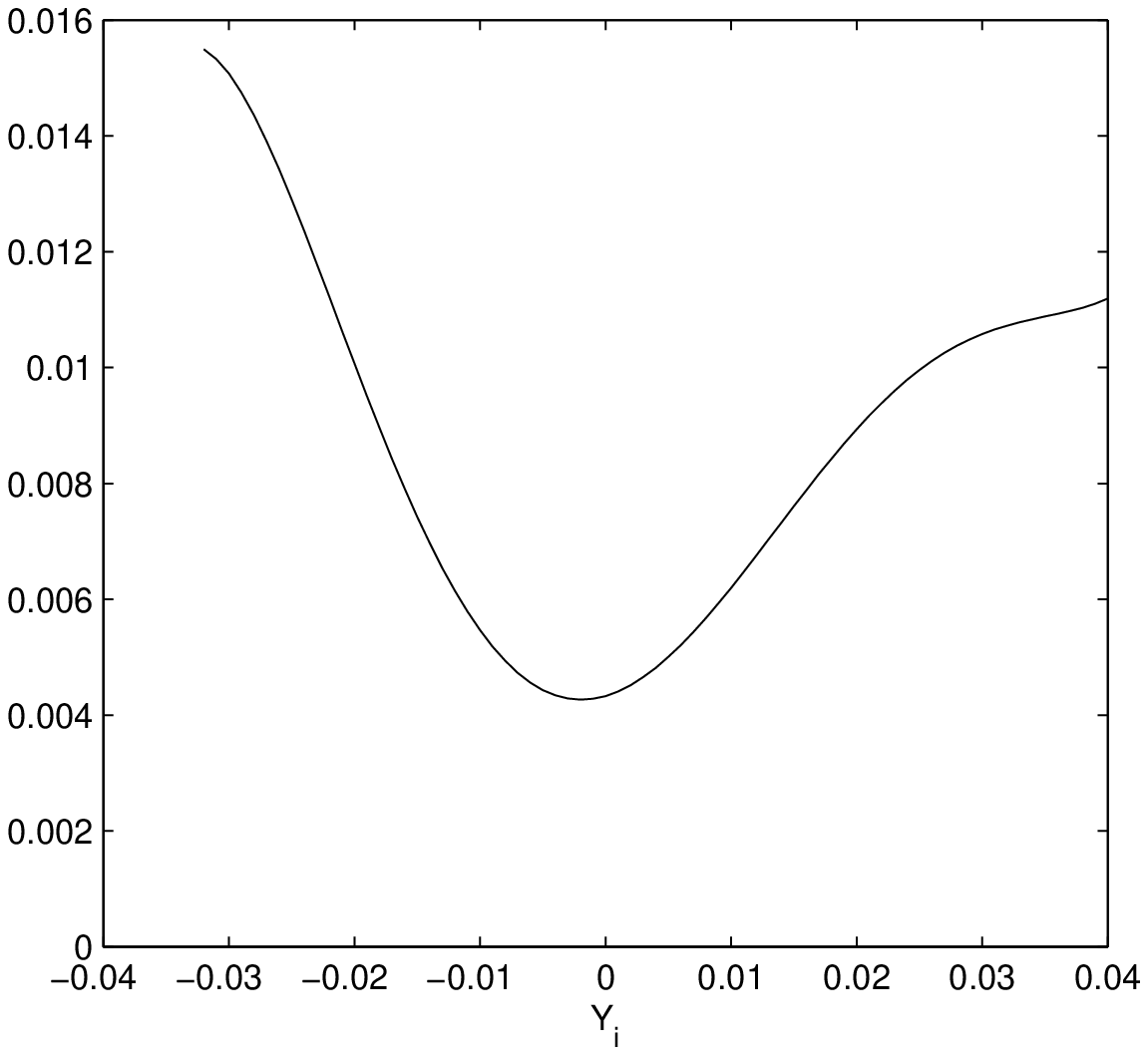}}
\subfigure[$\tilde{g}^2$]{\includegraphics[width=2.5in,height=2in]{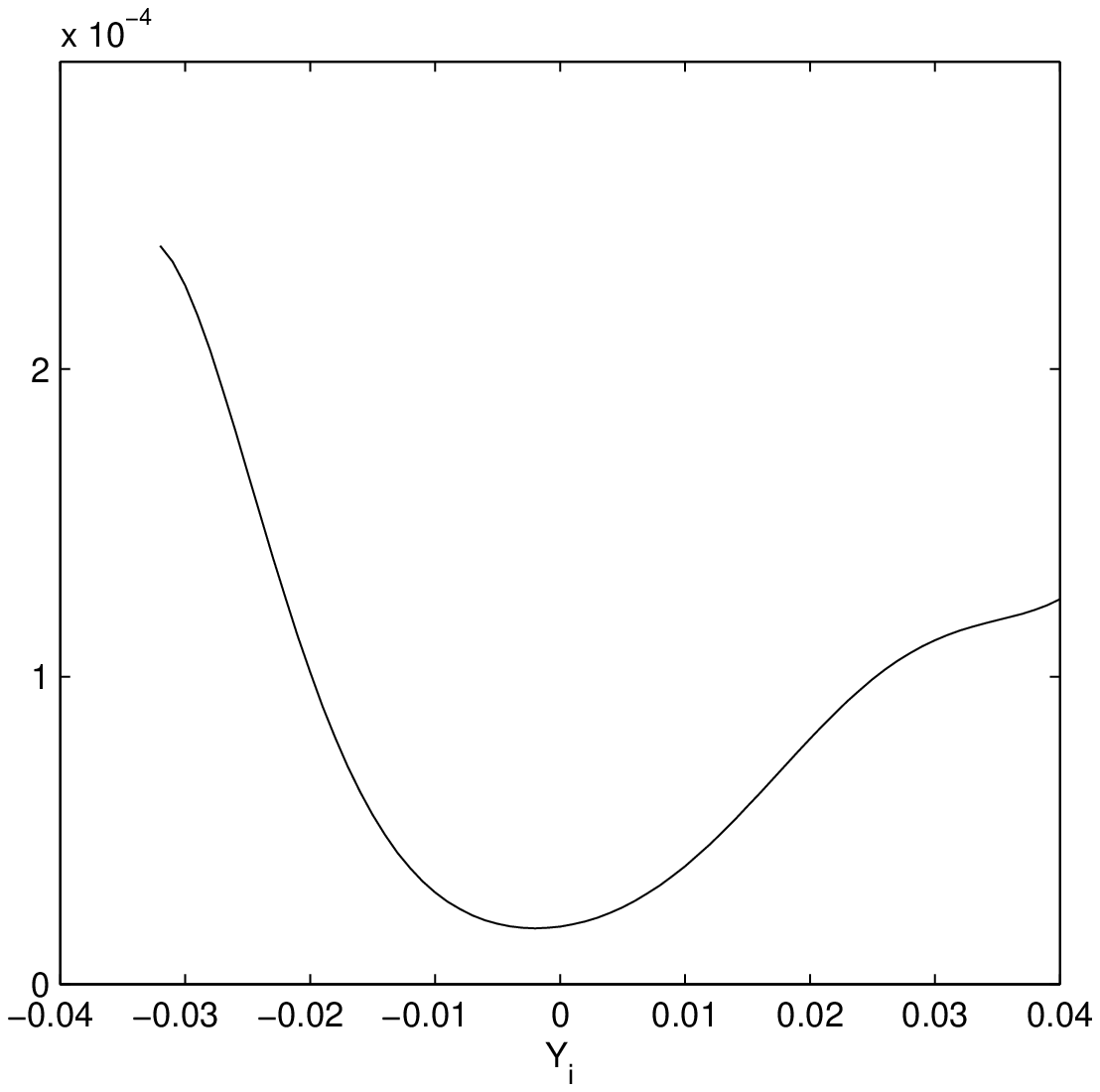}}
\caption{Polynomial approximations of $g,g^2$}\label{gstile_cur}
\end{figure}

In order to construct an ARCH model for $P_i$, we use polynomial
approximations of the functions $f$ and $g$. Let $\tilde{f}$,
$\tilde{g}$ be the polynomial approximations of the functions $f,g$
respectively, where $\tilde{f}(Y_i)=0.33Y_i$,
\begin{displaymath}\begin{split}\tilde{g}(Y_i)=&4.328\times 10^{-3}+6.422\times 10^{-2}Y_i+15.73Y_i^2\\
&-2.934\times 10^2Y_i^3-6.987\times 10^3Y_i^4+1.542\times
10^5Y_i^5,\end{split}\end{displaymath} then we may plot the graphs
of $\tilde{g},\tilde{g}^2$  in Figure \ref{gstile_cur}.

With this setting the following ARCH model is obtained:
\begin{equation}\label{archP}\begin{split}
\log P_{i+1}=&\log P_{i} +0.33Y_i\\
&+(4.328\times 10^{-3}+6.422\times 10^{-2}Y_i+15.73Y_i^2\\
&-2.934\times 10^2Y_i^3-6.987\times 10^3Y_i^4+1.542\times
10^5Y_i^5)\epsilon_i,
\end{split}
\end{equation} where $Y_i = \log P_i-\log P_{i-1}$.

\begin{figure}[h!]
\centerline{\includegraphics[width=3in,height=2.5in]{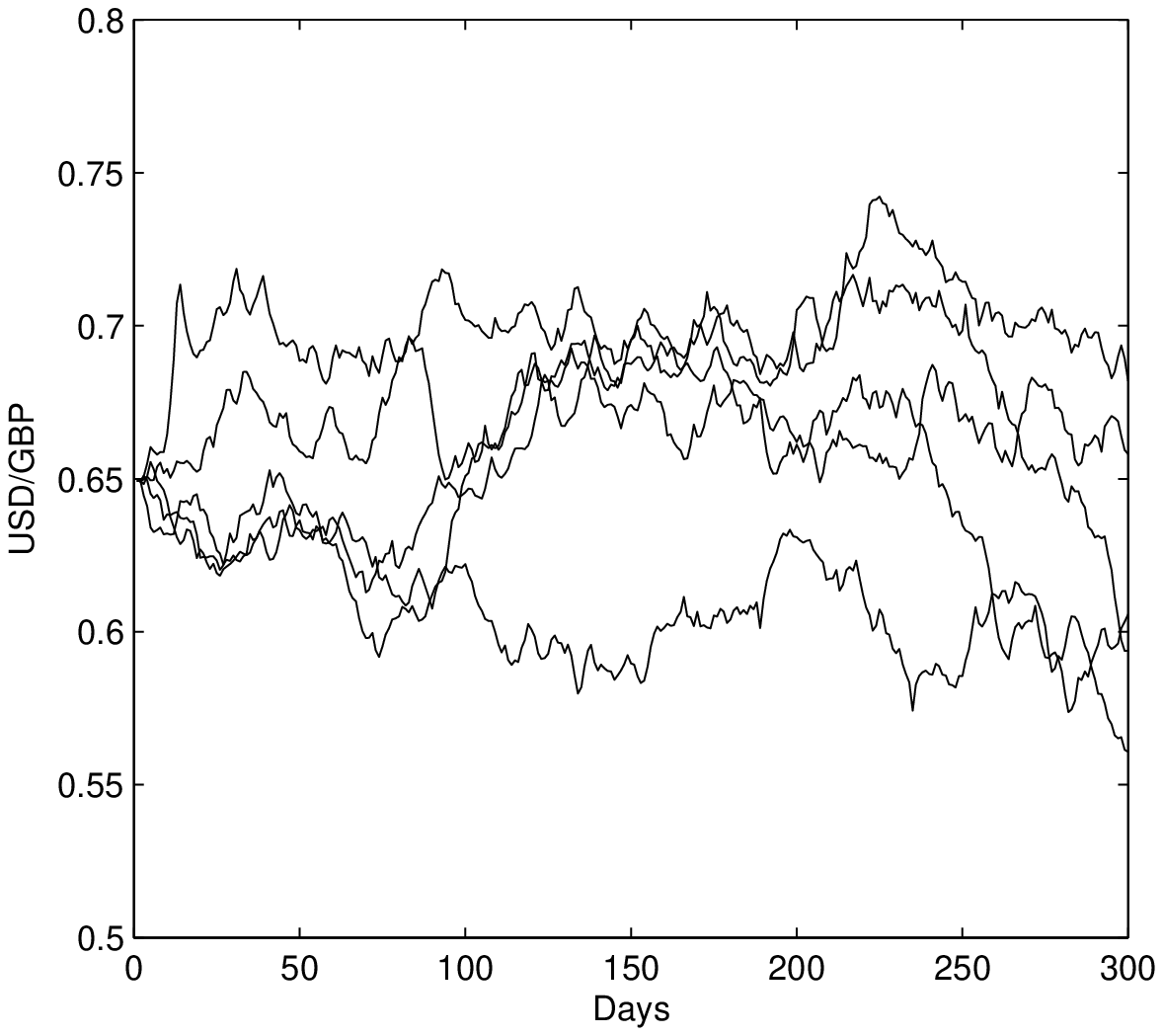}}\hspace{4cm}\caption{Sample
paths}\label{samplepath}
\end{figure}

To illustrate the properties we observed, we randomly generate some
sample paths of $P_i$ according to the model (\ref{archP}) with
starting points $P_1=0.6493$, $P_2=0.6492$, and the result is shown
in Figure \ref{samplepath}.

\section{Conclusions}\label{ending}
In this paper we introduced the  prospect agents as the market
participants. Based on the fact that asset price is generated by the
temporary balance between supply and demand, we model the price
fluctuation through the microeconomic approach. The prospect agents'
reactions to gains and losses affect the price process through the
feedback effect. If they experience gains of their investment based
on their expectation, the prospect agents illustrate a risk averse
behavior on their  demand, while if they experience losses, they
often show a risk seeking behavior. Then  ARCH models are
constructed for both the S\&P500 index process and the USD/GBP
exchange rate, respectively. Through this model we could reproduce
the volatility smile phenomenon on options that is well observed in
actual financial market. To the best of our knowledge, we are the
first to explain the volatility smile/skew through prospect theory.

This paper provides a new approach and a framework to construct
discrete and possibly continuous models for asset prices. Future
research may focus on parameter estimation and extending this method
to model the price fluctuations of other financial products.

\section*{Appendix} In this section we shall discuss a numerical issue on
recovering the implied volatility surface from Black-Scholes
formula. It is well known that the implied volatility of actual
option prices does not equal a constant, but shows a volatility skew
for equity options or volatility smile for foreign exchange
contracts. Many models have been developed to replicate this
phenomenon. For example, the local volatility model \cite{Dupire94},
the stochastic volatility model \cite{Heston93}, and the
Jump-Diffusion models \cite{Duffie00}\cite{Kou02}.

When a model is presented, very often one uses Monte Carlo
simulation to find the options prices and put them in Black-Scholes
formula to find the implied volatilities. Then an implied volatility
surface can be found, and if it mimics the actual volatility
surface, one claims that this model is a good one. However, there is
a numerical issue involved. Even if the classical geometric Brownian
motion model $dP_t = rP_tdt+\sigma P_t dW_t$ with constant
volatility is used in the Monte Carlo simulation, it is very hard to
recover $\sigma$ as a constant if the aforementioned procedure is
implemented. Of course the price generated by Monte Carlo simulation
can not perfectly replicate the Black-Scholes price due to the
computer generated pseudo random numbers. But even if the error
between these two prices is tiny, in some cases the implied
volatility is quite different from $\sigma$. The reason behind this
is due to $Vega$, and in some situations the option price is
extremely insensitive to volatility. That means, on the contrary,
even if there is a tiny error on the option price, there will be a
big error between the implied volatility and $\sigma$. In what
follows we shall study the cases where $Vega$ is small, i.e., option
price is insensitive to volatility.

\begin{figure}[h!]\centering
\includegraphics[width=4in,
height=3in]{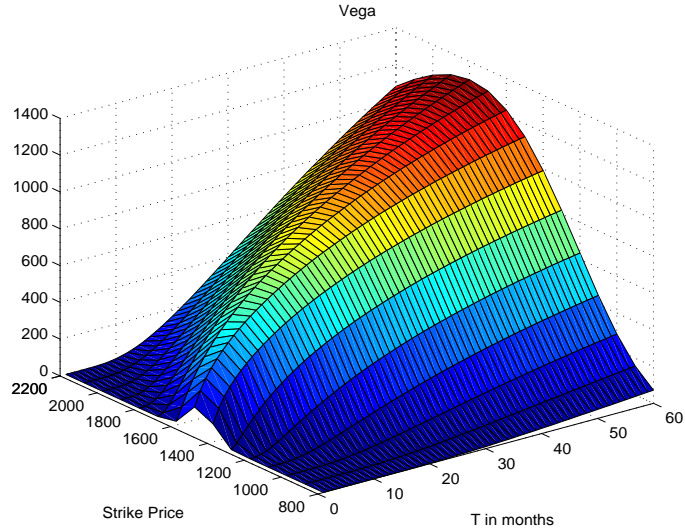}\caption{$Vega$, $\sigma=0.15$}\label{ve}
\end{figure}

\begin{figure}[h!]\centering
\includegraphics[width=4in,
height=3in]{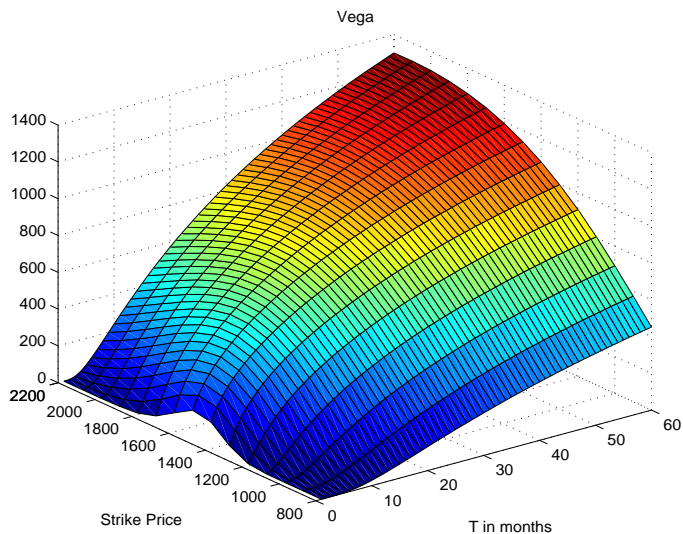}\caption{$Vega$, $\sigma=0.3$}\label{ve1}
\end{figure}

We first write out the Black-Scholes formula for European call
option,
\begin{displaymath}\begin{split}
C&=P\Phi(d_1)-e^{-rT}K\Phi(d_2),\\
d_1&=\frac{\ln(P/K)+(r+0.5\sigma^2)T}{\sigma\sqrt{T}},\\
d_2&=d_1-\sigma\sqrt{T},
\end{split}
\end{displaymath} where $r$ is the annual interest rate, $P$ is the
spot price, $K$ is the strike price, $T$ is the time to maturity,
and $\Phi$ is the standard normal distribution function. The $Vega$
is given by
\begin{equation}\label{vega}
\frac{\partial C}{\partial \sigma}=P\phi(d1)\sqrt{T},
\end{equation} where $\phi$ is the standard normal probability
density function. We pick the values $P=1459.37$, $r = 0.03$,
$\sigma=0.15$, and plot the $Vega$ as a function of both $T$ and $K$
in Figure \ref{ve}.
The strike price ranges from $800$ to $2200$ and $T$ ranges from $1$
to $60$ months. It can be easily seen that when $K$ is much lower or
much higher than the spot price, and $T$ is small, the $Vega$ is
very small. For example,
$Vega(K,T)=Vega(800,1)=7.824541295983192\times 10^{-041}$. That
means, even if there is a tiny error in the pricing of options, the
implied volatility could be very different from the true $\sigma$.
We then change the value of $\sigma$ to $0.3$ and plot the $Vega$ in
Figure \ref{ve1}, and we see a similar pattern.

In a typical research on implied volatility surface, unfortunately,
the region where the volatility skew/smile is observed coincides
with the region where  $Vega$ is small. That means, even if an
implied volatility surface is obtained which is similar to the
actual volatility surface of market data, the result is skeptical if
the effect of $Vega$ was not considered.

As a conclusion, in order to find the correct implied volatility
through the Black-Scholes formula, the computation should be
executed over the region where $Vega$ is not close to zero.
Typically this is the region where $T$ is not too small,  $K$ is
close to the curve of $Pe^{rT}$, and the true volatility is not too
small, see Figures \ref{ve} and \ref{ve1}.

\end{document}